UNIVERSITY OF CALIFORNIA

Los Angeles

Machine Learning in Materials Science---A case study in Carbon Nanotube field effect

transistors

A dissertation submitted in partial satisfaction of the requirements for the degree Doctor

of Philosophy in Materials Science and Engineering

by

Shulin Tan

2024

ABSTRACT OF THE DISSERTATION

Machine Learning in Materials Science---A case study in Carbon Nanotube field effect transistors

by

Shulin Tan

Doctor of Philosophy in Materials Science and Engineering

University of California, Los Angeles, 2024

Professor Dwight Christopher Streit, Chair

Carbon Nanotube has long been seen as a promising candidate for high-performance electronic material, yet its unique 1D structure leads to challenges in device fabrication. Many processing approaches have been proposed to produce better performing CNTFETs and this explosion of data needs an efficient way to explore. In this thesis, I explored the use of several machine learning techniques, including neural networks, simulation-based inference, and generative flow networks, on predicting CNTFETs performance, probing the conductivity properties of CNT network, and generating CNTFETs processing information for target performance.

In the beginning, I built up a neural network model for CNTFETs. I begin my work with simple cases where only certain continuous parameters like gate length are considered and developed a data cleaning method. It was shown that neural networks can work as a model for CNTFETs and reasonably perform as a device predictor for symmetric field effect transistors. I've also developed a neural network model that can incorporate processing information using encoding technique. The model can predict the performance of CNTFETs with various choices of processing methods and material combinations.

At the same time, I explored the conduction properties of non-aligned CNT networks. I built up a compact model for CNTFETs built on non-aligned CNT networks and used simulation-based inference to extract key parameters to fit the model to the experimentally observed data since extraction is impossible through traditional methods. The model with extracted parameters can fit well with the observed data. We show that simulation-based inference can be a powerful tool for building models in cases where a distribution, rather than a certain value, will be the result.

In the last step, I developed a generative model to generate device performance with target current performance. I first built a model to generate three key parameters and built the research on a compact model. The results show that this model can successfully generate multiple solutions that meet the goal. I've further developed a generative model that can generate device processing information at the same time. Though further improvement will be needed, some of the targets are met.

I hope my work can show the ability of machine learning to solve some of the material science problems. Neural network can be a good function approximator for experimental observations, though it doesn't provide understanding of the phenomenon. If probing of mechanism will be needed, simulation-based inference can be a good way to test human-created models and automatically generate parameters that humans can compare with experimental observations later. This is especially useful when the experiment input or result is a random variable described through the probability mass function or the probability density function. Generative models might be a way for experimental optimization, especially for engineering works like device fabrication, which usually requires testing out different combinations of parameters.

The dissertation is uploaded here just to protect my ideas.

University of California, Los Angeles

2024

**Dedication**

To my parents, Jun Tan and Xuxiu Zhuang

and everyone who is working on applying Machine Learning in their own field.

Table of Contents

# List of Abbreviations

Adam: Adaptive Moment Estimation

AFM: Atomic Function Microscopy

CNT: Carbon Nanotube specifically single-walled Carbon nanotube in this thesis

CNTFET: Carbon Nanotube Field Effect Transistor

GAN: Generative Adversarial Network

GFLowNet: Generative Flow Network

ML: Machine Learning

MOO: Multi-objective Optimization

MSE: Mean Squared Error

NN: Neural Network

PMF: Probability Mass Function

PDF: Probability Density Function

RL: Reinforcement Learning

SBI: Simulation-based Inference

SGD: Stochastic Gradient Descent

SWNT: Single-walled Carbon Nanotube

# List of Figures

# List of Tables

# Acknowledgement

First, I want to thank my advisor, Dr Dwight Streit. He gave me the chance to do research when I couldn't find anyone who's willing to support me, and had encouraged me so many times when I had no confidence of myself. We have also spent a lot of time discussing ideas and experimental results. I really admire his insights and knowledge in this field. Thank you so much, Professor Streit.

I want to thank my parents, Jun Tan and Xuxiu Zhuang, for their endless support, both financially and mentally. I can't remember how many times I broke down because I can't find a good idea or my experiment failed. They always support me, comfort me and try their best to help me. I'm so glad to be your child.

I also want to thank Dr Emmanuel Bengio for the discussion on GFlowNet. He helped me review the experimental setup and gave me suggestions to build a stack of environments. I would also like to thank for the discussion with Prof. Kang Wang for giving out suggestions to my thesis about the random variable part. He helped me put a better theoretical foundation of my work. I also would like to thank Prof. Yahong Xie for his suggestions to add discussion about avoiding overfitting and underfitting. This makes my work more complete.

At the same time, I really want to thank people in the West Coast Machine Learning Group, Sandiego Machine Learning and AI Frontiers Forum, including but not limited to Ted Kyi, Dr. Junling Hu, Jerry Kurata, Roger Stager, Dev, Dr. Julius Smith and David Selinger. I learned so much from them about machine learning techniques, and I had a great time discussing ideas with them.

In the end, I would like to thank my friends who supported me during my pursue of PhD. I want to thank my friends in the Board Game Night group of the ECE department, including but not limited to: Wojciech Romaszkan, Lev Tauz, Ankur Mehta, Alexander Johnson(AJ), Debarnab Mitra, Vivian Dao, Abdullatif Jazzar, Benjamin Domae, Richard Lin (Ducky), Christopher Chen, Joseph Hwang  Christopher Liu. I spend a lot of time playing board games with them, which I would call one of the best time of my life. I also want to thank my roommates and the friends I made along the way, including but not limited to Meixu Su, Ruimiao Wang and Jingzhu Kong. They gave me a lot of support during my PhD. I also want to thank people in Gradswe where I spent two years help organizing events. I have special thanks to Dr Andrew Webster, who healed my long time trauma, gave me so much suggestions about how to get along with other people and myself, and supported me through my PhD. Thank you, doctor, you made me a better person.

Thanks for so many people who showed me that the world has its kind side.

P.S. Special thanks to my cat, Mr. Quantum (Mr. Q), for offering his belly, fur and paws whenever I need it.

VITA OF SHULIN TAN

December 2024

EDUCATION

Bachler of Material Physics, Nanjing University, June 2016

Master of Science in Materials Science and Engineering, University of California, Los Angeles, June 2019

Doctor of Philosophy in Materials Science and Engineering, University of California, Los Angeles, Expected December 2024

PPROFESSIONAL EMPLOYMENT

UNIVERSIT TEACHING

2019 Fall: Teaching Assistant, Department of Asian Languages and Cultures, University of California, Los Angeles

2020 Winter: Teaching Assistant, Department of Materials Science and Engineering, University of California, Los Angeles

2020 Summer: Teaching Assistant, Department of Asian Languages and Cultures, University of California, Los Angeles

2020 Fall: Teaching Associate, Department of Life Science, University of California, Los Angeles

2020 Winter: Teaching Associate, Department of Materials Science and Engineering, University of California, Los Angeles

2020 Spring – 2022 Spring: Teaching Associate, Department of Life Science, University of California, Los Angeles

PUBLICATIONS

"A Physics-Based Neural Network Carbon Nanotube FET Model" DGM-AIMSE 2023

"Probing CNT network conductivity with Simulation-based Inference"AIMS Workshop July 2024

# Chapter 1

# Introduction and Goals

## 1.1 Introduction

With the rapid development of science, the amount of knowledge and data has exploded, and its speed has surpassed human's ability to learn.[1]–[4] Therefore, a more efficient method must be developed for processing experimental data, building models, and planning future research. Machine learning (ML) has been seen as an effective tool for scientific discovery because of its ability to process large amounts of data. It has already seen success in protein structure prediction,[5][6] drug discovery,[7] and quantum physics.[8]

Since their discovery in 1991,[9] carbon nanotubes (CNTs) have caught the eye of many researchers. Because of their high charge carrier mobility,[10] semiconducting CNTs have long been seen as a candidate to save Moore's law, and extensive research has been done on making devices from them.[11]–[17] Applications based on CNT devices like logic circuits[18]–[21] and radio-frequency circuits[22]–[26] have also seen mass research. Extensive studies on CNT devices have led to an explosion of data, and some of its properties are still not fully explained.

To facilitate the development of the CNT device and explore its properties, we applied some machine-learning techniques to this material to predict its properties, explore the CNT transport mechanism, and design experiments with CNT devices. This thesis aims to explore ways of combining machine learning techniques with materials science and use this technique to probe problems that are hard to solve with traditional methods.

## 1.2 Challenges and Opportunities

**Faster modeling of CNTFET performance and incorporating device processing information:** The development of models for new devices usually takes decades, and processing method selection, which is associated with device interface properties, is difficult to consider in traditional models. A new modeling method to generalize this information could help future researchers.

**Modeling the distribution of the performance of CNTFET with non-aligned CNTs:** Though models have been proposed for electrical conductance in random CNT networks, the exact resistance at the CNT-CNT junctions is hard to extract. Research on field-effect transistors based on it has shown a distribution of on and off currents, yet no model exists to explain it.

**Generating CNTFET processing information:** Though models have been built to predict device performance, no models have been proposed to create device process information for target design. A model could be constructed to make use of the explosion of new data and produce suggestions for future research.

## 1.3 Thesis structure

Within each part of this thesis, the chapters progress as described below

Chapter 1 provides an introduction to the thesis, outlines possible research directions, and describes the Chapter 2 provides an overview of the development of carbon nanotube (CNT), including its structure, characterization methods, current development of CNT processing, and development of carbon nanotube field effect transistor (CNTFET).

Chapter 3 introduced key concepts in machine learning and some cutting-edge techniques used in this thesis. It also briefly introduces probability theory and how it can be a new way of tackling scientific problems.

Chapter 4 shows how neural networks can model CNTFET performance and how processing information can be incorporated into this modeling.

Chapter 5 builds up a model for CNTFETs with random distributions and explores using simulation-based inference as a parameter extraction method for models with distribution as an output.

Chapter 6 builds up a generative model for CNTFETs and explores generating processing information with a targeted I-V curve.

# Chapter 2

# Background

## 2.1 Structures and basic properties of Carbon Nanotubes

With the approaching physical limit in silicon transistors, researchers in the semiconductor industry have been worrying about the end of Moore's law and keep looking for substitute materials. Carbon nanotubes (CNTs) have long been seen as a hopeful candidate because of their high charge carrier mobility, but electronics based on them are still far from being used in real life. Many factors, like semiconducting CNT sorting and a combination of fabrication methods, still restrict CNT device production. The conduction property of CNT is also still not fully understood, especially in the case of non-aligned CNTs.

**Structure of CNTs**

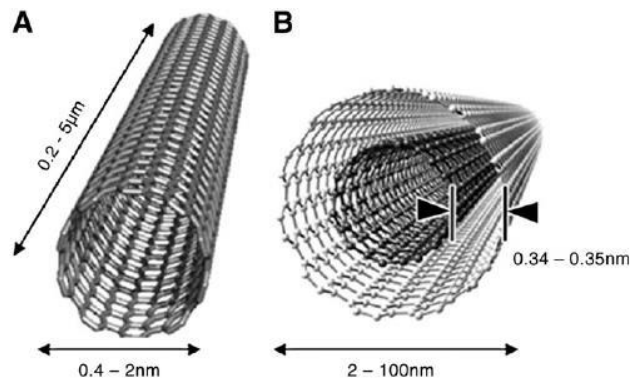

(Fig 2.1: Carbon Nanotube structure. A: single-walled Carbon Nanotube B: Multi-walled carbon nanotube. We can see that a single-walled carbon nanotube only consists of one layer of Carbon, and the arrangement of atoms resembles that of graphene.)

CNTs can be seen as graphene rolled up as a tube.[27] Based on the number of graphene layers the tube has, CNTs can be divided into single-walled carbon nanotubes (SWCNTs) and multi-walled carbon nanotubes (MWCNTs), with their structure shown in Fig 2.1. MWCNTs typically show no gate modulation since only the outermost carbon layer is involved in its electron transportation,[28] making them unsuitable for being fabricated into field-effect transistors, so we only focus on SWNTs in the rest of this thesis.

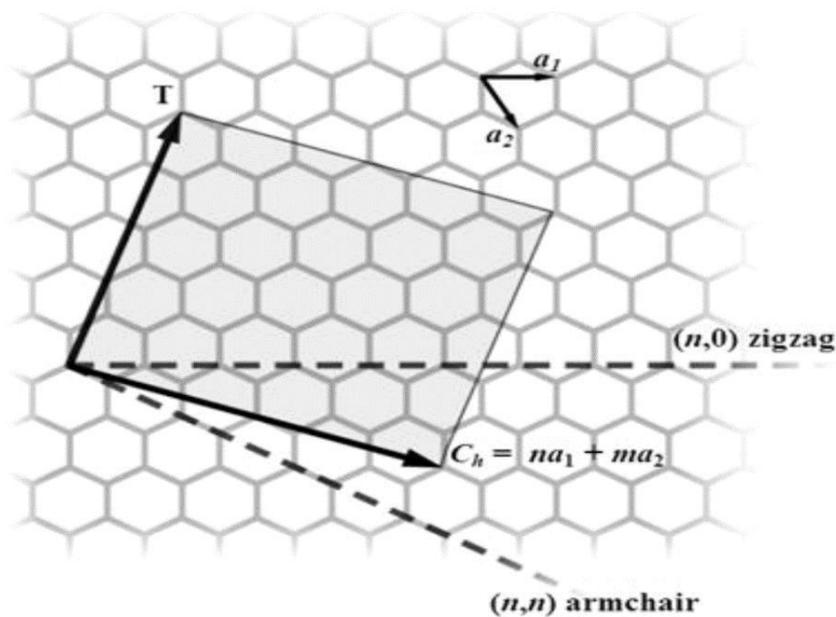

(Fig 2.2: Chirality of CNTs. $a_1$ and $a_2$ are the lattice vectors of a graphene, and $C_h$ is the direction of CNT roll up, whose value is also the perimeter of the CNT.)

The electronic properties of CNTs are determined by their chirality, which is the direction of the roll-up of graphene. If we denote the two lattice vectors of a graphene as $a_1$ and $a_2$, we can express the structure of a Carbon nanotube as $C_h = na_1 + ma_2$, where n and m are the number of the chiral vectors involved. The charity of CNTs determines the bending of C-C bonds and the alignment of carbon atoms, thus determining the electrical properties of CNTs.

Based on chirality, CNT can be either semiconducting or metallic. If 2n+m=3q (where q is any integer), the CNT is metallic, while the CNT is semiconducting in other cases. We can see that theoretically, only two-thirds of the SWCNTs are semiconducting, and semiconducting CNTs need to be sorted out before being fabricated into electronic devices since a metallic carbon nanotube may cause a circuit shortcut. The chirality of CNT also determines the diameter of it and thus determines its bandgap.[27] The diameter of a single-walled carbon nanotube is

$\frac{\sqrt{3}}{\pi} a_{c-c} \sqrt{m^2 + m \cdot n + n^2}$, where $a_{c-c} = 0.14 nm$ is the Carbon-Carbon bond length in graphene. For semiconducting CNTs, their diameter also determines their bandgap, which goes as

$$E_g = \frac{2 E_p a_{cc}}{d}$$

## 2.2 Characterization of CNTs

**Raman Spectroscopy – CNT diameter and type**

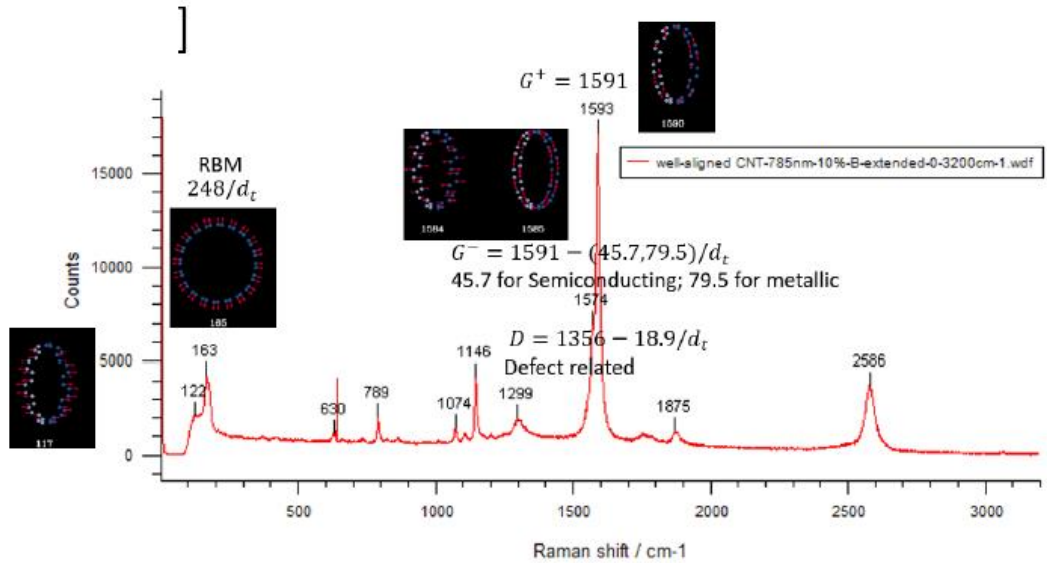

(Fig 2.3: Typical Raman spectroscopy of SWCNT, vibration modes cited from [33])

Raman spectroscopy is the typical way of characterizing CNT diameter and types.[30]−[32] Diameter of CNT can be seen from the radio breathing mode (RBM) peak, which is the coherent vibration of the C atom in the radial direction, as if the tube is breathing. RBM occurs in a frequency range of 120-350$cm^{-1}$ for nanotubes with a diameter of 0.7nm-2nm. The association between the diameter of CNT and the RBM peak is $248/R_t$, where $R_t$ is the diameter of the carbon nanotube. The G band at ~1590$cm-1$ is a good indicator of the existence of carbon nanotube because of its high intensity and is useful as an indicator of CNT existence. The G band originates from the vibration of the C-C band along the nanotube direction. Graphene has a similar Raman spectra peak, so other peaks need to be considered to determine whether the band is graphene, such as the G' band. The G' band around 1570 $cm-1$ is associated with the vibration along the circumferential direction along the CNT. This band is also very strong, and it is a good indicator of whether the material is carbon nanotube or other types of carbon materials.

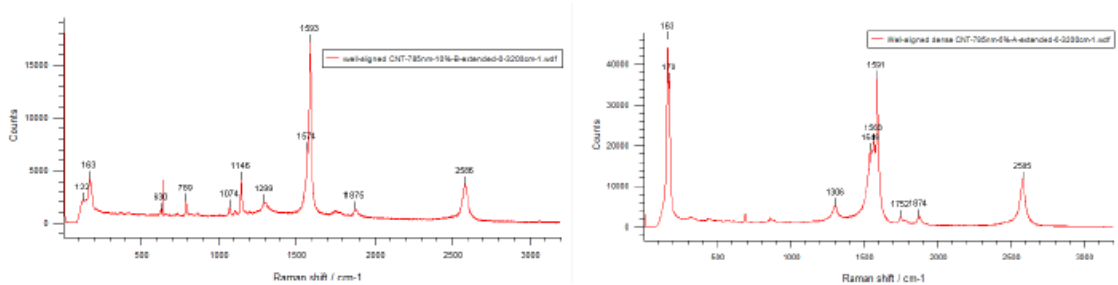

(Fig 2.4: Raman Spectra of semiconductive CNT (left) Metallic CNT (right))

Atomic force microscopy (AFM) is usually the best way to characterize spectroscopy of CNTs. Techniques like conducting AFM can also be used to characterize CNT conducting properties. Scanning electron microscopy (SEM) is also a good way to characterize CNTs.

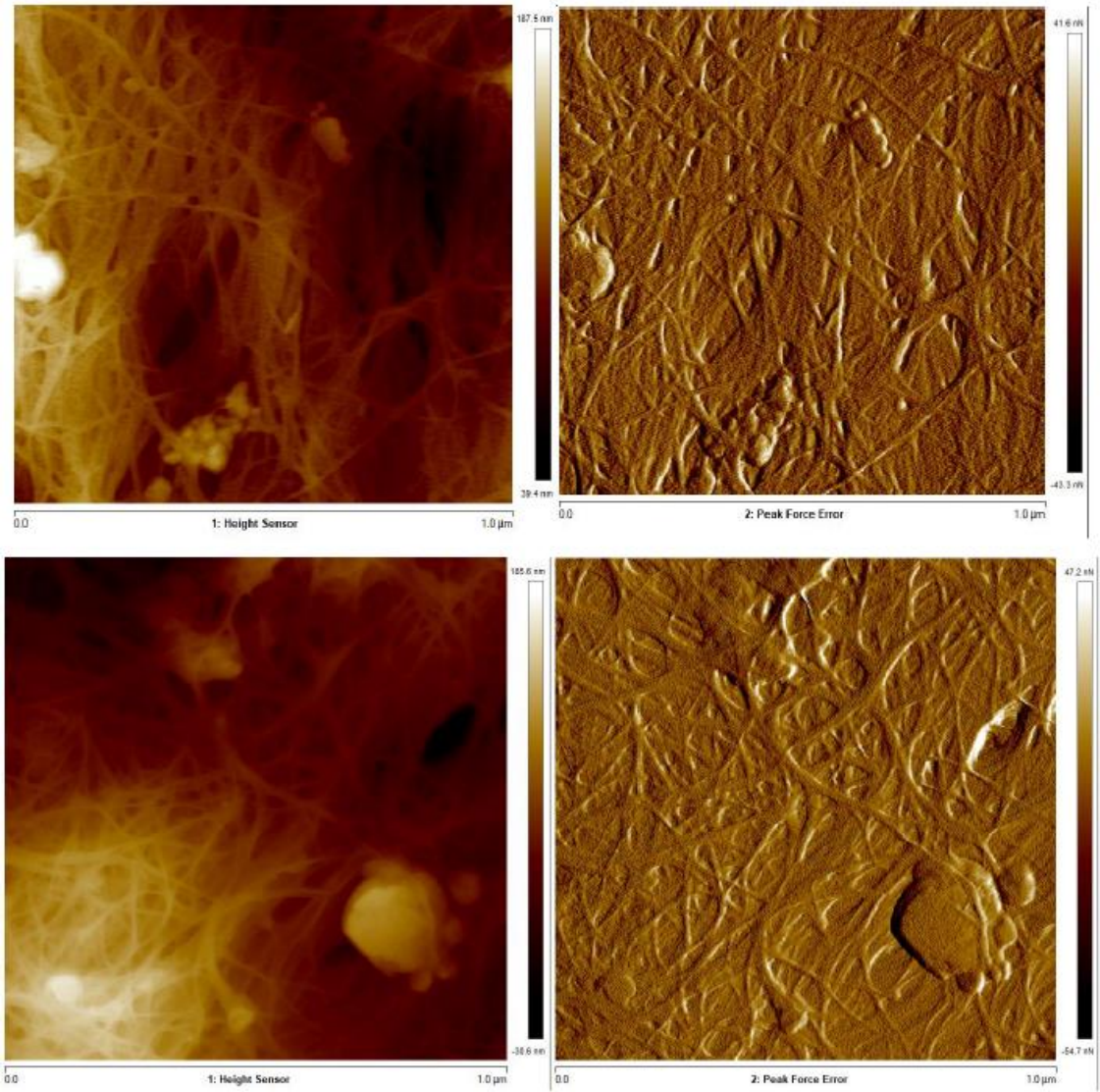

(Fig 2.5: AFM image on CNT deposited on Si substrate)

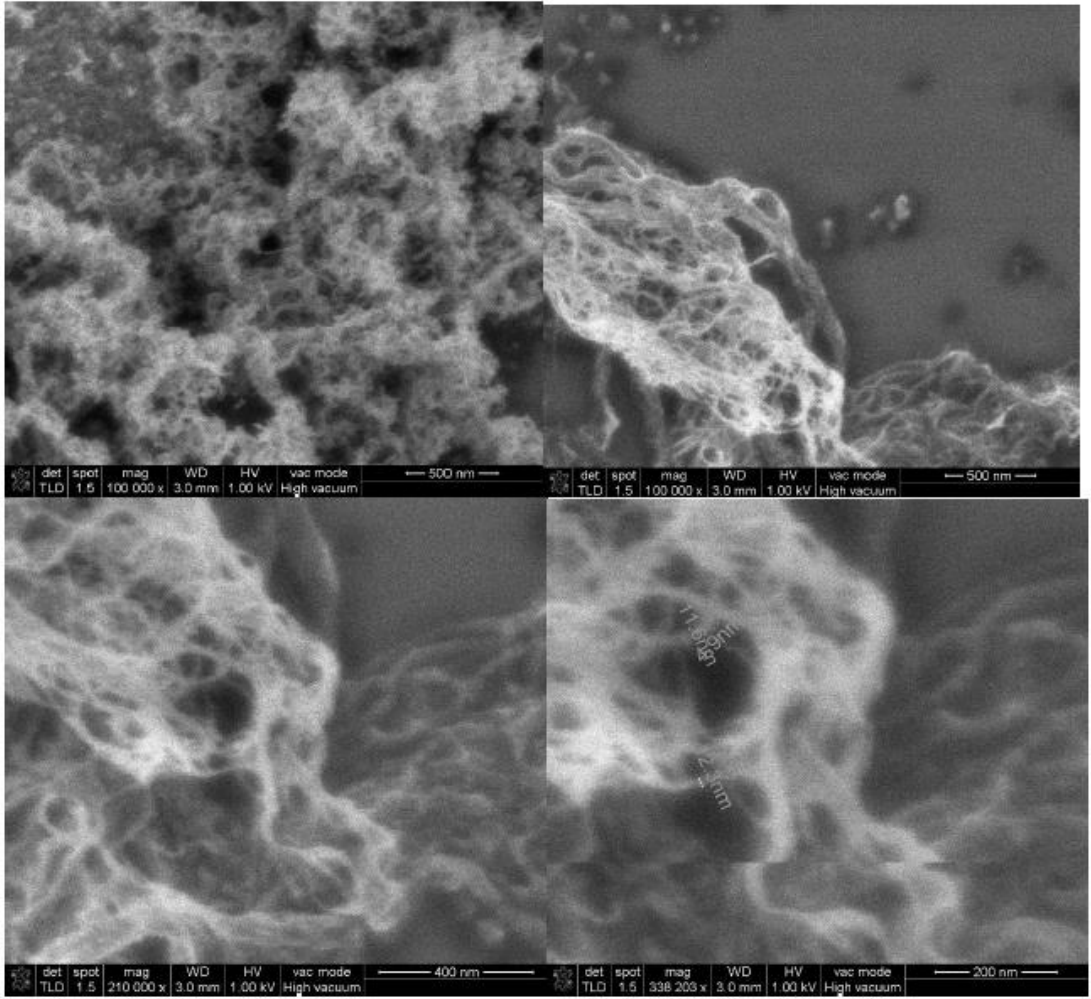

(Fig 2.6: SEM image on CNT powder)

## 2.3 Carbon Nanotube Field effect transistors (CNTFETs)

Since CNTs are predicted to have much better conductivity than silicon, they have been a focus for substitute semiconducting materials for electronics. Early research has shown that semiconducting carbon nanotubes can exhibit great charge carrier mobility and be made into ballistic electronic devices. Ballistic electronic devices made from carbon nanotubes have experimentally demonstrated superior electron and thermal conductivity and show similar I-V characteristics to traditional semiconductor electronics. The mean free path of a single-walled carbon nanotube is estimated to be larger than 1μm.

Most CNTFETs use the MOSFET structure, since the chemical doping of carbon nanotube is unstable. Compared with silicon, CNTFETs take advantage in the following aspects:[100]

1. Lower passive power consumption. Passive power consumption denotes the energy consumed by transistors when they are in the off state with $V_{gs} = 0\ V$. It grows rapidly with the reduction of device size. An effective way to reduce passive power consumption without harming on-state current is to increase the charge carrier velocity of the channel material.[101] Compared with silicon, which has a saturated charge carrier velocity of $1 \times 10^7\ cm/s$, carbon nanotube has a much higher one measured $3 - 4 \times 10^7\ cm/s$ [101] with a gate length of 10-15nm.
2. Less short-channel effects. A critical limiting factor for silicon-channel MOSFET scaling is the short channel effect, where charge carriers in a MOSFET with too short channel length $L_g$ may directly penetrate underneath the depleted region underneath the gate between the source and drain, and the gate electrode fails to control the channel. The drain-induced-barrier-lowering (DIBL) effect may also appear in nanometer-size devices, where threshold voltage $V_{th}$ becomes reliable on drain bias. One way to reduce the short-channel effects is to reduce the channel thickness along with the decrease of $L_g$,[103] where CNTs with a diameter of 1-2nm have an obvious advantage. FinFET structure may also suppress the short-channel effects and enable further scaling of Si-based MOSFET $L_g$. However, quantum confinement appears when scaling $L_g$ to 10nm, where conduction and valence bands are separated into subbands and thus widens the effective bandgap.[104] The greater confinement of electrons with thinner films also leads to enhanced scattering and decreased the charge carrier mobility.

3. Shorter contact length. Metal contact in MOSFETs must be long enough to efficiently collect charge carriers. Si or II-V transistors' metal contact is connected to the semiconductor through relatively weak van der Waals interactions. However, the CNT under metal contact is usually open-ended since they were etched to the channel length before metal contact deposition. These quasi-zero-dimension open ends can be directly welded to the metal contacts where the metal atoms of the contacts and the carbon atoms of the CNTs are bonded directly.[106] This strongly coupled interface enables charge carriers to be collected more efficiently.

An easy way to express the behavior of CNTFET is through the virtual-source compact model, where the source-drain current goes as:

$$I_{dS} = Q_{xo} v_{xo} F_S$$

Here, $Q_{xo}$ is the charge carrier density in the channel, which is affected by the capacitance of CNTFET. $v_{xo}$ is the virtual source velocity of charge carriers, which is affected by the gate length and the diameter of CNTs. $F_S$ is a shape factor of the current output associated with the drain-source voltage and gate-source voltage. This factor reflects the effect of electric field distribution on the conductance of CNTFETs.

The capacitance of CNTFET is a combination of gate oxide capacitance and CNT capacitance. Since the gate oxide capacitance and CNT capacitance are in parallel, the total capacitance of CNTFET goes as

$$\frac{1}{C_{inv}} = \frac{1}{C_{ox}} + \frac{1}{C_{qe}}$$

The gate oxide capacitance $C_{ox}$ is a function of its thickness $d$ and its dielectric constant $k_{ox}$, which is a property of the gate oxide material chosen. An easy case of gate oxide is the cylinder oxide gate, that we assume that gate oxide is deposited evenly around the CNTs. The gate capacitance will go as

$$C_{ox} = N \times \frac{2\pi k_{ox}\varepsilon_0}{ln[(2t_{ox} + d)/d]}$$

The capacitance of Carbon Nanotube $C_{qe}$ is mainly a function of the bandgap of CNTs, which is associated with the diameters of CNTs.

$$C_{qe} = N \times \left[0.64\sqrt{E_g} + 0.1\right] (fF/\mu m)$$

The virtual source velocity $v_{xo}$ is mainly affected by the gate length. The gate length is the length of charge carriers that flow from source to drain. Though theoretically, charge carriers in CNTs can do ballistic transport, in reality, charge carriers will inevitably be scattered by factors like surface defects or CNT defects. Charge carrier velocity is usually used to describe this phenomenon, which goes as:

$$v_{xo} = \frac{\lambda_v}{\lambda_v + 2L_g}$$

$$v_B = v_{B0}\sqrt{d/d_0}$$

Where $v_B$ is the carrier velocity in the ballistic limit, and $\lambda_v$, $v_{B0}$ and $d_0$ are empirical parameters. The charge carrier mobility is

$$\mu = \mu_0 \frac{L_g}{\lambda_\mu + L_g}\left(\frac{d}{1nm}\right)^{c_\mu}$$

Where $\mu_0 = 1350 cm^2/V \cdot s$, $\lambda_\mu = 66.2 nm$, and $c_\mu = 1.5$ are empirical parameters.

Another thing to be taken into consideration is the metal contact.

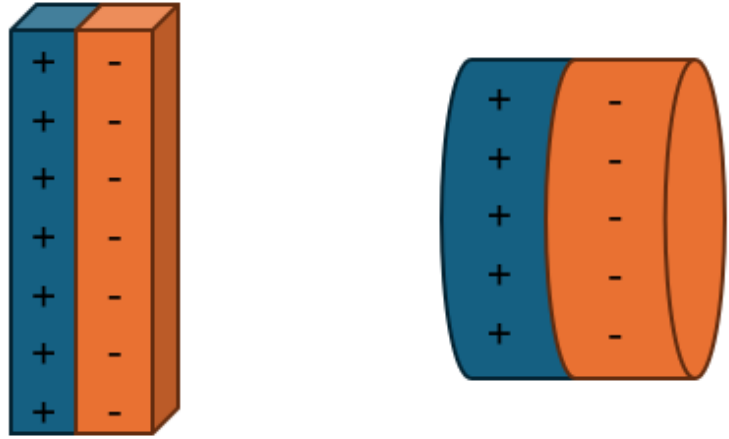

(Fig 2.7: Dipole structure of Si and CNT metal contacts. Left: dipole sheet in Si metal contact; Right: dipole ring in CNT metal contact)

It is usually considered that the work function difference between metal and semiconductor determines the behavior of metal-semiconductor junction. The surface at the metal/semiconductor interface introduces boundary conditions, creating metal-induced gap states (MIGS) in the middle of the semiconductor band gap which decay exponentially away from the interface. Compared with traditional 3D semiconducting materials like Si, the MIGS charge takes the form of a dipole ring rather than a dipole sheet in CNTs. This creates a difference for metal-CNT contact since the electrostatic potential is a constant far from a dipole sheet, but decays as the third power of distance far from a dipole ring.[107] For a typical CNT with a bandgap of 0.6eV, and for the CNT mid-gap 4.5eV below the vacuum level, metal work functions larger than 4.8eV (or less than 4.2 eV) would thus lead to a negative Schottky barrier, i.e., the metal contacts the CNT in the valence (conduction) band, giving an Ohmic contact. Thus, one may expect that gold (Au) and Palladium (Pd) would give Ohmic contacts. For CNT transistors with Pd contact, the device conductance is close to the maximum conductance of $4e^2/h$, indicating that no barrier exists at the contact. For Au, the as-deposited metal contact will behave like a Schottky barrier but will resume as an Ohmic contact after annealing. This is likely due to the poor wettability of Au on CNTs. Most n-CNTFETs fabricated in recent years choose Palladium as metal contact materials and deposit Au on top of it.[108] fabricate p-FETs, the work function of Sc is more applicable.

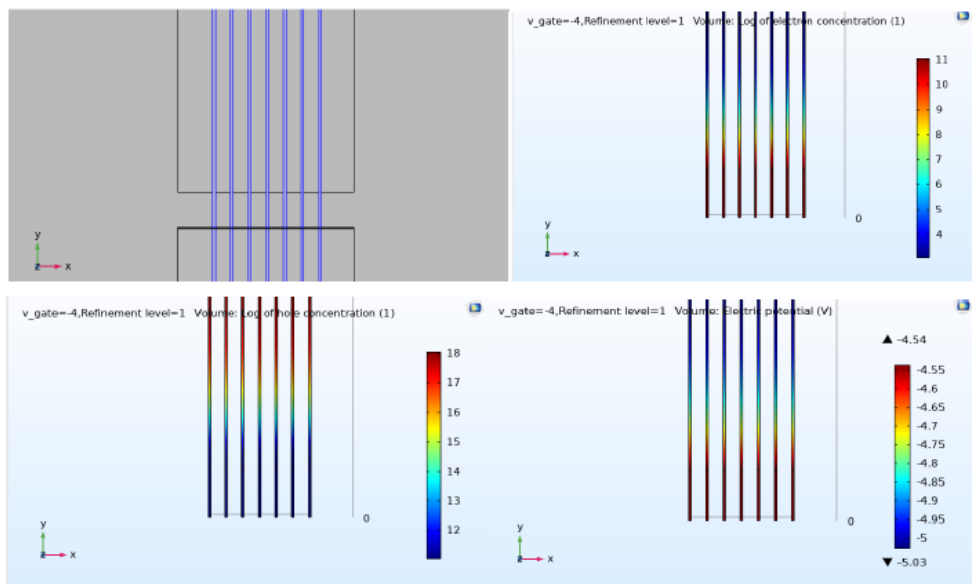

(Fig 2.8 COMSOL modeling of CNTFETs near source-gate contact region)

At the same time, the development of CNTFETs faces a lot of problems, and one of them is the choice of device structure. Unlike other semiconducting materials, CNT is just one layer of atoms and is around 1nm in diameter. This super-thin body means that the electric field in CNTs is almost the same as that of the interface around it. As the simulation shows in Fig, if there is a gap between metal contact and gate oxide, which means gate length $L_g$ is smaller than channel length $L_{ch}$, then there will be an abrupt change in electric voltage for the CNTs in this gap. This may lead to extra resistance, so many device structures were proposed. Some choose to leave no gap between the metal contact and the gate, while some apply additional gate material only around source-gate.

**Sortation of semiconducting CNTs**

One of the challenges in CNTFET production is the separation of semiconducting CNTs since as-synthesized CNTs are a mixture of metallic and semiconducting CNTs. A commonly used way is to apply surfactants like sodium dodecyl sulphate (SDS) [34]–[35] and poly[9-(1-octylonoyl)-9H-carbazole-2,7-diyl] (PCz))[36] which attach on CNTs through hydrophobic or $\pi - \pi$ interactions. The difference in chirality affects the number of surfactants CNTs are encapsulated with, so CNTs can be separated after centrifugation. This technique can selectively sort semiconducting CNTs over a diameter range of $0.7 - 1.6 nm$.

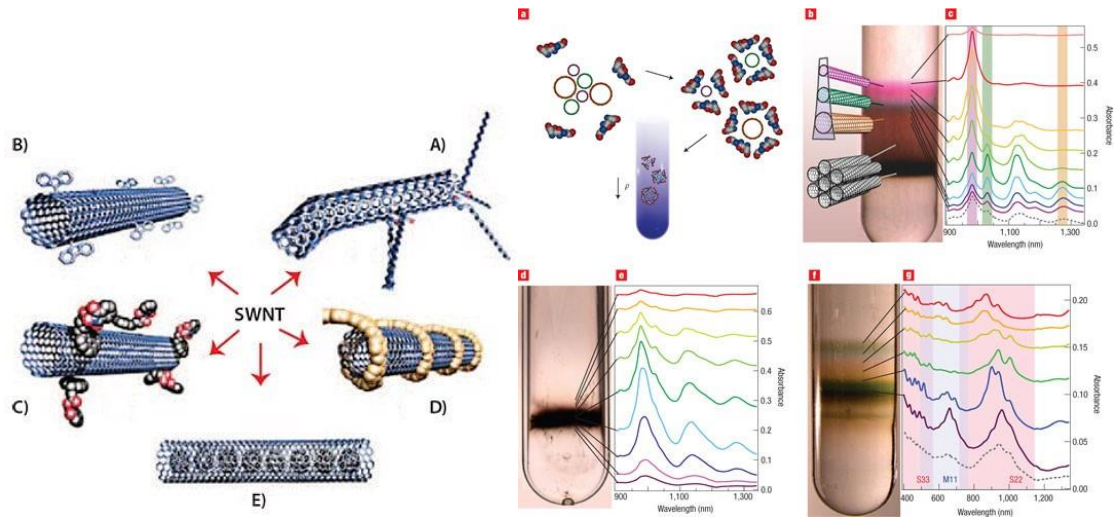

(Fig 2.9: Surfactant sorting of CNTs (a): Surfactant wrapping of SWNTs, (b): separation of CNT solution after centrifugation[37])

**Post-treatment of CNTs**

Some surfactants may remain on CNT after cleaning and may affect CNT performance. To further clean CNTs, Yttrium Oxide Cleaning (YOCD)[39] has been proposed to clean remaining surfactant by first deposit around 2.5 nm Yttrium Oxide and then remove it with HCl solution followed by repeated rinsing in OPA. This step is usually done after the deposition of CNTs.

**Deposition of CNTs and alignment**

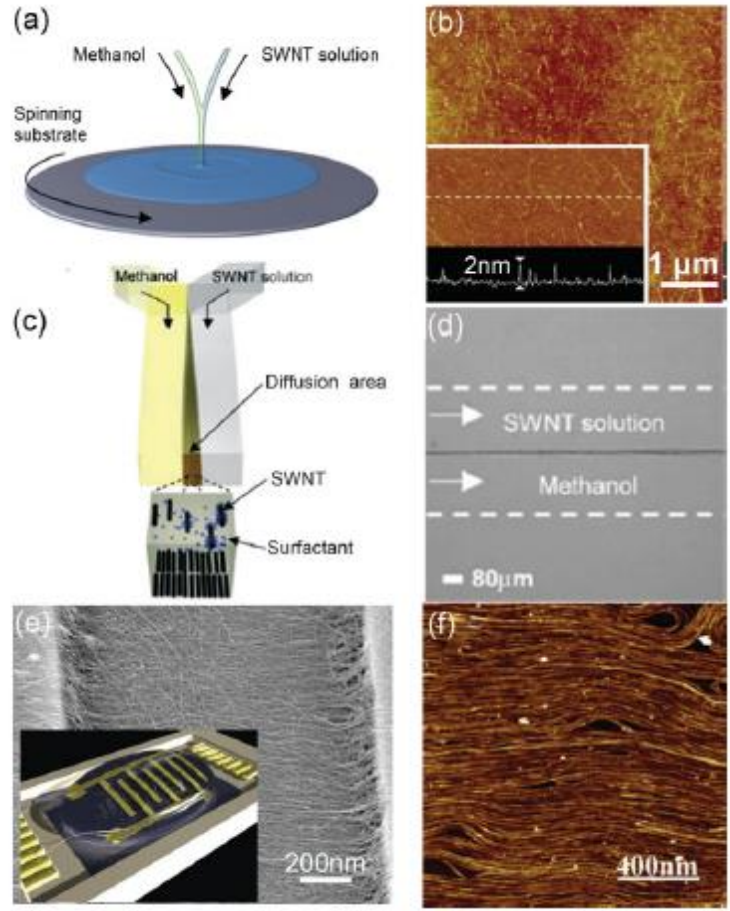

(Fig 2.10: Deposition of CNT films (a) Illustration of spin-casting method of CNT film; (b) resulting film from spin-casting. We can see from this AFM image that there's no alignment of CNTs deposited. (c) deposition of CNTs through Langmuir–Blodgett. CNTs are deposited through the water-substrate interface. (d) interface of water-substrate (e)(f): SEM and AFM films of aligned deposited CNTs)

At the same time, the carbon nanotube is a 1D material, which means electrons can only propagate and be reflected in one direction. Therefore, the positional distribution, along with the length of CNTs, has a significant effect on CNTFET performance. If the CNT solution is spin-casted on the substrate, the deposited CNTs are usually randomly distributed. A useful way to deposit aligned CNTs is through the Langmuir–Blodgett (LB) technique. Silicon substrates are inserted in water, and the CNT solution is dropped near the water-substrate interface. The surface tension aligns CNT at the interface, so when the substrate is pulled out of water, CNTs are left aligned on its surface. A variation of this method is the dimension-limited self-alignment method (DLSA)[42] which further improves the density of CNT deposited.

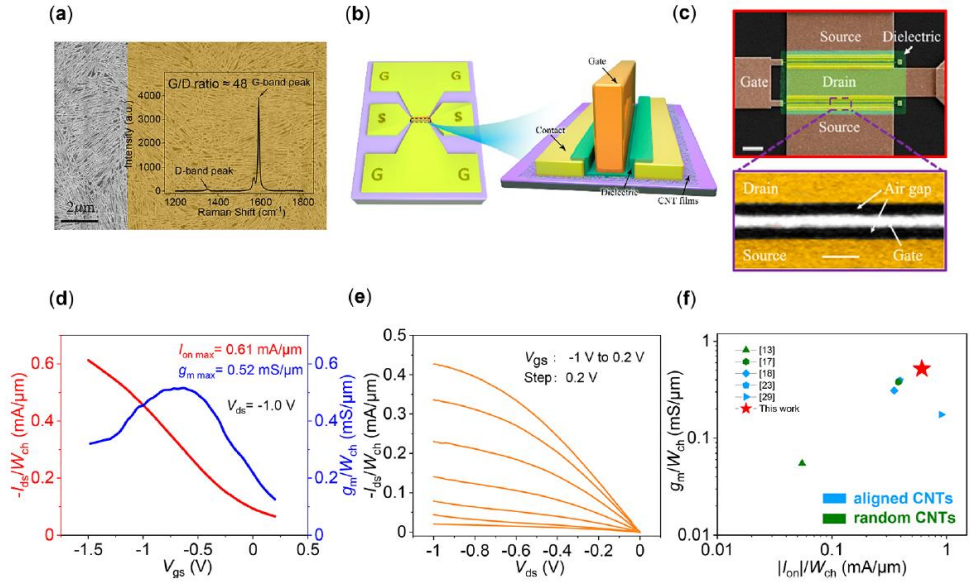

(Fig 2.11: Example of CNTFET built on non-aligned $CNTs$[43])

However, CNT aligning usually takes hours, and the resulting CNT density is restricted, so some devices are made with non-aligned CNTs. These devices have also shown high I-V curve output and are widely applied in devices. But, at the same time, the modeling of CNT network conduction is still not understood. Experiments have shown that charge carriers can hoop between two close-by CNTs, and AFM has been conducted on it. It has been proposed to use resistance to characterize CNT-CNT junction conductance. However, a random CNT network contains hundreds of these junctions, and traditional parameter extraction methods cannot calculate the resistance of this resistance incorporated in a network. Therefore, we developed a method to tackle this problem in chapter 4 using simulation-based inference to extract parameters and validate conduction models.

**Other issues with CNTFETs**

As a nanomaterial, CNTs easily absorb molecules in air and get n-doped. P-doping of CNTs can be achieved with specific molecules, but it is usually unstable. Therefore, most CNT devices are CNTFETs. CNT also absorbs moisture in the air, and it causes large hysteresis.

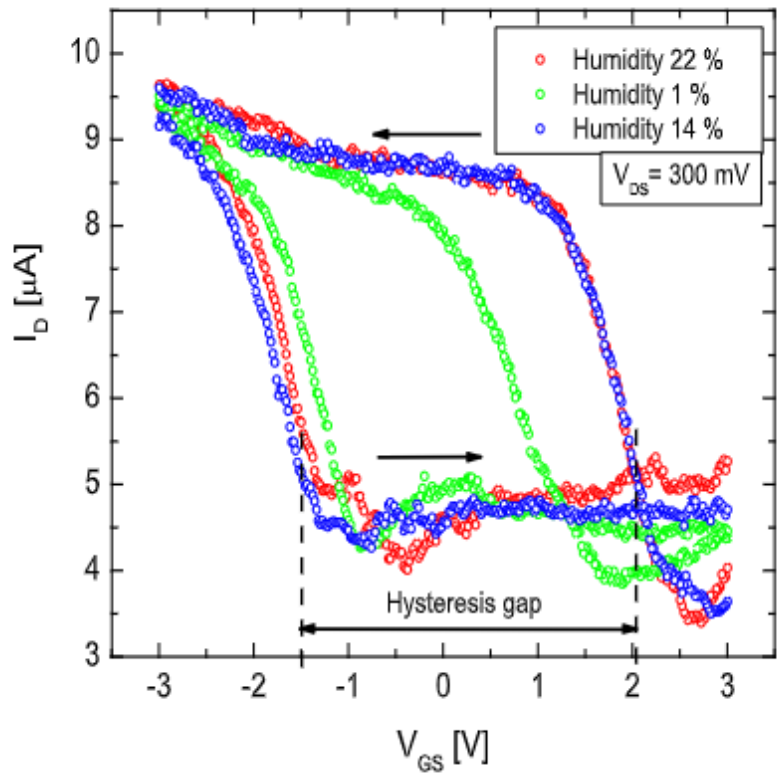

(Fig 2.12 Hysteresis effect on CNTFETs caused by moisture[44])

Metal contact is another issue with CNTFETs.[45] - [47] Since CNTs have a high working function, the metal contact for CNTs need also have a high working function up to 4.7 - 5.0 eV to form an ohmic contact with CNTs, which in most cases is Pd. However, Pd does not have good enough conductivity as a metal contact, so a gold layer is typically deposited above it to enhance conductivity. The thickness of these two metal layers may slightly impact CNTFET performance.

# Chapter 3 Introduction to Machine Learning

## 3.1 Introduction

With the rapid development of modern science, there will inevitably be an explosion of experimental data. Traditionally, scientists make observations of natural phenomena and make theories from them; however, with the development of science and the accumulation of past knowledge, the speed of knowledge accumulation already exceeds the speed for most people to master. Therefore, a more efficient method will be needed to facilitate scientific discovery. Machine learning (ML), a commonly used tool to treat large amounts of data, can be a good candidate for this problem. Viewing scientific problems as probability may also help develop ML tools for materials science problems.

## 3.2 Introduction of Machine Learning

Machine learning is the study of algorithms that improve their performance P at some task T with experience E.[48] There are three categories in machine learning: Supervised learning, Unsupervised learning, and Reinforcement learning, which differ from each other in their training tasks and methods. The difference between supervised learning and unsupervised learning is that supervised learning uses labeled data, where the input data corresponds with one or several output data sets. In contrast, unsupervised learning uses unlabeled data where only input data is involved. Data labeling can be categorical, such as whether a picture is a dog or cat, which often discriminates different inputs in classification tasks. It can be continuous data like the current flow in the device under particular bias, which is usually used in regression tasks to predict output with unknown input. Supervised learning aims to predict the output with a specific input. An example is a neural network (NN), which uses a network of interconnected units to predict output data with input data. Unsupervised learning learns the distribution of the input. A common example of unsupervised learning is the large language model (LLM), which predicts the probability of the next word given the previous context.

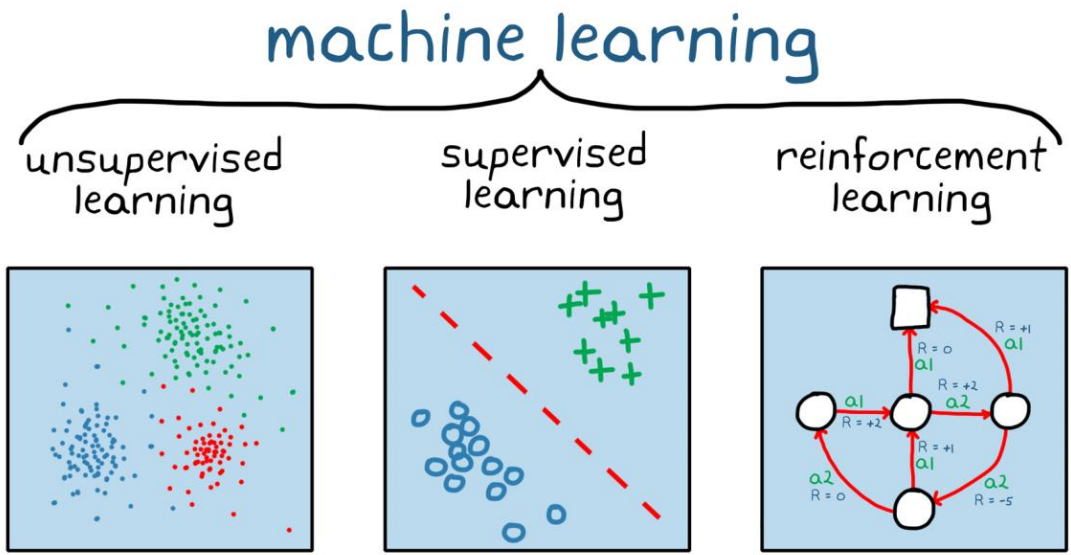

(Fig 3.1: Three categories of machine learning. In the unsupervised learning part, we can see that the model collects data with similar traits together. In supervised learning, the learning is done by taking actions)

However, reinforcement learning (RL) trains a model to make decisions to maximize rewards in an environment to achieve the most optimal reward. The problem studied in RL is set up as an environment that rewards different actions, and an agent is created to take a series of actions in the environment and learn the reward. The model of action taken and reward is called policy. RL algorithms use a reward-and-punishment paradigm as they process data. They learn from the feedback of each action in the policy model and discover the best processing paths to achieve final outcomes.

**Neural Network**

Deep feedforward neural networks, called feedforward neural networks or multilayer perceptrons (MLPs), are the quintessential deep learning models. Its name, neural network, comes from its original idea to mimic the human brain system.[51] The essential components of the neural network are perceptrons, or called neurons, which multiplies the income signal $x$ with weights $w$, add bias $b$, and pass the result through a step function $h$ to get an output value $f(x)$.

$$f(x) = h(w \cdot x + b)$$

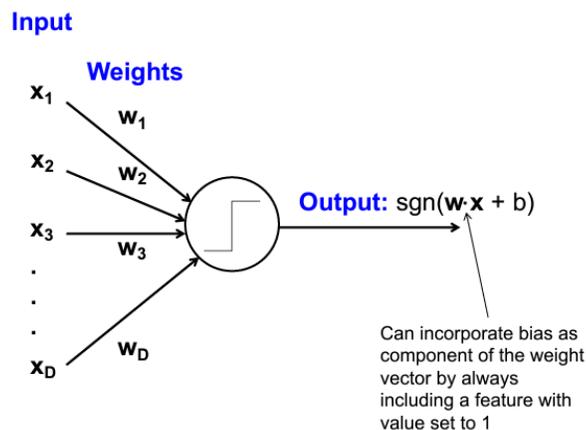

(Fig 3.2: Structure of neuron)

Though a single neuron has a limited ability to process data, an interconnected system of thousands of neurons can represent complex functions. In a deep neural network, several layers of neurons are used, and information passes from one layer to the next. The layers consist of one input layer receiving inputs, several hidden layers as information processors, and a final layer called the output layer, which gives output predictions. When predicting, the information flows only in the direction from the input layer to the output layer, so these models are called feedforward models. Neural networks can behave as complex functions because they are typically represented by combining many different functions. For example, in three layers of the neural network, we may have their functions as $f^1$, $f^2$ and $f^3$ connected in a chain, and the function of them linked together will be $f^3\left(f^2(f^1(x))\right)$. With proper choice of hyperparameters like layer and neuron numbers and a good structure of neuron connections, neural networks can be used almost as a predictor or a classifier. Though the explainability of neural networks is still under research, it is widely used as a key component in many other ML techniques.

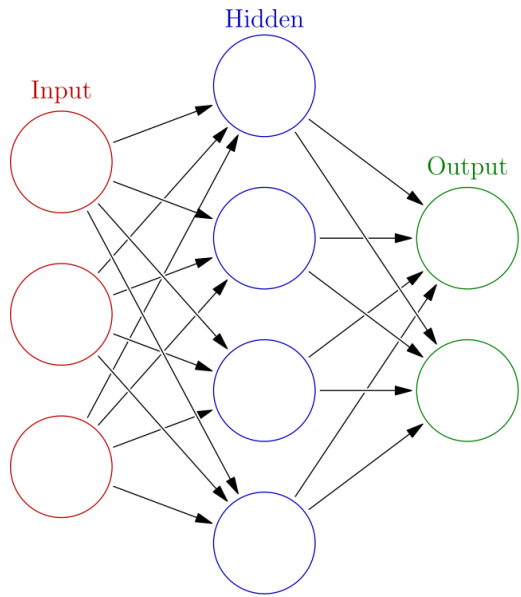

(Fig 3.3: Structure of neural network)

The training of neural networks is done by backward propagation, a gradient estimation technique that works by moving backward from the output layer to the input layer. During training, a labeled data set is used as an example to teach the neural network by letting the neural network predict outputs based on the input data. The difference between predicted outputs and real outputs is calculated as loss, and the loss gradient is passed to all parameters in the neural network. This process is called optimization. The parameters are updated with a technique called gradient descent, which is a way to minimize an objective function $f(\theta)$ parameterized by a model's parameters $\theta \in R^d$ by updating the parameters in the opposite direction of the gradient of the objective function $\nabla_\theta f(\theta)$ w.r.t to the parameters. The learning rate $\eta$ determines the size of the steps we take to reach a (local) minimum. In other words, we follow the direction of the surface slope created by the objective function downhill until we reach a valley.

$$\theta_{k+1} = \theta_k - \eta \cdot \nabla f(\theta_k), \quad k = 0, 1, \ldots$$

$\eta \cdot \nabla f(\theta_k)$, is the called the incremental step. The process of updating parameters to reach optimal model performance is called training, and how incremental steps are calculated is called optimization. The most used optimization methods are Stochastic Gradient Descent (SGD)[49] and Adaptive moment Estimation (Adam). If we simply use the gradient $\nabla f(\theta_k)$ as an incremental step, the optimization method is SGD.

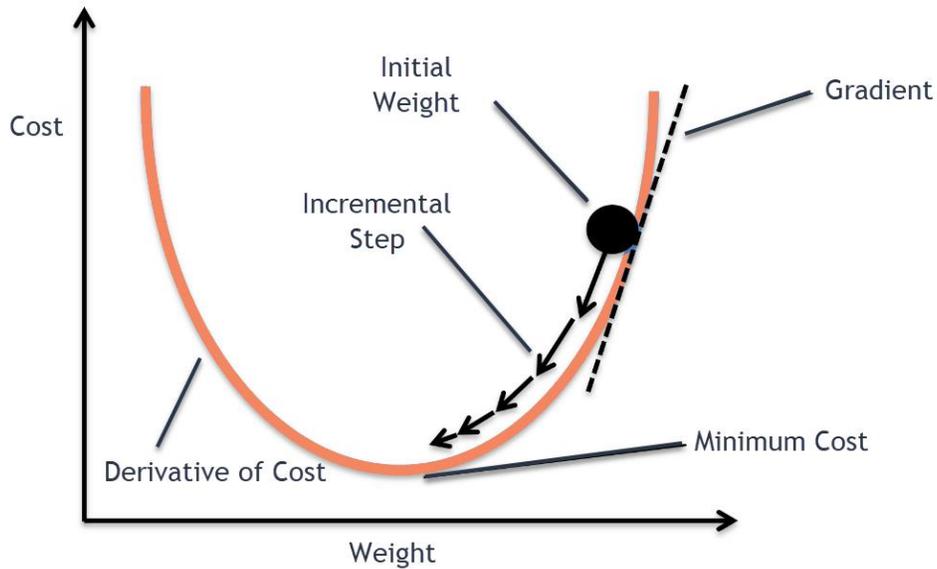

(Fig 3.4: Illustration of optimization. The model starts at the initial weight point. In each training step, the gradient of the loss function is calculated, and the system does an increment step to update parameters in the neural network. Ideally, the training will lead to the global minimum of loss functions, as is shown at the bottom of the valley in the figure. At a global minimum, the loss derivative is zero, and parameters will stop updating.)

Though SGD achieves good convergence in training, it is usually slow. Incorporating gradient momentum will fasten the training process, and one of the examples is (Adam)[50]. Adam computes individual adaptive learning rates for different parameters from estimates of first and second gradients of the loss. In addition to storing an exponentially decaying average of past second gradients $v_t$, Adam also keeps an exponentially decaying average of past first gradients $m_t$.

$$m_i \leftarrow \beta_1 m_i + (1-\beta_1)\nabla_\theta$$

$$v_i \leftarrow \beta_2 v_i + (1-\beta_2)\nabla_\theta^2$$

As $m_t$ and $v_t$ are initialized as vectors of. 0's, the authors of Adam observe that they are biased towards zero, especially during the initial time steps, and

$$\hat{m}_t = \frac{m_t}{1 - \beta_1^t}$$

$$\hat{v}_t = \frac{v_t}{1 - \beta_2^t}$$

Then use these to update the parameters just as we

$$\theta_{t+1} = \theta_t - \frac{\eta}{\sqrt{\hat{v}_t} + \epsilon} \hat{m}_t$$

Rather than using a constant learning rate, Adam computes individual learning rates for each parameter and speeds up convergence and improve the quality of the final solution. It performs well in cases with noisy gradients and is straightforward to implement in deep neural networks.

When the training process reaches a stable state that loss stops decreasing, we call the training is converged. Adam converges faster than SGD, but SGD usually leads to better training result.[49] Another factor to convergence is the number of input training data, which is called batch size. A smaller batch size leads to better convergence, but more time will be needed for training.

## 3.3 Introduction of probabilities

**Probability space**

Before talking about probability estimation, let's first define the probability space to describe the instances. We define a probability space to be a triple $(\Omega, F, P)$, where $\Omega$ is the sample space, which is the set of possible outcomes from an experiment; F is the event space, which is the set of all possible subsets of $\Omega$; and P is the probability measure, which is a mapping from an event $E \subseteq \Omega$ to a number in $[0, 1]$ (i.e., $P : F \to [0, 1]$, which satisfies certain consistency requirements. The simplest setting is where the outcome is discrete variables, like $\Omega = \{A, B, C\}$, where A, B and C are all the possible outcomes of the experiment. When the outcomes are continuous, we assume the sample space is a subset of the reals, $\Omega \subseteq R$

$$P([a,b]) = \int_E dP = \int_a^b p(x)dx$$

Consider two events $E_1$ and $E_2$. If $P(E_2) \neq 0$, we define the conditional probability of $E_1$ given $E_2$, or say the probability that $E_1$ happens when we know that $E_2$ has happened, will be

$$P[E_1|E_2] = \frac{P[E_1 \cap E_2]}{P[E_2]}$$

Here $P[E_1 \cap E_2]$ denotes the probability that $E_1$ and $E_2$ happen at the same time. From this, we can get the multiplication rule:

$$P[E_1 \cap E_2] = P[E_1|E_2]P[E_2]$$

If $E_1$ and $E_2$ are independent, that says the $P[E_1]$ will not be affected by the occurrence of $E_2$ and vice versa, the probability of $E_1$ and $E_2$ happen together will can be simplified as

$$P[E_1 \cap E_2] = P[E_1]P[E_2]$$

From the definition of conditional probability, we can derive the law of total probability, which states the following: if $\{A_1, \ldots, A_n\}$ is a partition of the sample space $\Omega$, then for any event $B \subseteq \Omega$, we have

$$P[B] = \sum_{i=1}^{n} P[B|A_i]P[A_i]$$

From the definition of conditional probability, we can derive Baye's rule,

$$P[E_1|E_2] = \frac{P[E_2|E_1]P[E_1]}{P[E_2]}$$

For discrete random variables X with K possible states, we can write Baye's rule as follows, using the law of total probability:

$$p(X = k|E) = \frac{p(E|X = k)p(X = k)}{p(E)} = \frac{p(E|X = k)p(X = k)}{\sum_{k'=1}^{K} p(E|X = k')p(X = k')}$$

Here, $p(X = k)$ is the prior probability, $p(E|X = k)$ is the likelihood, $p(E|X = k')$ is the posterior probability, and $p(E)$ is a normalization constant, known as the marginal likelihood.

**Estimating probabilities**

In the probabilistic approach to machine learning, all unknown quantities—be they predictions about the future, hidden states of a system, or parameters of a model—are treated as random variables and endowed with probability distributions. The process of inference corresponds to computing the posterior distribution over these quantities, conditioning it to whatever data is available.

A popular method for sampling from high-dimensional distributions is Markov chain Monte Carlo (MCMC). The basic idea behind MCMC s is to construct a Markov chain on the state space X whose stationary distribution is the target density $p^*(x)$ of interest. In Bayesian inference, this is usually the posterior $p^*(x) \propto p(x|D)$. That is, we perform a random walk on the state space, in such a way that the fraction of time we spend in each state x is proportional to $p^*(x)$. By drawing correlated samples $x_0, x_1, x_2, ...$ from the chain, we can perform Monte Carlo integration $p^*$. One of the simple MCMC algorithms is the Metropolis-Hastings algorithm (MH algorithm). The basic idea is that at each step, we propose to move from the current state x to a new state x' with probability $q(x'|x)$, where q is called the proposal distribution (also called the kernel). The user is free to use any kind of proposal they want.

The other method, Hamiltonian Monte Carlo (HMC), leverages gradient information to guide the local moves. HMC sees parameters $\theta$ as position and v as speed. The set of possible values for $(\theta, v)$ is called the phase space. We define the Hamiltonian function for each point in phase space as:

$$H(\theta, v) = \varepsilon(\theta) + K(v)$$

Where $\varepsilon(\theta)$ is the potential energy, $K(v)$ is the kinetic energy, and Hamiltonian $H(\theta, v)$ is the total energy. The momentum of

$$\varepsilon(\theta) = -log\tilde{p}(\theta)$$

Where $\tilde{p}(\theta)$ is possibly unnormalized distribution, such as $p(\theta, D)$, and the kinetic energy to be

$$K(v) = \frac{1}{2}$$

The simplest way to model the time evolution is to update the position and momentum simultaneously by a small amount, known as the step size ŋ:

$$v_{t+1} = v_t + \eta \frac{dv}{dt}(\theta_t, v_t) = v(t) - \eta \frac{\partial \varepsilon(\theta_t)}{\partial \theta}$$

$$\theta_{t+1} = \theta_t + \eta \frac{d\theta}{dt}(\theta_t, v_t) = \theta_t + \eta \frac{\partial K(v_t)}{\partial v}$$

A slightly more accurate way is through a modified Euler's method, where we first update the momentum, and then update the position using the new momentum:

$$v_{t+1} = v_t + \eta \frac{dv}{dt}(\theta_t, v_t) = v(t) - \eta \frac{\partial \varepsilon(\theta_t)}{\partial \theta}$$

$$\theta_{t+1} = \theta_t + \eta \frac{d\theta}{dt}(\theta_t, v_{t+1}) = \theta_t + \eta \frac{\partial K(v_{t+1})}{\partial v}$$

**Random Variable** [99]

Not all experimental results are bound to be definite. Sometimes, random variables will be a better choice for describing it. A random variable is an abstraction of an outcome from a randomized experiment. The random process involves some element of chance, so we cannot be sure about its outcome. The opposite of it is a "deterministic process", where the same actions will always lead to the same result. Based on the output data types, the random variables can be categorized into discrete random variables and continuous random variables. A random variable is discrete if its domain consists of a finite set of values and is continuous if its domain is uncountably infinite. An example of the discrete random variable is the number of heads up when flipping a coin for n times.

For continuous random variables, we can use the example that we spin the hand of a clock and observe where it stops.

Probability mass and density functions are usually used to describe random variables. If the random variable is discrete, the function to describe it is the probability mass function (PMF), which returns $P(X = x)$ with each x in the sample space S. Any PMF must define a valid probability distribution, with the properties:

$$f(x) = P(X = x) \geq 0 \text{ for any } x \in S$$

$$\sum_{x \in S} f(x) = 1$$

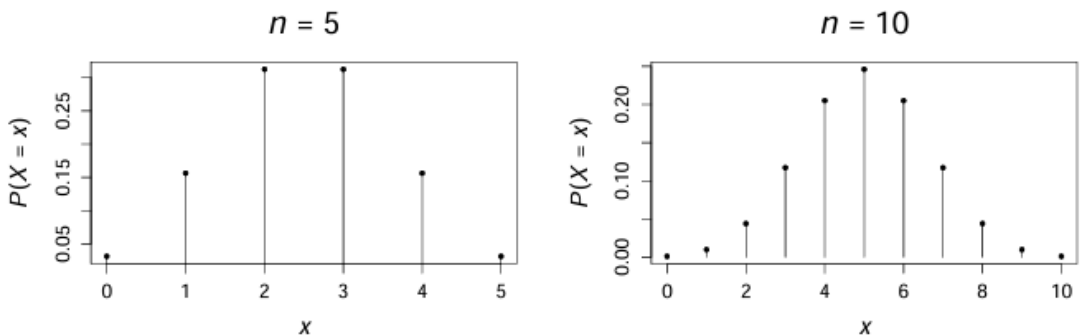

(Fig 3.5: Example of probability mass function using the case of flipping a coin)

The probability density function (PDF) of a continuous variable X is the function $f(\cdot)$ that associates a probability with each range of realizations of X. The area under the PDF between a and b returns $P(a < X < b)$ for any $a, b \in S$ satisfying $a < b$.

Any PDF must define a valid probability distribution, with properties

$$f(x) \geq 0 \text{ for any } x \in S$$

$$\int_a^b f(x)dx = P(a < X < b) \geq 0 \; for \; any \; a, b \in S \; satisfying \; a < b$$

$$\int_{x \in S} f(x)dx = 1$$

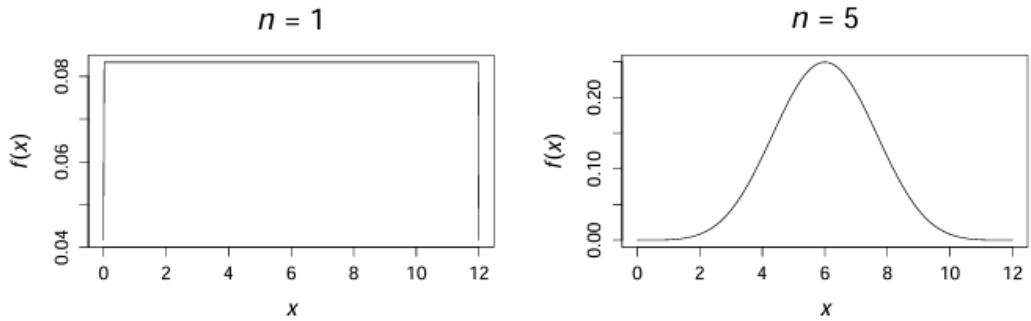

(Fig 3.6: Example of probability mass function using the case of flipping a coin)

**A change of view ---- seeing scientific problems as probabilities.**

Models and simulations are a good way to test theory and predict future situations in scientific research. For hypothesis-building, we often want to decide which of several candidate models provides the best explanation of empirical data. Usually, several parameters are involved in the research. Though some of the parameters can be extracted theoretically, lots are empirical. Though these models typically don't seem to have probability, they are implicit statistical models.[52] Let us suppose we have a model $f(\cdot)$, a set of input $x$ and a vector of parameters $\theta$. With different $\theta$, model outcome $f$ will be different given the same $x$. Our preferred approach is to estimate the likelihood function from the model simulation results. Since probability is equal to or smaller than 1, we can construct a log-likelihood function:

$$L(\theta) = \log f(y; \; \theta)$$

Our goal is to maximize $L(\theta)$ to select the correct set of $\theta$ with the given outcome y. As an example, if we generate data $y_i: I = 1, \ldots 25$ as an independent random sample from the distribution.

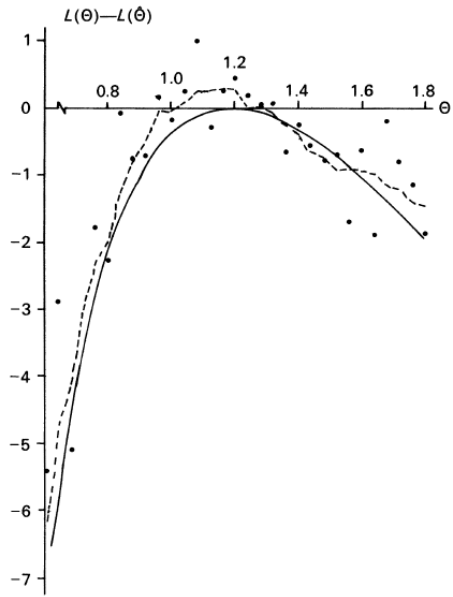

(Fig 3.7: Example of finding optimal parameter for a model.[52])

Given candidate models $m_i$ with parameters $\theta$ and observed data $x_0$ the posterior of a model is

$$p(m_i|x_0) \propto p(m_i)p(x_0|m_i) = p(m_i) \int p(x_0|\theta, m_i)\, p(\theta|m_i) d\theta$$

Where $p(x_0|\theta, m_i)$ denotes the likelihood of the data given the model. So, the problem of finding parameters for a model can be seen as maximizing the probability of getting correct results by choosing parameters and the proper model. We can further reform scientific problems into probability problems, in which instead of using functions to describe the relation between experimental procedures and outcomes, we can see it as the probability of getting experimental outcomes with certain experimental procedures.

## 3.4 Special machine learning techniques used in this thesis

**Simulation-based Inference**

Simulation-based Inference (SBI)[53]–[56] is a method to infer the parameters of a model given its output distribution. The theoretical base of SBI is Bayesian Inference that calculates the probability of one instance to happen when several instances happen together. Suppose we have parameters $\theta$ and experimental observation $x$, obviously the choice of model parameter $\theta$ won't affect the real-life observation $x$, so these two instances are independent. We can estimate the posterior $p(\theta|x)$ with Bayes' rule using $p(x|\theta)$ and a prior $p(\theta)$:

$$p(\theta|x) = \frac{p(\theta)p(x|\theta)}{p(x)}$$

Typically, SBI consists of 3 parts: a simulator that can generate numerical samples, a posterior estimator, and a sampler. During the process, we first assume a possible distribution of parameter $\theta$. Then we draw samples from the $\theta$ distribution and calculate the mode, to estimate the $p(x|\theta)$ distribution. After we build the $p(x|\theta)$, we can infer the distribution of $\theta$ given the output distribution.

There are many ways to perform SBI. In this research, we used Sequential Neural Posterior Estimation (SNPE)[56] which generates parameter samples $\theta_n$ from a proposal $\tilde{p}(\theta)$ instead of the assumed prior $p(\theta)$. This method shrinks the range of possible parameters and makes generated data $x_n$ more likely to be close to the observed data point $x_o$. SNPE finds a good proposal $\tilde{p}(\theta)$ by training the estimator $q_\phi$ over several rounds, whereby in each round $\tilde{p}(\theta)$ is taken to be the approximate posterior obtained in the round before. SNPE finds a good proposal $\tilde{p}(\theta)$ by training the estimator $q_\phi$ over several rounds, whereby in each round $\tilde{p}(\theta)$ is taken to be the approximate posterior obtained in the round before.

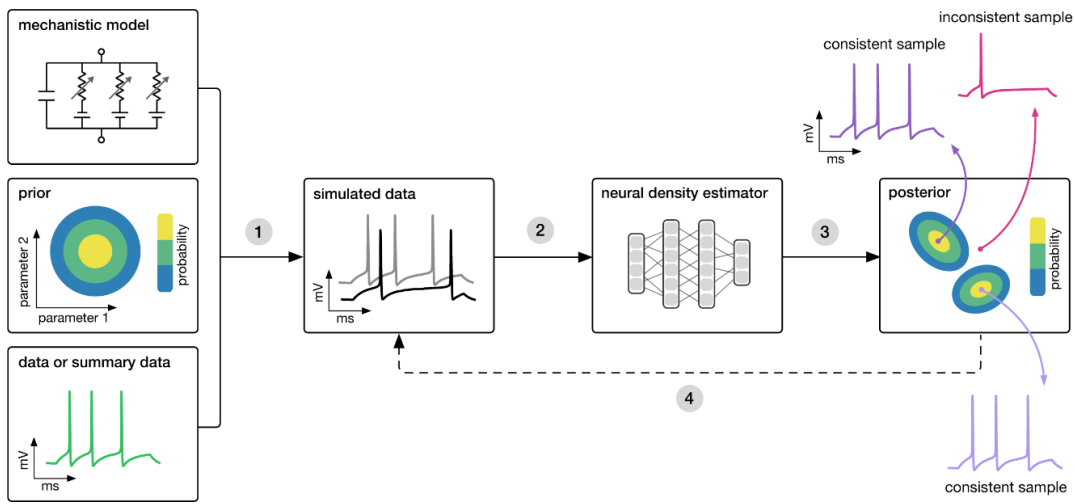

(Fig 3.8: Structure of simulation-based Inference)

The fundamental difficulty in inferring the parameters of a simulator given data is the unavailability of the likelihood function. In Bayesian Inference, we multiply the likelihood $p(x|\theta)$ with prior beliefs $p(\theta)$. However, calculating the likelihood $(x|\theta)$ of a simulator model for given parameters $\theta$ and data $x$ is computationally infeasible in general, thus traditional likelihood-based Bayesian methods, such as variational inference or Markov Chain Monte Carlo, are not directly applicable.

Several methods for likelihood-free inference have been developed to overcome this difficulty, such as Approximate Bayesian Computation and Synthetic Likelihood, which require only the ability to generate data from the simulator. Such methods simulate the model repeatedly and use the simulated data to build estimates of the parameter posterior. In general, the accuracy of likelihood-free inference improves as the number of simulations increases, but so does the computation cost.

**Sequential Neural Likelihood (SNL)**

The main idea of SNL [57] is to train a Masked Autoregressive Flow on simulated data to estimate the conditional probability density of data given parameters, which then serves as an accurate model of the likelihood function. During training, a Markov Chain Monte Carlo sampler selects the next batch of simulations to run using the most up-to-date estimate of the likelihood function, reducing the number of simulations of several orders of magnitude.

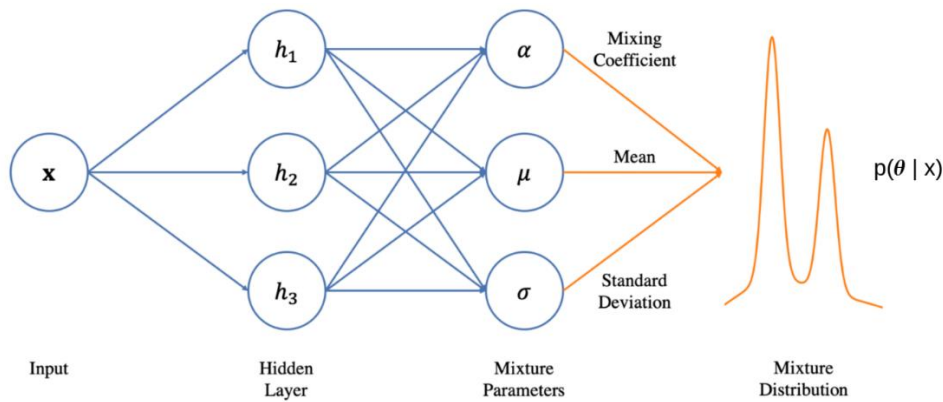

(Fig 3.9: Structure of SNL)

**Generative Flow Network**

Generative models have recently seen wide applications, especially in text and image generation. Generative models create the distribution of the training data they see and can generate new data similar to the training data. A famous example of a generative model is Generative Adversarial Networks (GAN)[59] , which use a generator to create samples and a discriminator to see if the generated samples have the same distribution as the training data. The goal of training is to minimize this difference. However, while generating, this model may need to explore more possible options, which may have potential restrictions on its application to scientific and engineering tasks.

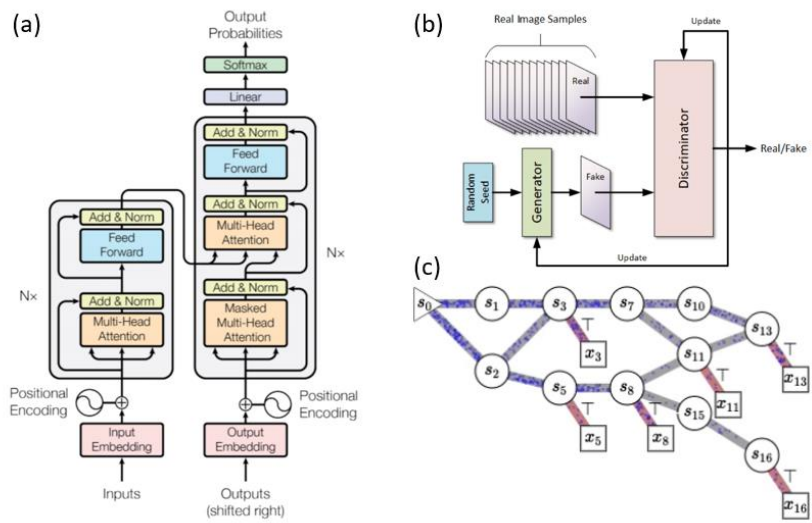

(Fig 3.10: Structure of generative models. (a): transformer [58] (b): GAN (c): GFLowNet)

Generative Flow Network (GFLowNet)[60]–[66] is a new method for generative AI models. Rather than encoding the input data into a more straightforward representation, GFlowNet trains a model that samples a distribution of trajectories whose probability is proportional to a given positive return or reward function. GFlowNet combines flow network and reinforcement learning. The structure of GFLowNet resembles that of RL, which includes an environment that returns rewards based on the series of actions taken, an agent that creates random actions and explores the environment, and a policy model that models the expected rewards for every action given the previous actions.

Unlike typical generative models that learn the probability distribution of states, GFLowNet amortizes its object over the trajectory of forming the final state. The sampling in GFLowNet takes place at training time, while run-time sampling or computations of marginalized quantities can be done in a single pass through a sequence of constructive stochastic steps. GFlowNets trained a sampling policy to make the probability $P_T(s)$ of sampling an object $s$ approximately proportional to the value $R(s)$ of a given reward function applied to that object. The reward value is usually a positive value within (0,1). For probability calculation, GFlowNet used the energy-based model that used an energy function $\varepsilon(s) = -\log E(s)$, i.e., the reward function $E(s)$ is non-negative and corresponds to an unnormalized probability.

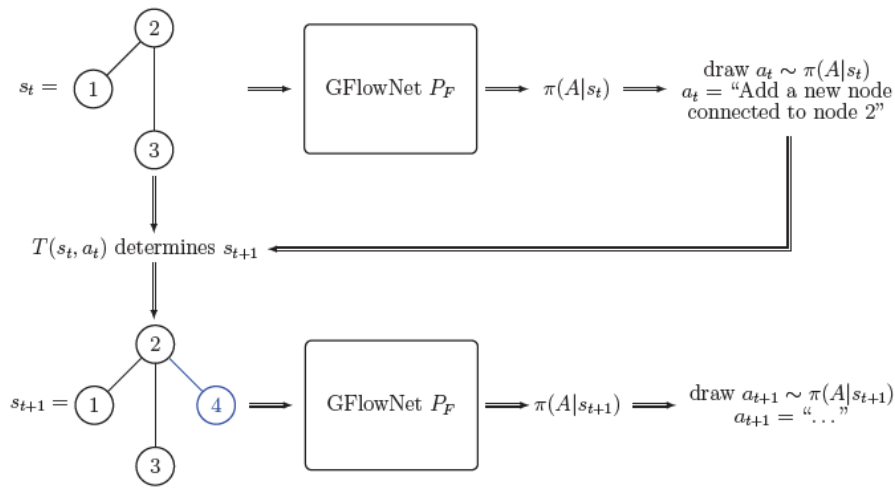

(Fig 3.11: States generation process of GFLowNet)

Whereas one typically trains a generative model from a dataset of positive examples, a GFLowNet is trained to match the given energy or reward function and convert it into a sampler. Compared with the usual RL that pursues a single highest-reward sequence of actions, GFlowNet can explore multiple possible actions. It may give it an advantage in scientific discovery and engineering solution generation since more options can be probed.

The structure of GFLowNet combines flow network and reinforcement learning. The flow of GFlowNet is proportional to the reward of the trajectory, and the trajectory should be a directed acyclic graph (DAG), meaning no loop can be involved in the trajectory. Suppose we have a trajectory $(s_0, s_1, ..., s_{n+1})$ generated, and we call $P_F$ to be the forward probability which denotes the probability from a step to its next step, and $P_B$ to be the backward probability that denotes the probability from a step to its previous one. From Bayesian inference, we have

$$P_F(s'|s) = \frac{P(s \to s')}{P(s)}$$

$$P_B(s|s') = \frac{P(s \to s')}{P(s')}$$

It is easily seen from the Markov chain that

$$P(\tau) = P(s_0 \to s_1 \to \cdots \to s_{n+1}) = \prod_{t=1}^{n+1} P_F(s_t|s_{t-1}) = \prod_{t=1}^{n+1} P_B(s_{t-1}|s_t)$$

In GFlownet, we usually create an initial state $s_0$ denotes the beginning of trajectories, and no step can be taken before it. We also created an end state $s_\perp$ marks the end of the trajectory that no action can be taken afterwards. All trajectories go from the initial state to the end state, so the sum of the forward probability of each trajectory should be 1, and so does the sum of backward probabilities.

$$\sum \prod_{t=1}^{n+1} P_F(s_t|s_{t-1}) = \sum \prod_{t=1}^{n+1} P_B(s_{t-1}|s_t) = 1$$

Now, let's define the flow in the model. In GFlownet, we design the flow of each trajectory $\tau$ to be proportional to its reward.

$$P(\tau) = \frac{1}{Z} F(\tau)$$

Z is the total flow in the function, which is set up as a trainable target. The sum of flow of all possible trajectories $\tau$ from initial state to sink state will be

$$Z = \sum_{\tau \in T} F(\tau) = \hat{Z} \prod_{t=1}^{n+1} \hat{P}_B(s_{t-1}|s_t) = \hat{Z} \prod_{t=1}^{n+1} P_B(s_{t-1}|s_t) = \hat{Z}$$

Suppose that a model with parameters $\theta$ outputs estimated forward policy $P_F(-|s;\theta)$ for state $s$ (just as for detailed balance above), as well as a global scalar $Z_\theta$ estimating

$$Z \prod_{t=1}^{n} \hat{P}_F(s_t|s_{t-1}) = F(x) \prod_{t=1}^{n} \hat{P}_B(s_{t-1}|s_t)$$

Where we have used that $P(s_n = x) = \frac{F(x)}{Z}$

For a trajectory $\tau = (s_0 \to s_1 \to \cdots \to s_{n+1})$, define the trajectory loss

$$L_{TB}(\tau) = \left( \log \frac{Z_\theta \prod_{t=1}^{n} \hat{P}_F(s_t|s_{t-1};\theta)}{R(x) \prod_{t=1}^{n} \hat{P}_B(s_{t-1}|s_t;\theta)} \right)^2$$

If $\pi_\theta$ is a training policy – usually that given by $P_F(-|s;\theta)$ or a tempered version of it – then the trajectory loss is updated along trajectories sampled from $\pi_\theta$, i.e., with stochastic gradient

$$E_{\tau \sim \pi_\theta} \nabla_\theta L_{TB}(\tau)$$

# Chapter 4

# Neural Network–based model for CNTFETs

## 4.1 Introduction

Transistor models are indispensable for circuit simulation and essential for the efficient analysis and design of integrated circuits (ICs). The most common model for devices is the compact models, which predict the current behavior of the devices and help determine biased circuits and amplified circuits for a successful design. Standard compact models are combinations of physics-based equations chosen based on device structure. Some empirical parameters like need to be extracted to fit the model to reality. Though these models are accurate, they usually take a long time to be set up since they need to be both physically sound and fit with all experimental observations. The explosion of new materials may also make this task harder since new materials and their unique electrical properties need to be researched before a valid model can be set up. A more convenient device modeling method may help plan research before the explicit model is studied. Here, we take CNTFET as an example to show that neural networks can be a model for semiconductor devices.

## 4.2 Structure of CNTFETs:

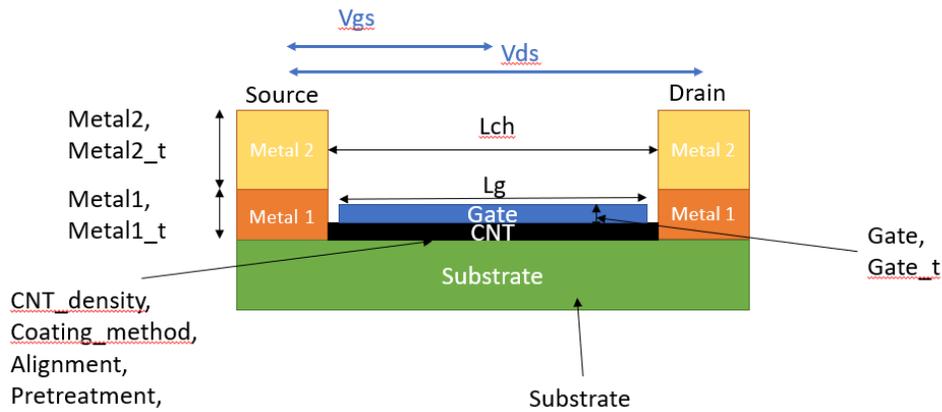

(Fig 4.1: Structure of a typical CNTFET. Selected categorical and continuous device parameters are shown in the graph, except for device structure, which is shown in the appendix)

A field-effect transistor (FET) usually consists of the following parts: a substrate as a base to build the device, the semiconductor material itself, two metal contacts named source and drain to let current flow, and a gate to control the charge carrier density. When operating, an electrical potential $V_{ds}$ is applied between the two metal contacts source and drain, and an electrical bias $V_{gs}$ is used on the gate to control current. The distance between the source and the drain is often described as channel length Lch, and Lg characterizes the length controlled by the gate. Except for these parameters, the choice of materials in each section and processing methods also significantly impact the device's performance.

## 4.3 Neural network with experimental CNTFET data.

A good model for device behavior should capture both $I_{ds} - V_{ds}$ behavior and $I_{ds} - V_{gs}$ behavior. Usually, $I_{ds} - V_{ds}$ behavior is better captured by using $I_{ds}$ as the model output, while $I_{ds} - V_{gs}$ relation is better trained with the logarithm of Ids. This is because $V_{gs}$ usually have an exponential influence on drain current, while $V_{ds}$ affect device performance more linearly. We first tried to create a model structure to solve this problem by using a two-step model that we first train $log(I_{ds})$ then $I_{ds\_}ratio$. The first model ensures that the relation between $I_{ds}$ and $V_{gs}$ will be captured. The training data of the second model is the ratio between the real $I_{ds}$ and the exponential of the predicted result of the first model, which goes:

$$I_{ds\_}ratio = I_{ds}/exp(model1(inputs))$$

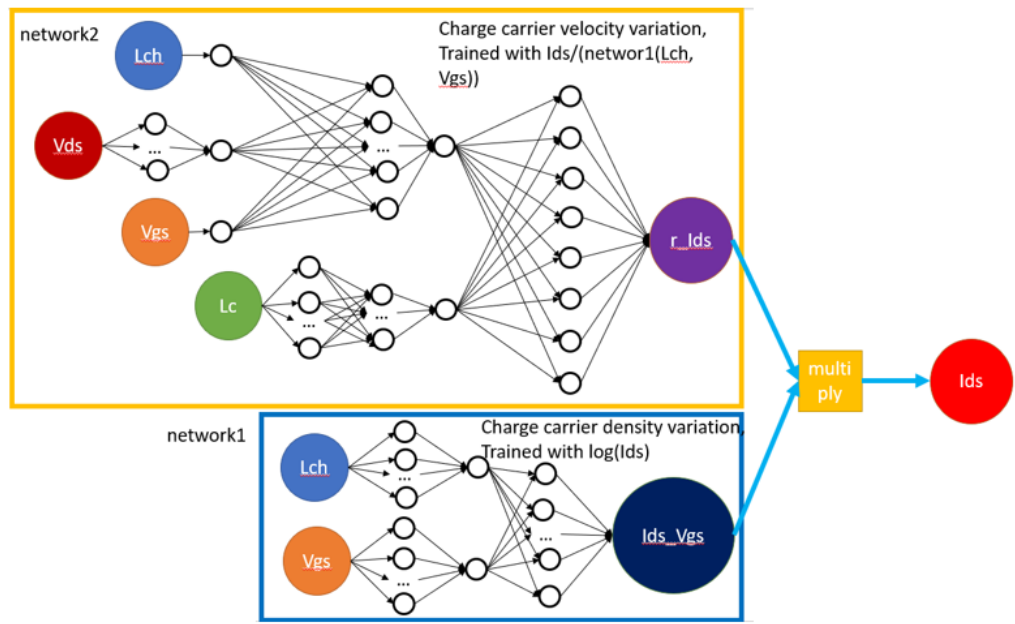

(Fig 4.2: Two-step neural network model for CNTFET modeling)

We explore the setup of the CNTFET model, starting with devices with one single CNT using experimental data from [ ] and PlotDigitier as a data abstraction method. Since the experimental data only discussed the effect of channel length $L_{ch}$ and metal contact length $L_c$ on CNTFET performance, we take only $V_{ds}$, $V_{gs}$, $L_c$ and $L_{ch}$ as input of the neural network and use $I_{ds}$ as output of the model.

Before training, we performed data cleaning for these data. The first step is to remove the hysteresis effect of the data. Due to moisture in the air, early CNTFETs usually have severe hysteresis, that $I_{ds}$ measured for the same device is likely different when measured forward and backward, and the device performance will also vary under different times of measurement. Therefore, when we see the obtained data, we found that $I_{ds}$ data under the exact condition will usually be different. Since we have a minimal amount of data here, this caused difficulty for convergence in training, so we cleaned the hysteresis effect before training. We clean out the deviation of $I_{ds}$ under the same condition by shifting the $V_{gs}$ position of $I_{ds} - V_{gs}$ data to align them with $I_{ds} - V_{ds}$ data under the same condition, as is shown in Fig 4.3.

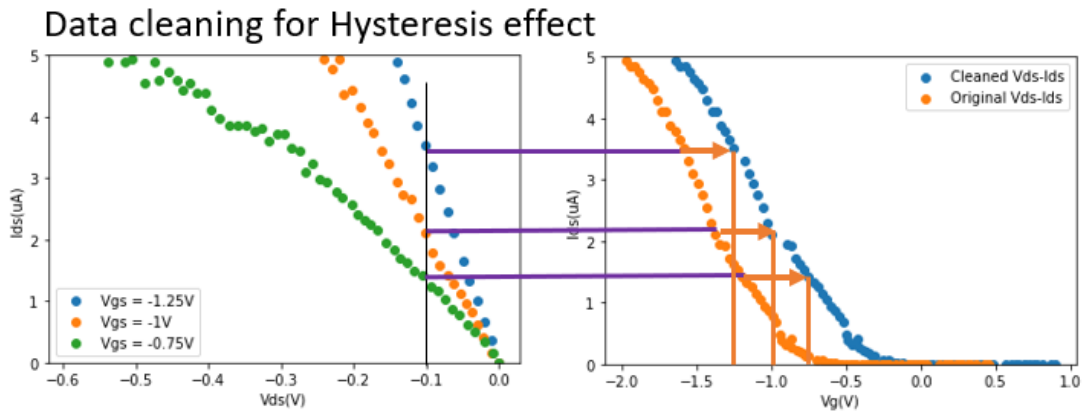

(Fig 4.3: Hysteresis effect cleaning: Left: original experiment data, Right: cleaned data)

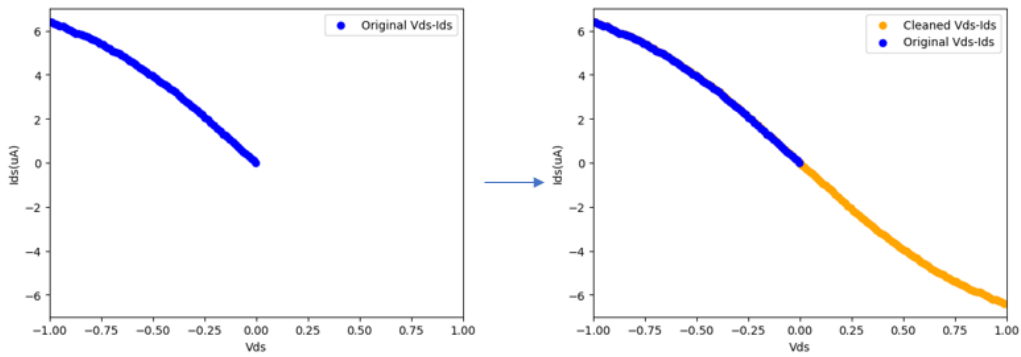

(Fig 4.4: Data generation for $V_{ds}$ symmetry)

After removing the hysteresis effect, we duplicated the data with reversed VDs and IDs data. This is because, for MOSFETs with symmetric structures, Ids should be symmetric for $V_{ds}$. A typical way to test the model is the Gummel test, where the $I_{ds} - V_{ds}$ and its derivatives are plotted. We've also added Ids=0, data with a random combination of other conditions to make sure that $I_{ds} = 0$ when $V_{ds} = 0$. After data set cleaning, we normalized the inputs and output data in the following way:

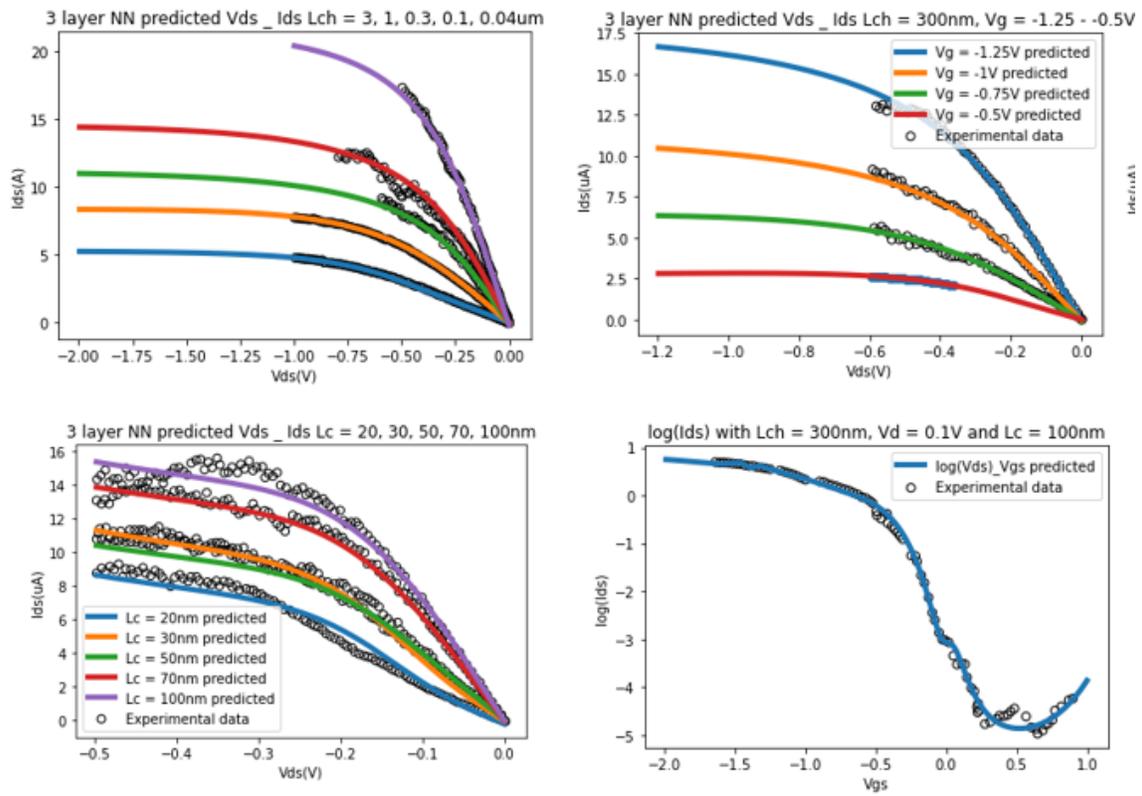

(Fig 4.5: $I_{ds} - V_{ds}$ and $I_{ds} - V_{gs}$ prediction)

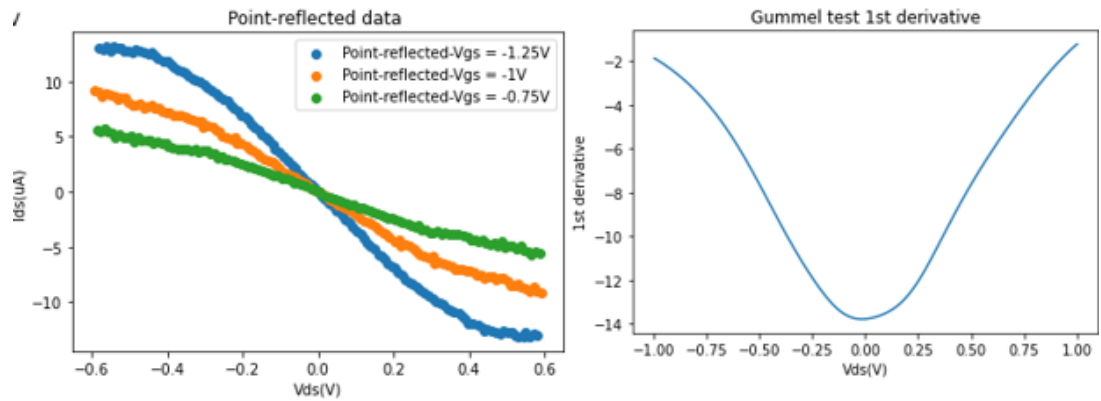

(Fig 4.6: Symmetry of $I_{ds} - V_{ds}$ prediction and Gummel test)

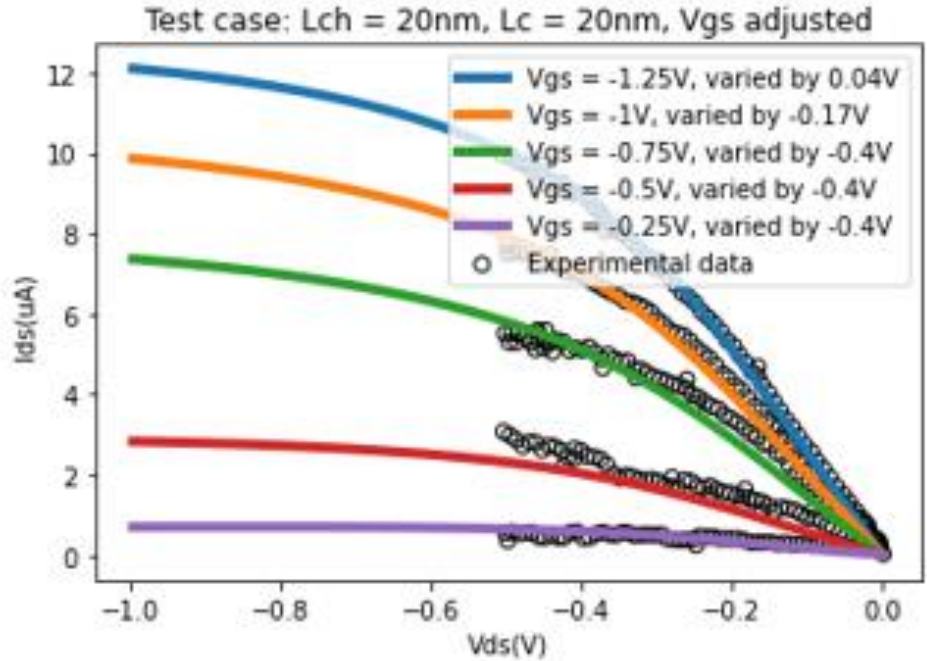

(Fig 4.7: Predicted result of unseen cases)

As is shown in Fig 4.5, the two-step model successfully predicts both the $I_{ds} - V_{ds}$ behavior and the exponential $I_{ds} - V_{gs}$ behavior under different biases. Fig 4.6 shows that the model is symmetric to $V_{ds}$ and passes the Gummel test. We've also shown that the model can predict cases it never sees successfully.

## 4.4 Neural Network model incorporating processing methods

As is discussed in chapter 1, CNTFET performance is not only affected by device structure parameters, like channel length and gate width, but is also affected by the selection of processing methods and choice of gate, metal contact, and substrate materials. Therefore, it is essential to consider them when building models for CNTFETs.

This task will be challenging for compact models since these phenomena may be complex to express in equations.

Different from parameters like $L_{ch}$, the choice of fabrication methods and materials cannot be expressed as continuous values. We can represent them as categorical values and assign them different integer values to process them in a neural network.

In a neural network, continuous data can be directly fed into input layers and multiplied with weights. This may not be a good idea for categorical data since they are represented by integers whose value is only an identity rather than carrying meaningful information. Encoding is usually used to contain features of categorical data in training. One of the most widely used encoding methods is one-hot encoding. In one-hot encoding, a parameter matrix of $E = n \times m_{dim}$, where $n$ is the total number of categories, and $m_{dim}$ the dimension of hidden layer matrix. When a categorical sequence $[a_1, a_2, \cdots, a_i]$ pass to the input layer, instead of multiplying $[a_1, a_2, \cdots, a_i]$ with the input layer weights, the agent form extracts the lines of matrix E with the corresponding index and form a matrix $[E[a_1], E[a_2], \cdots, E[a_i]]$. This encoded matrix now represents the effect of categorical parameters and passed to the next layer for processing.

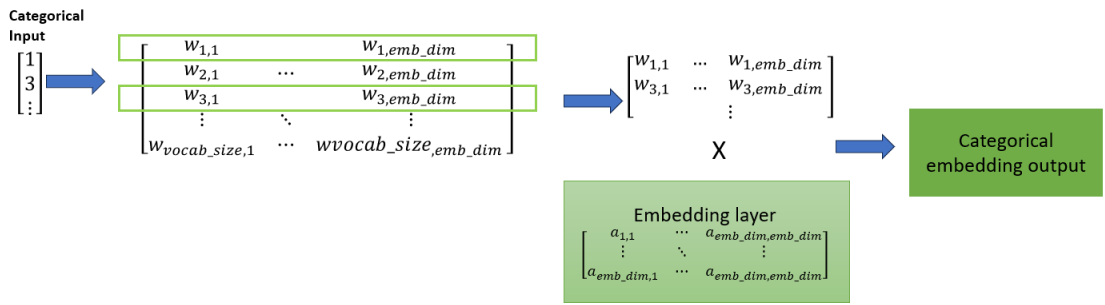

(Fig 4.8: Encoding of categorical parameters)

In this research, we considered the following parameters:

**Categorical Parameters**

**Substrate**: The material used for substrate. Si/SiO2 substrate is used in most CNTFETs, but a few tried soft materials like

**Gate material and gate thickness (Gate_mat1, Gate_mat2 and Gate_t1, Gate_t2):** Gate contact applies gate potential $V_{gs}$ on the device and controls the maximum current allowed in the channel. Both gate material type and gate thickness may affect device performance. Some articles used two layers of different materials. We describe the layer directly in contact with CNT as Gate_mat1 and Gate_t1. If no second oxide layer is used, we use None for Gate_mat2 and set Gate_t1 as 0.

**Metal contact and thickness (Metal1, Metal1_t, Metal2, Metal2_t):** Most CNTFET apply a two-layer metal contact. The first layer (Metal1) comes into direct contact with CNT and is used to change the doping type of CNT since metals have their unique working potential. When using Pd as Metal 1, CNTFET is p-doped, and when using Sc as metal contact, the device is n-doped. The doping type changes the direction of current flowing through the device and is expressed as positive or negative current. The second metal layer is added over metal 1 to increase the conductivity of the metal contact since the resistivities of Pd and Sc are large.

**CNT properties (CNT_density, Coating method, Alignment, Pretreatment):** CNT_density is how many CNTs there are per um; thus, a higher value of this will lead to higher current IDs. Depending on coating methods, CNTs can be randomly deposited on the substrate or **Structure**: Structure describes the shape of the device and the position of each material. Though all the articles in the data source used top-gated structures, they differ in detail and can be put into three categories. The first type has a symmetric top-gated structure, but the gate covers the gap between the gate-source and the gate-drain. The second one is also symmetrical but with gaps uncovered. The third type denotes the

**Coating method**: In this research, two coating methods are used. The first is dip-coating, dipping the silicon wafer in the CNT solution. This usually leads to a randomly distributed CNT. The other way is DLSA, which used

**Alignment:** Whether CNT is aligned in the device.

**Pretreatment:** In some research, YOCD is used to clean CNTs.

**Sub Pretreatment:** In some of the research,

Continuous parameters

**Lch:** Channel length of CNTFET

**Lg:** Gate length of CNTFET

**Metal2, Metal2_t:** Thickness of metal contact layer 1 and 2.

**CNT_density:** The number of CNTs per um of channel width.

**Vgs:** Gate-Source voltage

**Vds:** Drain-Source voltage

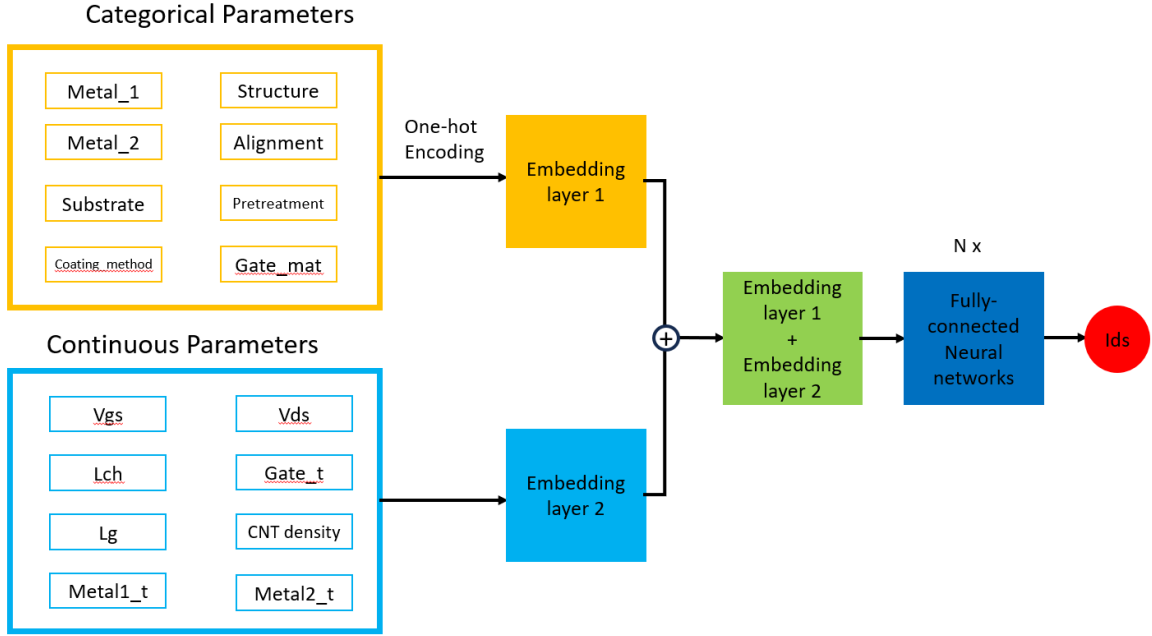

(Fig 4.9: Structure of Neural Network for describing the effect of both categorical and Continuous parameters on device performance)

Here is the structure of the neural network used in this research. We first separate inputs into two kinds: Categorical and Continuous. We first use One-hot Encoding to encode the effect of Categorical Parameters into Embedding layer 1 with an output of size $8 \times d_{model}$, then multiply Continuous with a matrix of $8 \times d_{model}$ to get an output matrix of the same dimension. After that, we concatenate these two matrices together and feed this $16 \times d_{model}$ matrix to the afterward training steps, which are eight layers of fully connected $d_{model} \times d_{model}$. To capture the feature of both $I_{ds} - V_{ds}$ curve and $I_{ds} - V_{gs}$ curve, we separate the training process into two steps. Its details are described in the Appendix.

| Categorical Parameters | Encoding number |
|---|---|
| Metal_1 | Pd: 0, Sc: 1 |
| Metal_2 | Au: 2, Al: 3 |
| Gate Material | HfO2:4 |
| Coating Method | DLSA:5, dip-coating: 6 |
| Alignment | Aligned:7, Random:8 |
| Pretreatment | No:9, YOCD:10 |
| Substrate Material | SiO2:11, parylene:12, quartz:13 |
| Gate_metal_1 | Pd:14, Ti:15 |
| Gate_metal_2 | Au:16, None:17 |
| Sub Pretreatment | Etch:18, None:19 |
| Device Structure | Structure 1: 20, Structure 2: 21, Structure 3: 22 |

(Table 4.1: Categorical Parameter values)

We use the same two-step model in 3.1 that trains log ($I_{ds}$) and $I_{ds}$ ratios sequentially. However, we use a fully connected neural network here to ease construction. We collected data from 9 articles [43],[67]–[74] using Plot-digitizer to collect I-V curve data. During training, we randomly select 80% of the data as training data and leave the rest as testing data to ensure the model won't overfit. In the first step, though most data can be predicted within a range of 0.1-10 times the original data, there are always a few data that cannot be fitted, and their deviation can be as high as 10^5 compared with the original data. We can remove those data with a range above 0.1-10 times, about 1% of the original data, to prepare training data for our next step since a too large range will make most data indistinguishable for the model to tell apart. For the rest of the data, we again randomly selected 80% of the data as training data and 20% as testing data.

We tried different combinations of hyperparameters. An initial learning rate of $10^{-5}$ was used until testing loss stopped decreasing, then $10^{-7}$ was used until convergence. During training, we monitored the testing loss to ensure the data kept decreasing to prevent overfitting. The model produced the lowest loss with an embedding size of 512 and 6 layers and this model is used as the logs model. We've also noticed that a too-small embedding size and number of layers may cause underfitting, that some features of the model are not captured.

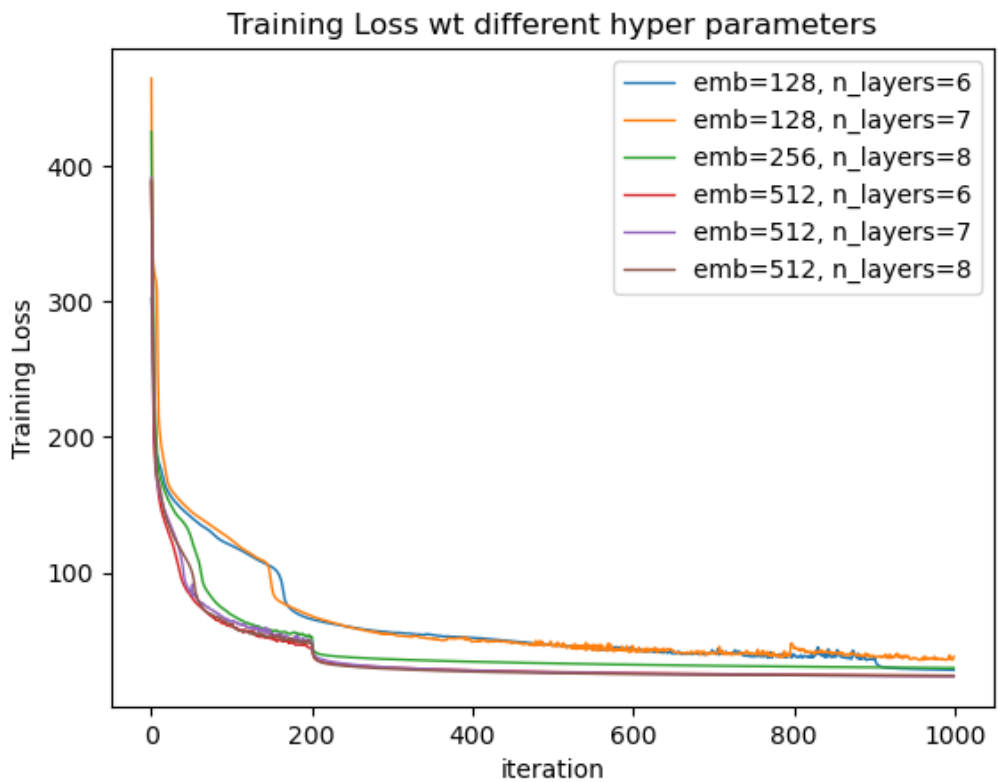

(Fig 4.10: Training Loss for logIds model with various combinations of embedding size and embedding layer numbers)

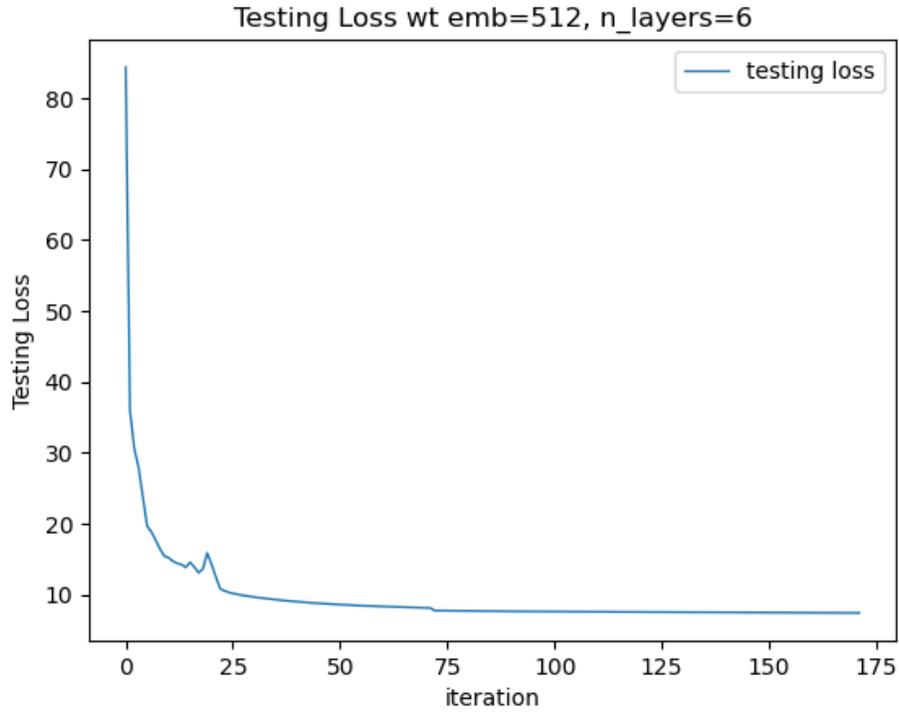

(Fig 4.11: Training Loss for logIds model)

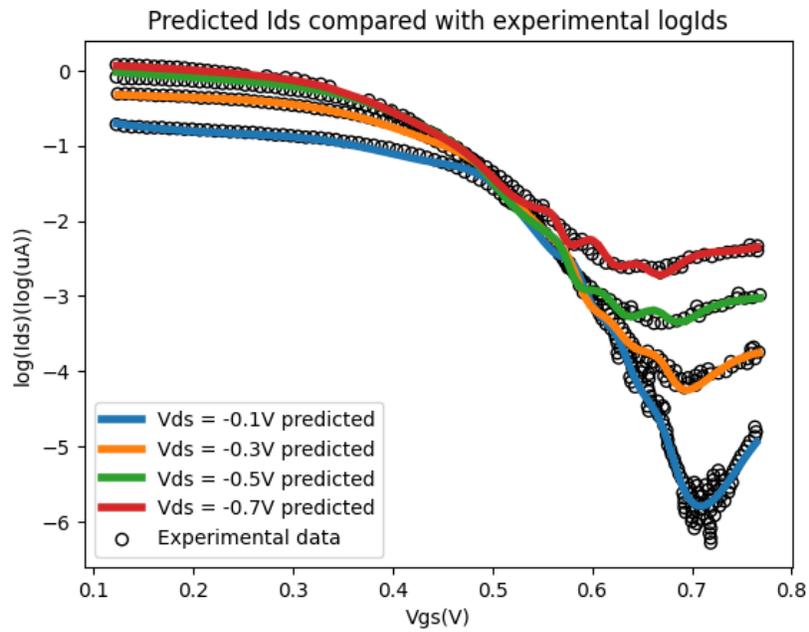

(Fig 4.12: Predicted $\log(I_{ds}) - V_{gs}$ for processing information incorporated model)

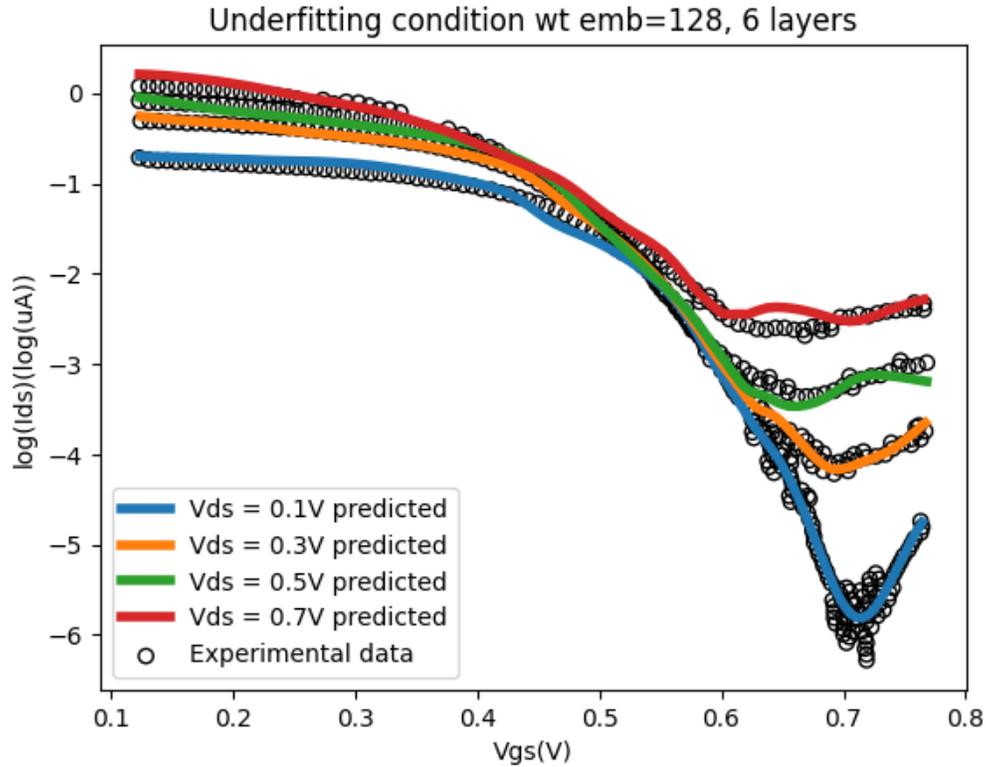

(Fig 4.13: Underfitting condition for $\log{(I_{ds})} - V_{gs}$ with smaller embedding size and number of layers.)

The model can capture the trend of device performance and make a relatively good prediction for both $I_{ds}$ and $\log{(I_{ds})}$ relations. It can also provide reasonable predictions for different combinations of processing information. However, we've also observed that the model may fail at some specific points, which are likely data in the testing data set. This is probably due to the small amount of data the model doesn't have enough to learn from.

| Lch | Lg | CNT_density | Metal_1_t | Metal_2_t | Gate_t | Gate_metal_1_t | Gate_metal_2_t |
|---|---|---|---|---|---|---|---|
| 0.12 | 0.1 | 150 | 0.03 | 0.05 | 0.0073 | 0.01 | 0.02 |

| substrate | Metal_1 | Metal_2 | Gate_mat | Coating_Method | structure |
|---|---|---|---|---|---|
| SiO2 | Pd | Au | HfO2 | DLSA | 2 |
| Alignment | Pretreatment | Gate_metal_1 | Gate_metal_2 | Sub_Pretreatment | |
| Aligned | YOCD | Pd | Au | None | |

(Table 4.2: Process information for condition 1)

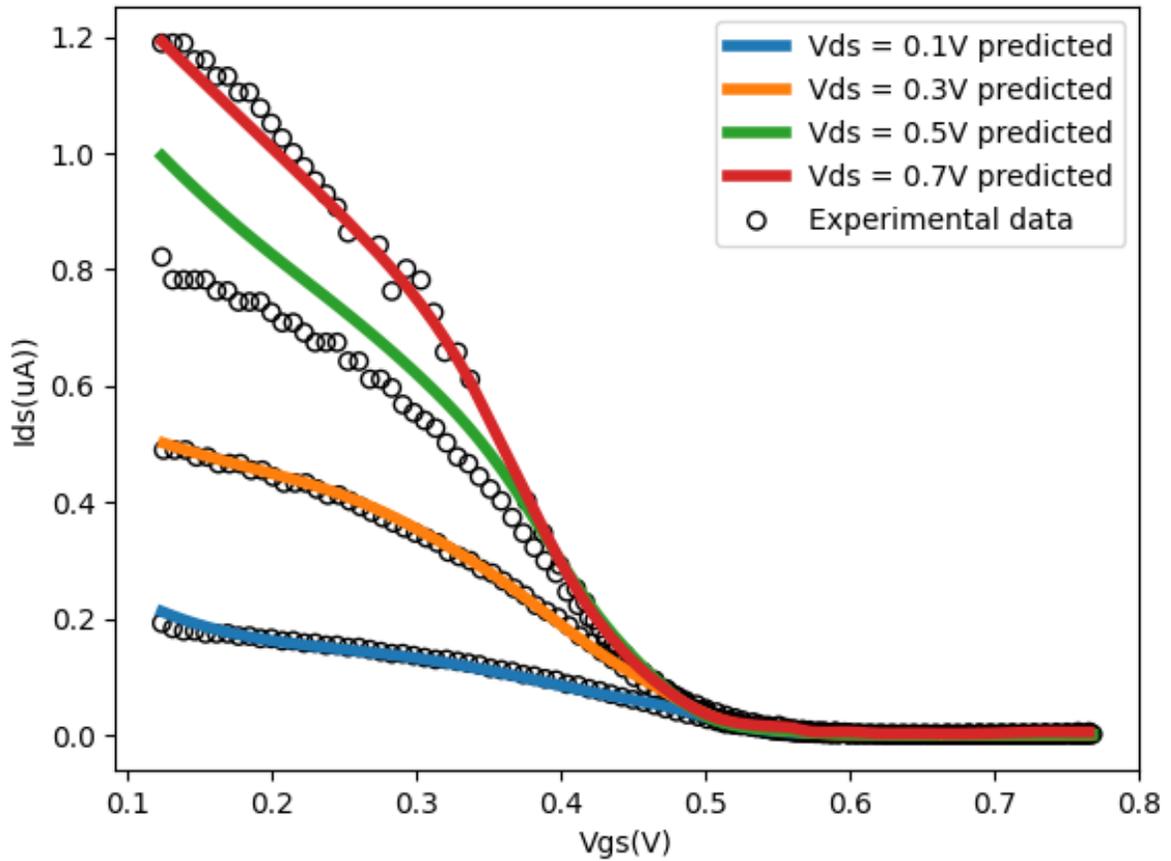

(Fig 4.14: Predicted $I_{ds} - V_{gs}$ under condition 1)

| Lch | Lg | CNT_density | Metal_1_t | Metal_2_t | Gate | Gate_metal_1_t | Gate_metal_2_t |
|---|---|---|---|---|---|---|---|
| 0.101 | 0.035 | 60 | 0.01 | 0.02 | 0.0048 | 0.005 | 0.18 |

| substrate | Metal_1 | Metal_2 | Gate_mat | Coating_Method | structure |
|---|---|---|---|---|---|
| SiO2 | Pd | Au | HfO2 | DLSA | 3 |
| Alignment | Pretreatment | Gate_metal_1 | Gate_metal_2 | Sub_Pretreatment | |
| Aligned | YOCD | Ti | Au | None | |

(Table 4.3: Process information for condition 2)

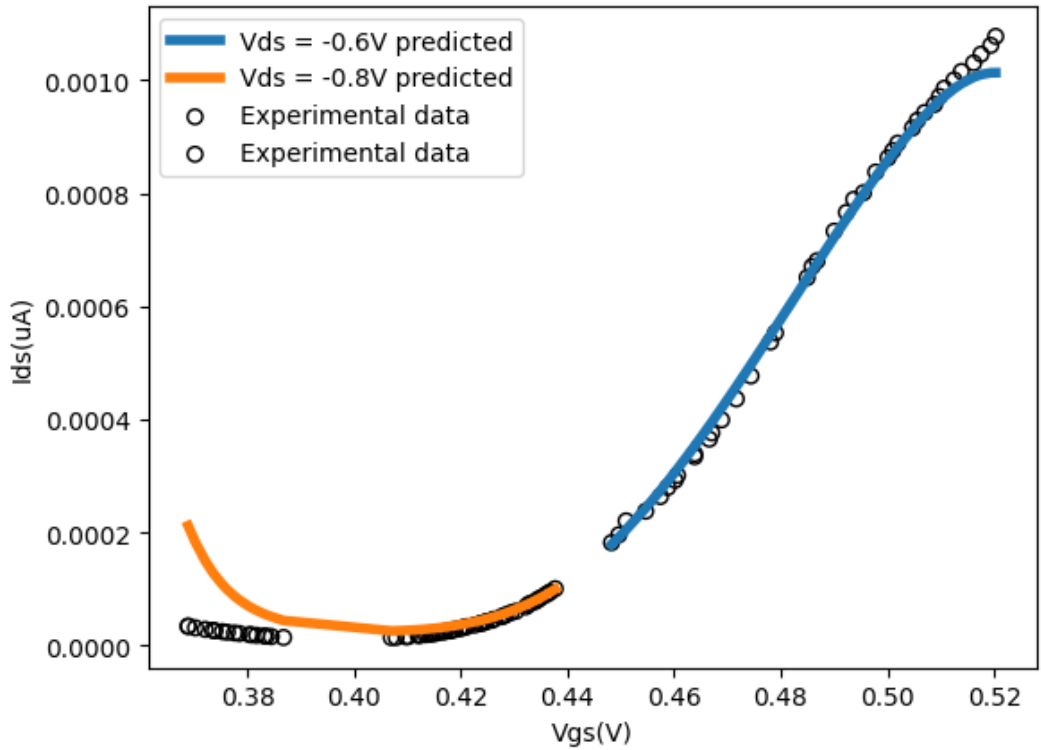

(Fig 4.15: Predicted $I_{ds} - V_{gs}$ under condition 2)

| Lch | Lg | CNT_density | Metal_1_t | Metal_2_t | Gate | Gate_metal_1_t | Gate_metal_2_t |
|---|---|---|---|---|---|---|---|
| 0.45 | 0.6 | 60 | 0.06 | 0.02 | 0.005 | 0.005 | 0.12 |

| substrate | Metal_1 | Metal_2 | Gate_mat | Coating_Method | structure |
|---|---|---|---|---|---|
| parylene | Pd | Au | HfO2 | dip-coating | 3 |
| Alignment | Pretreatment | Gate_metal_1 | Gate_metal_2 | Sub_Pretreatment | |
| Random | YOCD | Ti | Au | None | |

(Table 4.4: Process information for condition 3)

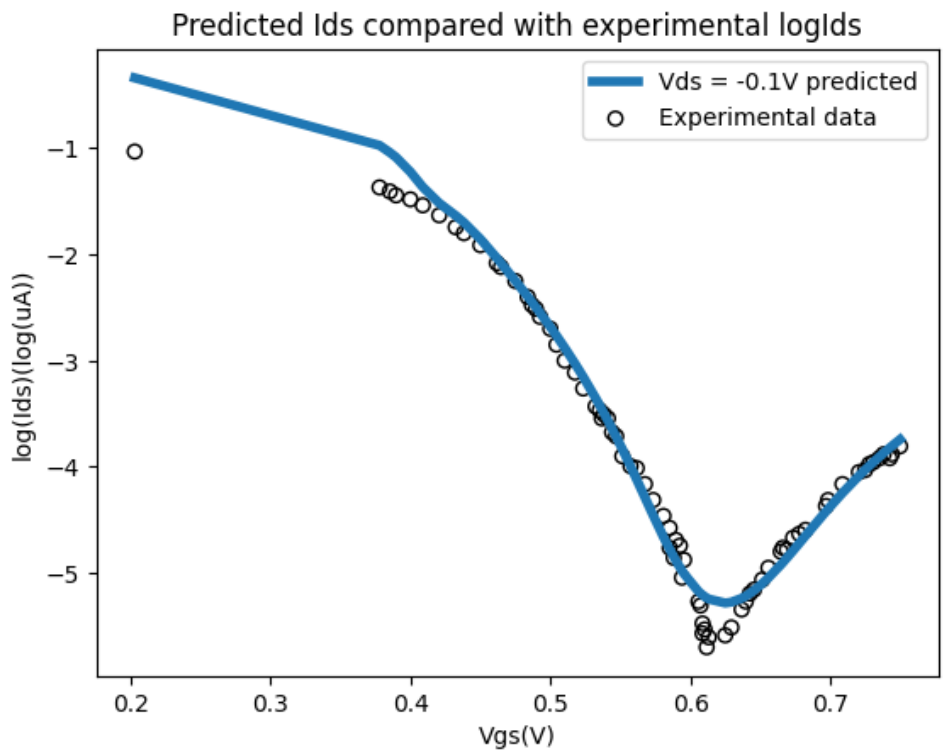

(Fig 4.16: Predicted $I_{ds} - V_{gs}$ under condition 3)

| Lch | Lg | CNT_density | Metal_1_t | Metal_2_t | Gate | Gate_metal_1_t | Gate_metal_2_t |
|---|---|---|---|---|---|---|---|
| 0.14 | 0.08 | 200 | 0.02 | 0.01 | 0.0048 | 0.005 | 0.35 |

| substrate | Metal_1 | Metal_2 | Gate_mat | Coating_Method | structure |
|---|---|---|---|---|---|
| SiO2 | Pd | Au | HfO2 | DLSA | 3 |
| Alignment | Pretreatment | Gate_metal_1 | Gate_metal_2 | Sub_Pretreatment | |
| Aligned | YOCD | Ti | Au | None | |

(Table 4.5: Process information for condition 4)

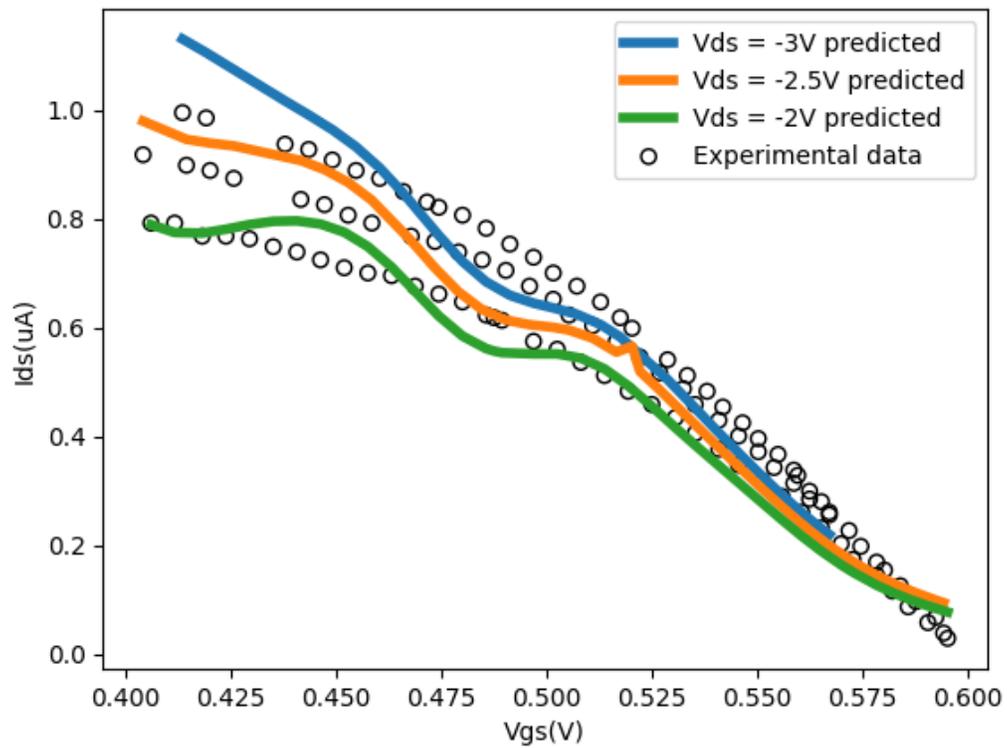

(Fig 4.17: Predicted $I_{ds} - V_{gs}$ under condition 4)

## 4.5 Theoretical issue: Extrapolation and Interpolation

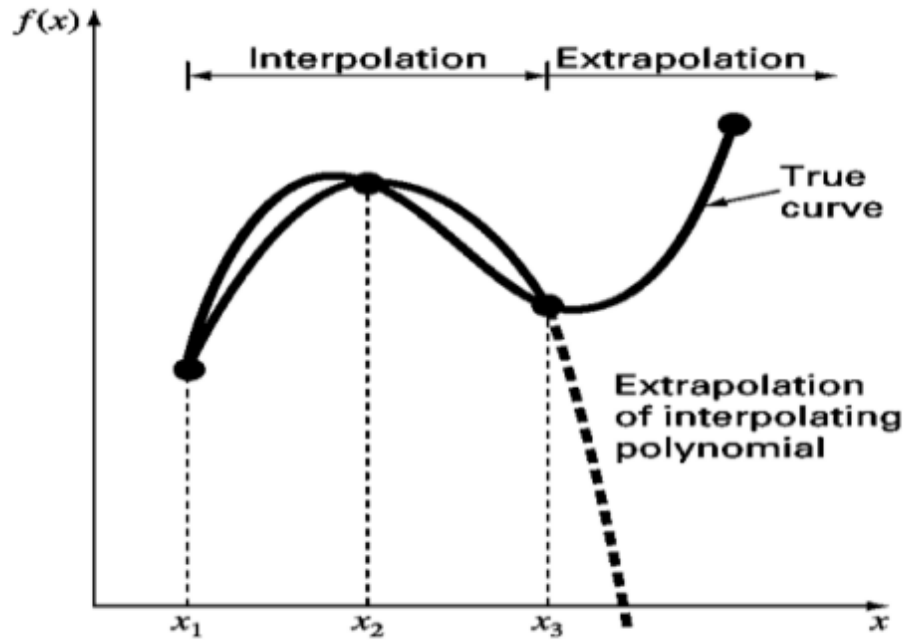

(Fig 4.18 : Extrapolation issue with interpolation models [109][110])

As an interpolation technique, neural network is able to capture patterns of the training data and fit the curve of input-output. However, in extrapolation tasks, where prediction need to be made on inputs beyond the training data, the interpolation model may fail. As is seen in Fig 4.18, though the interpolation model can give out reasonable prediction within the interpolation range, the prediction significantly failed when the model is doing an extrapolation task. Therefore, using only neural network for FET modeling may not be able to provide accurate results when the input value falls out of the training value.

At the same time, since the neural network model used here is a curve-fitting method, the accuracy of neural network prediction is restricted by the interval of the input data. Theoretically, the prediction between the interval of the input data may fail. A more accurate model will be achieved by using more input data with smaller intervals. .Further study should also been done on estimating the amount of data needed to train a good enough model and how small the interval should be for enough accuracy. Different intervals may be needed in areas with different slopes.

The electronic devices all follow physical rules. For example, in Silicon MOSFETs, when $V_g > V_t$, the drain-source current $I_{ds}$ is driven by:[111]

$$I_{ds} = \mu_{eff} C_{ox} \frac{W}{L} \left[ \left( V_g - V_{fb} - 2\psi_B - \frac{V_{ds}}{2} \right) V_{ds} \right.$$

$$\left. - \frac{2\sqrt{2\varepsilon_{Si} q N_a}}{3 C_{ox}} \left[ (2\psi_B + V_{ds})^{3/2} - (2\psi_B)^{3/2} \right] \right]$$

Where $\mu_{eff}$ is the effective mobility of charge carriers, $V_{fb}$ and $\psi_B$ are associated with band diagram of silicon, $\varepsilon_{Si}$ is the dielectric constant of silicon, $C_{ox}$ is the gate-oxide capacitance, which is related to gate oxide thickness and dielectric constant. For carbon nanotube, similar expression can also be expressed, as is shown in the compact model discussion in chapter 1. These parameters are related to material used, and a change of material may lead to a drastic change in the expression. For example, a change of metal contact material may change the metal contact type from Ohmic contact to Schottky contact, and these two types of contact behaves differently. Therefore, the neural network trained on a few materials may likely fail when it sees data with materials it was never trained on. Models built on physical expressions have a better extrapolation ability and may provide better results.

At the same time, some of the parameters in the compact model may be affected by hard-to-predict conditions, and that may use the help of neural network. Carbon Nanotube is notorious for being sensitive to surrounding environments. Since it only consists of one layer of Carbon atom, absorption of molecules and surface conditions may greatly change the electronic properties of CNTs. Therefore, a better way to build up models might be to incorporate past physical knowledge in the model built up. I think neural networks could be used as a simulator of some of the parameters in a physically-built model when the extrapolation models of them are hard to be extracted, but for unknown physics parts, it might be better to use physical equations rather than a neural network. These case might be processing method chosen, as lot of unpredicted condition might be introduced in the process. Further work need to be done on building a more reliable model that also fits the reality.

## 4.6 Conclusion

We developed a two-step NN model for CNTFET performance and successfully predicts device performance. The main contribution here is that we created a data cleaning method for correcting hysteresis effect, so the training data will have less noise. In the second part, we created a NN model that can take device processing method and materials combination into consideration. Though the training data amount is not enough and further work should be done one the extrapolation technique, we have shown that encoding technique can be a way to incorporate non-numerical information for neural networks of electronic devices. Further work should be done on the extrapolation ability of neural network and the amount of data needed to provide a precise enough prediction.

A physically model incorporating neural network might be a way to solve the problem.

# Appendix

Structure 1:

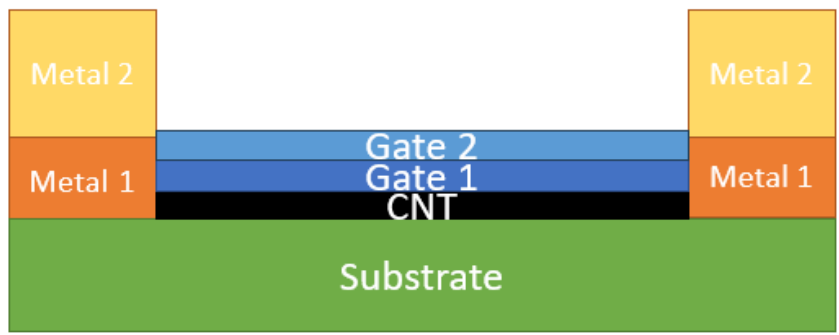

As is shown in the graph, structure 1 denotes

Structure 2:

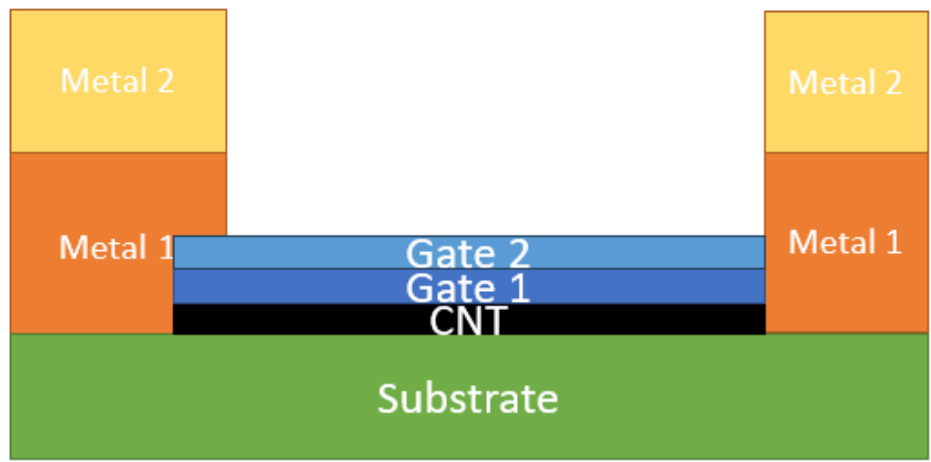

Structure 3:

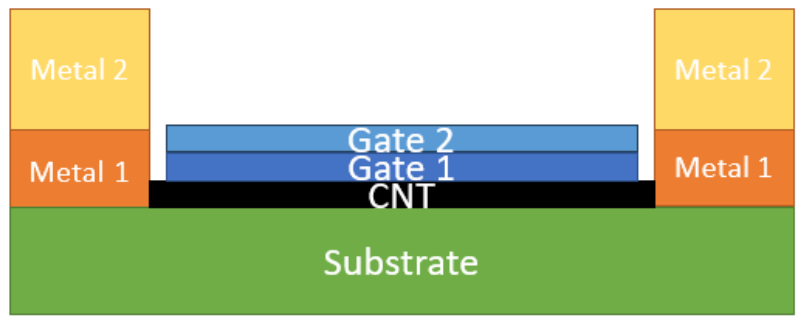

Hyper parameters of training of processing information incorporated model

| number of embedding layers | 6 |
|---|---|
| Encoding dimension | 512 |
| Embedding dimension | 512 |
| Learning rate | $10^{-5} - 10^{-7}$ |
| Batch_size | 1 |
| Optimizer | Adam |
| training epochs | 2000 |

(Table 4.6: Hyper parameters of training of processing information incorporated $log(I_{ds})$ model)

| number of embedding layers | 7 |
|---|---|
| Encoding dimension | 256 |
| Embedding dimension | 256 |
| Learning rate | $10^{-5} - 10^{-6}$ |
| Batch_size | 1 |
| Optimizer | Adam |
| training epochs | 1300 |

(Table 4.7: Hyper parameters of training of processing information incorporated $I_{ds}$ ratio model)

# Chapter 5

# Compact model for CNTFETs with non-aligned CNTs and SBI-based extraction of resistivity parameters

## 5.1 Introduction

With the development of the Carbon Nanotube (CNT) sorting technique, sorting out semiconducting CNTs with high purity (98%) becomes possible, making CNTFETs fabrication much easier. The sorted CNTs are usually in solutions when used for device fabrications; CNTs are usually randomly distributed and form a network if no specific aligning process is applied. Though CNTFETs with aligned CNTs tend to perform better, those with non-aligned CNTs have also demonstrated a decent performance and have the advantage of easy fabrication. CNT network FETs can achieve a value as high as $10^7 - 10^8$, which is good enough for lots of cases.

However, for the further application of non-aligned CNTFETs, models need to be set up to predict device performance. DFT-related calculations have been done on the resistance of two CNTs intersecting, but the time cost would be unimaginable if we used it for a CNTFET that contains hundreds of CNTs. A simpler way to model CNT networks, which lots of work does, is to treat CNT as straight sticks or hollow cylinders where the contributing resistance resides in the lengths and junctions. These works can predict how CNT density can affect the successful conduction of the SWNT network from source to drain and predict output current variation with different gate lengths. However, since SWNTs are randomly distributed in the network, the output current is bound to have a distribution, and only some works have tried to explain it. Empirical models are also complex to establish due to the difficulty of extracting necessary parameters since electrons can hop between CNTs, and the parameters associated with them are hard to extract through traditional ways like linear approximation or exponential transformation.

However, with the new development of artificial intelligence, we may have more tools to solve these problems. Here, we used simulation-based inference [82][83][[84]][85] as a new tool to extract critical parameters from models. The advantage of using SBI for parameter extraction is that it does not require models to be simple expressions and can tolerate the case where the model produces a distribution of outputs rather than specific numbers. Simulation-based inference models the probability of outputs of a model with different combinations of parameters, which can later be used to infer the most likely parameters combo for the real-world data distribution. This makes it a good candidate for the case where parameters are difficult to extract.

In this work, we probed a way of using SBI to extract parameters for a model observing CNTFET performance distributions. We first developed a compact model for non-aligned CNFETs based on the compact model of aligned CNTFETs. Our model can create a current distribution rather than producing a single current value. We then used simulation-based inference to infer critical parameters in the model and successfully inferred the parameters that fit the experimentally observed distribution. Our research shows that SBI can be successfully applied to assist in setting up a compact model with distributed outputs.

## 5.2 Background

**Previous research for CNT intersections**

In SWNT networks, current can not only flow in CNTs but can also flow between two intersecting CNTs. This makes it an interesting subject for many researchers. For a low-density CNT network, percolation theory [86] [87] may be a good way to explain its conductivity. However, when the density of CNTs is higher, like above 10 CNTs per um, the conductivity of CNTs is more affected by the conductivity of CNTs themselves and the resistance of CNT-CNT intersections. The resistance of CNT-CNT intersections were calculated through DFT, and shows that it have a value of around 700 kΩ [78].At the same time, conducting AFM has also been used to characterize SWNT-SWNT junctions and shows that the resistance is around 200kΩ [88].However, conducting AFM may not be an good way to observe the conductance in CNTFETs due to the large number of CNTs involved, and thus, it will become highly time-consuming. The sensitivity of conducting AFM to the experimental setup environment may also mean that the observed data may have deviations.

**Compact model for aligned CNTFET**

The compact model is a widely used way to characterize semiconductor device performance. Compact models are built upon physical rules, like gate-voltage-induced charge accumulation and drift-diffusion current driven from source-grain bias ($V_{ds}$). Compact models have already been set up for aligned CNT FETs [89][90][91][92]. An easy way to set up one is through the virtual source method. An example can be written as follows [91][92]:

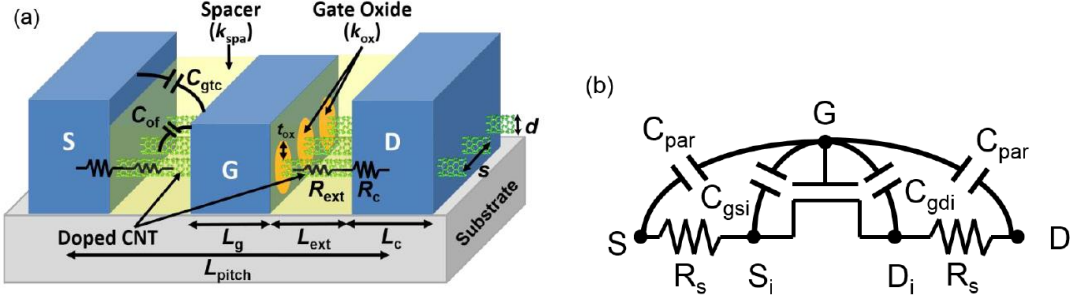

(Fig 5.1: Setup of compact model for CNTFETs (a) Structure of an aligned CNTFET; (b) Setup of the circuit)

$$V_t = V_{to} - \delta \cdot V_{dsi}$$

$$F_f = \frac{1}{1 + exp\left(\frac{V_{gsi} - [V_t - \alpha \cdot \phi_t/2]}{\alpha \cdot \phi_t}\right)}, \phi_t = \frac{k_B T}{q}$$

$$Q_{xo} = C_{inv} \cdot n_{ss} \cdot \phi_t \ln\left(1 + exp\frac{V_{gsi} - [V_t - \alpha \cdot \phi_t \cdot F_f]}{n_{ss} \cdot \phi_t}\right)$$

$$V_{DSATs} = \frac{v_{xo} L_g}{\mu}$$

$$V_{DSAT} = V_{DSATs}(1 - F_f) + \phi_t \cdot F_f$$

$$F_s = \frac{V_{dsi}/V_{DSAT}}{[1 + (V_{dsi}/V_{DSAT})^\beta]^{1/\beta}}$$

$$I_{dS} = Q_{xo} v_{xo} F_S$$

Here, $I_{ds}$ is the current flow from source to drain. $V_{to}$ is the threshold voltage without Drain-induced barrier lowering (DIBL) effect. $k_B$ is the Boltzmann constant, T is the temperature, and q is the elementary charge. The inversion gate Capacitance $C_{inv}$ is determined by the thickness and dielectric constant of the gate materials, and the CNT diameter. $\alpha$ and $\beta$ are empirical constants where $\alpha = -3.5, \beta = 1.8$.

## 5.3 Experimental design

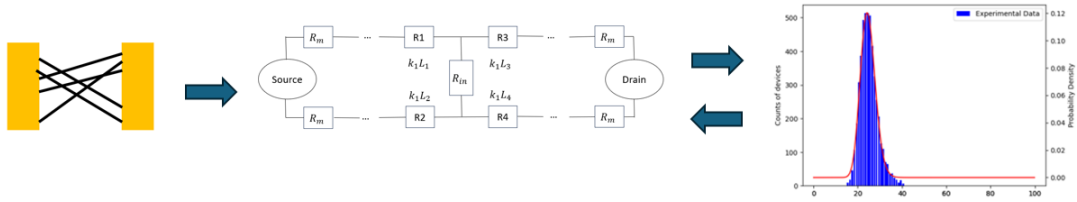

(Fig 5.2: Illustration of experiment setup. From left to right are: random CNT network, corresponding equivalent circuit and current output distribution)

The reason for the random current output of a CNTFET with random CNT network is because the length of CNTs in conductance and the way they overlap with each other is randomized. In other words, the conductance of a random network CNTFET is a random variable, since the length of CNTs in conduction and the way they interconnect are random variables. We designed a method of transforming the probability distribution of the length of CNTs in conduction and the way they interconnect are random variables in a random CNT network into the probability density function of its possible current output under a fixed voltage bias using a compact model based function. We adjust the three resistance related parameters in the compact model function so that the function can convert the random variable of CNT network to the probability density of the experimentally observed current output.

## 5.4 Methods

**CNT network generation**

To simulate non-aligned CNTFETs, we first need to sample random CNT networks. For a device with gate length of $L_g$ and a CNT density n per gate width, we first create a device area of $L_g \times 1$. In this device area, we randomly draw n points as the center of CNTs and assigned these CNTs with random orientations $\Theta$ in $(-\frac{\pi}{2}, \frac{\pi}{2})$ with lengths from the CNT length distribution. Since we only consider the conduction contribution of CNTs in the device area, we only keep the CNT parts inside the device area. By setting the source-to-drain direction as the x-axis, and metal contact direction as the y-axis, we can express CNTs in the form $y = k * x + b$ and calculate the position of intersection and the length of each CNT.

We construct a compact model for non-aligned CNTFETs to calculate the current flow in a CNTFET based on the CNT position information. Three types of resistances are used to build the model, which are CNT sections resistance $R$, CNT-CNT percolation resistance $R_{intersect}$ and metal contact resistance $R_m$. The model first constructs a circuit netlist basing on the CNT position information using the length of CNT sections and positions of connections.

**Circuit netlist setup**

1. Create resistors for each sector of CNT.
2. Create a resistor with a fixed value resistance for each position where two CNTs intersect (marked by a pair of position_A and position_B).

3. create a metal-CNT resistor with fixed resistance for each CNT-metal connection.

4. Apply voltage bias between source and drain and calculate current.

5. For each device, draw threshold voltage from the $V_{th}$ distribution, and calculate the current factor of $V_{th}$.

6. Multiply the $V_{th}$ current ratio with the calculated device current to produce the final current.

Here we use a simple case when only two CNTs are in the CNTFETs. Metal-contact resistance $R_m$ are added to both source and drain contact of the CNT, and an intersection resistance $R_{in}$ will be created to connect the CNTs at the point they intersect. R1, R2, R3 and R4 are CNT sections resistances that is related to their lengths. The value of these resistances will be discussed in the following section.

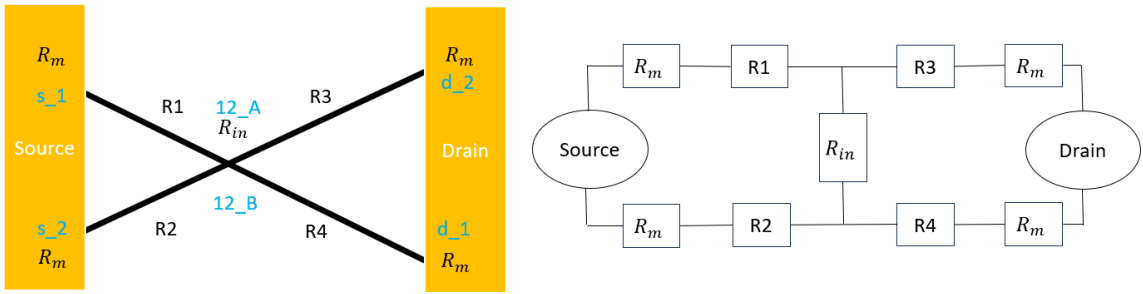

(Fig 5.3: Example with only 2 CNTs in the CNTFET (left) and corresponding circuit set up (right))

**Compact model for non-aligned CNTFETs**

In the aligned CNTFETs, the resistance between source and drain can be written as

$$R_{ds} = \frac{V_{ds}}{Q_{xo}v_{xo}F_S}$$

$$= \frac{k_1}{C_{inv} \cdot n_{ss} \cdot \phi_t \ln\left(1 + exp\frac{V_{gsi} - [V_t - \alpha \cdot \phi_t \cdot F_f]}{n_{ss} \cdot \phi_t}\right) \cdot v_{xo} \cdot F_S}$$

$$= \frac{k_1}{C_{inv} \cdot v_B} \cdot \frac{\lambda_v + 2L_g}{\lambda_v}$$

$$\cdot \frac{1}{n_{ss} \cdot \phi_t \ln\left(1 + exp\frac{V_{gsi} - [V_t - \alpha \cdot \phi_t \cdot F_f]}{n_{ss} \cdot \phi_t}\right) \cdot F_S}$$

If we suppose the resistance of CNTs in the non-aligned CNT network changes in the same way as those of aligned ones, we can construct the compact model for non-aligned CNTFETs in the following way:

$$I_{dS} = \frac{V_{ds}}{R_{network}} \cdot V_{xo} \cdot F_S$$

where

$$R_{network} = F(R_{section}, R_{in}, R_m)$$

$$R_{section} = \frac{k_1}{C_{int}v_{B0}} \cdot \frac{2\,l_{section} + \lambda_v}{\lambda_v} \cdot T_d$$

$$F_f = \frac{1}{1 + exp\left(\frac{V_{gsi} - [V_t - \alpha \cdot \phi_t/2]}{\alpha \cdot \phi_t}\right)}, \phi_t = \frac{k_B T}{q}$$

$$V_{xo} = n_{ss} \cdot \phi_t \cdot \ln\left(1 + exp\left(-\frac{V_{gsi} - [V_t - \alpha \cdot \phi_t \cdot F_f]}{n_{ss} \cdot \phi_t}\right)\right)$$

$$V_{DSATs} = \frac{v_{xo} L_g}{\mu}$$

$$V_{DSAT} = V_{DSATs}(1 - F_f) + \phi_t \cdot F_f$$

$$F_s = \frac{V_{dsi}/V_{DSAT}}{[1 + (V_{dsi}/V_{DSAT})^\beta]^{1/\beta}}$$

Here, $R_{CNT\_network}$ is the total resistance of the CNT network between source and drain, which is calculated based on the connections of CNTs and metal contacts. $R_{intersect}$ refers to the resistance for current exchange between two CNTs when they cross with each other, and $R_m$ refers to the metal contact resistance of one CNT with the metal contact.

$l_{section}$ is the length of SWNT sections before it crosses with another SWNT or metal contact. Here, we treat each CNT section as an individual virtual source system, and the current flowing in each CNT section is driven by the electric potential difference between its two sides. Constants $\alpha = -3.5$, $\beta = 1.8$. Due to the limitation of CNT sorting techniques, the diameters of the semiconducting CNTs used for fabricating CNTFETs are usually between 1 to 2 nm. Since SWNT properties strongly depend on their diameters, we added a diameter-related resistance ratio $T_d$ to describe the effect. The expression of $T_d$ is discussed in the appendix, which goes as follows:

$$T_d = \frac{\left(\frac{\mu}{C_{inv\_in} \cdot v_{xo}}\right)\Big|_d}{\left(\frac{\mu}{C_{inv\_in} \cdot v_{xo}}\right)\Big|_{1nm}}$$

$C_{inv\_in}$ is the capacitance related with one CNT, which writes as

$$\frac{1}{C_{inv\_in}} = \frac{1}{C_{ox}} + \frac{1}{C_{qe}}$$

Here $C_{ox}$ is the capacitance from the gate oxide per CNT and $C_{qe}$ is the capacitance of one CNT. Some simplification has been made on $C_{ox}$, which is discussed in Appendix.

## 5.5 Experimental setup

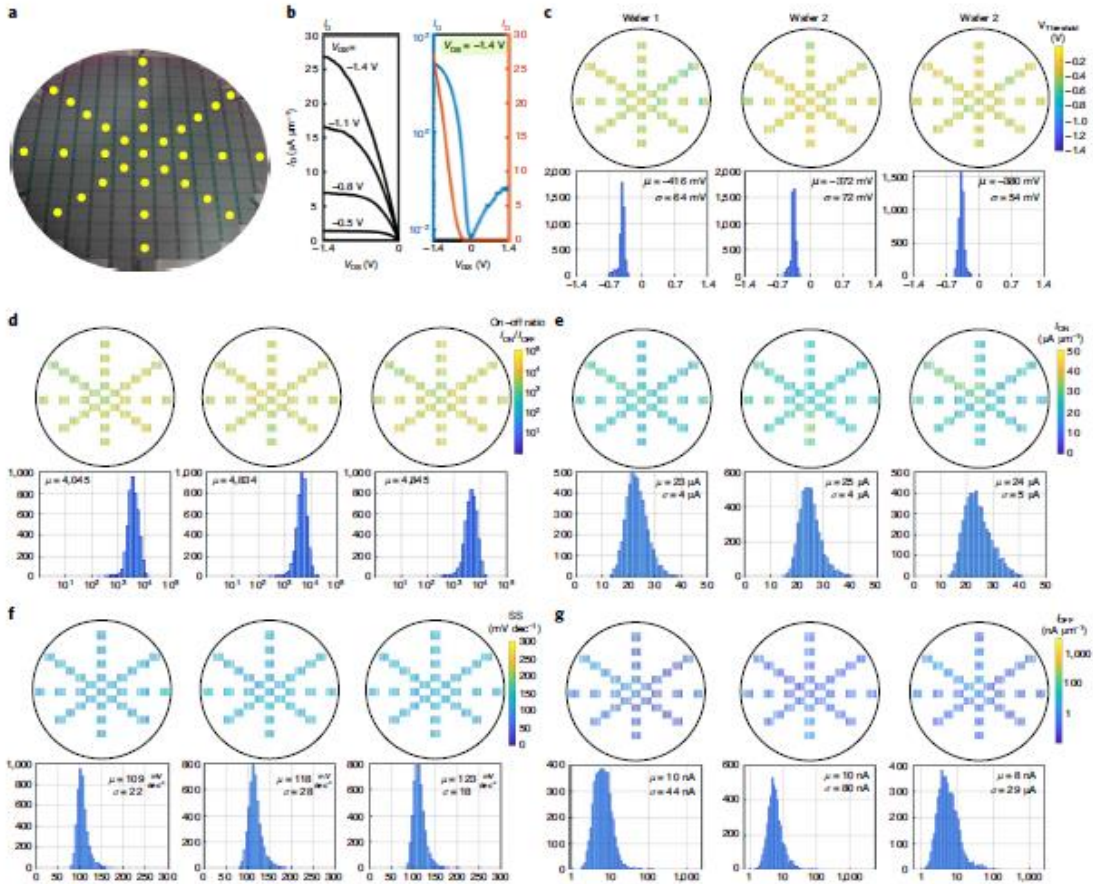

(Fig 5.4: Original data of CNTFET with randomly deposited CNTs.[94] The experiment was done on three wafers, and distributions of On-off current, On-off ratio, threshold voltages and subthreshold swing are given)

We used the data from [94] as the source of real-world observation. In its measurement result, 5 distributions are given, including threshold voltage $V_{th}$, subthreshold slope SS, on-current $I_{on}$, off-current $I_{off}$ and on-off ratio. To obtain an accurate model, all these distributions need to be considered. However, some of these distributions can be easily expressed with other distributions, so they don't need to enter the model by themselves. The first one is the on-off ratio, which directly correlates with $I_{on}$ and $I_{off}$. The other one is $V_{th}$, which can be derived from SS distribution through the following way

$$n_{ss} = -\frac{\partial E_{cmax}}{\partial V_{gs}}\bigg|_{V_{ds}=0} = \frac{1}{1-e^{-\eta}}$$

$$\delta = \frac{\partial E_{cmax}}{\partial V_{ds}}\bigg|_{V_{ds}=0} = e^{-\eta}$$

$$-\Delta V_t = (2E_{fsd} + E_g)e^{-\eta}$$

$$\eta = \frac{L_g + 2L_{of}}{2\lambda}$$

Therefore

$$\Delta V_t = -(2E_{fsd} + E_g)\left(1 - \frac{1}{n_{ss}}\right)$$

So, we suppose that $V_t = k * \frac{1}{n_{ss}} + b$, and infer the parameters k and b. The $V_t$ distributions are successfully generated with their corresponding SS distributions.

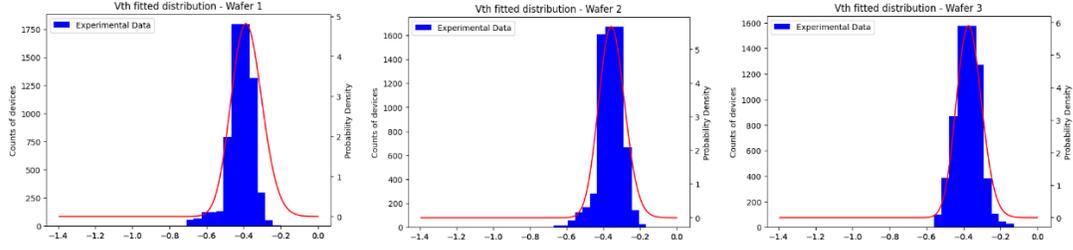

(Fig 5.5: $V_t$ distribution generated by SS distribution compared with the 3 measured wafers)

Since SS and $V_{th}$ can be affected by various factors and are hard to simulate, we expressed $V_{th}$ with SS and treated them s an input of the model. The on-off ratio was determined by the distribution of $I_{on}$ and $I_{off}$, so we only chose $I_{on}$ and $I_{off}$ distribution and simulation targets.

Simulator setup

Since the fabricated devices have a channel length of 285 nm and the current distribution is expressed in the way of current per um of gate width, we constructed a device area with a length of 285nm and a width of 1um. The CNT density of the original experiment is around 45 CNTs per um of the gate. To observe the distribution of device performance, we generated 100 devices in each run of the simulator and fitted the output current with gamma distribution, and the distribution is used as the output of the model. The experiment's on and off current distributions are also fitted with gamma distribution and are used as the sampling target. The gamma distribution is the correct choice here since it can model unsymmetric distribution, which is the case for current distributions here. The gamma distribution writes as

$$f(x, \alpha, \beta) = \frac{\beta^\alpha x^{\alpha-1} e^{-\beta x}}{\Gamma(\alpha)}$$

Where x is the input variable. $\alpha$ and $\beta$ are the two parameters determining the distribution of x, and thus are used as the output of the simulator and target of sampling.

Thus, we created a device performance distribution simulator that outputs $\alpha$ and $\beta$ of the gamma distribution of the device performance. The simulator will create devices 100 times and calculate their current flow with the input parameters. Then, the current results will be fitted with a gamma distribution, and give out $\alpha$ and $\beta$.

Inference of model parameters

We created prior distributions for the parameters as uniform distribution, and set their ranges as $k: [0.3, 1]$, $R_m: [1, 20]$, $R_{in}: [1, 500]$. The unit of $R_{metal\_contact}$ and $R_{intersect}$ are $k\Omega$, and k is 1. The range of the parameters are selected close to their real-world observations to make them physically sound, with the metal contact resistance range chosen from [83]. The circuit netlist is set up with Pyspice [96] [97] with voltage applied between source and drain and source grounded. The SBI agent is imported from the SBI-toolkit [98] and SNPE method was used. For the posterior estimator training step, we let the posterior simulator call samples from the simulator for 600 times and trained an estimator. The training time is around 26 hours with one CPU. After training, 100,000 samples are drawn from the posterior, and a parameter distribution is plot.

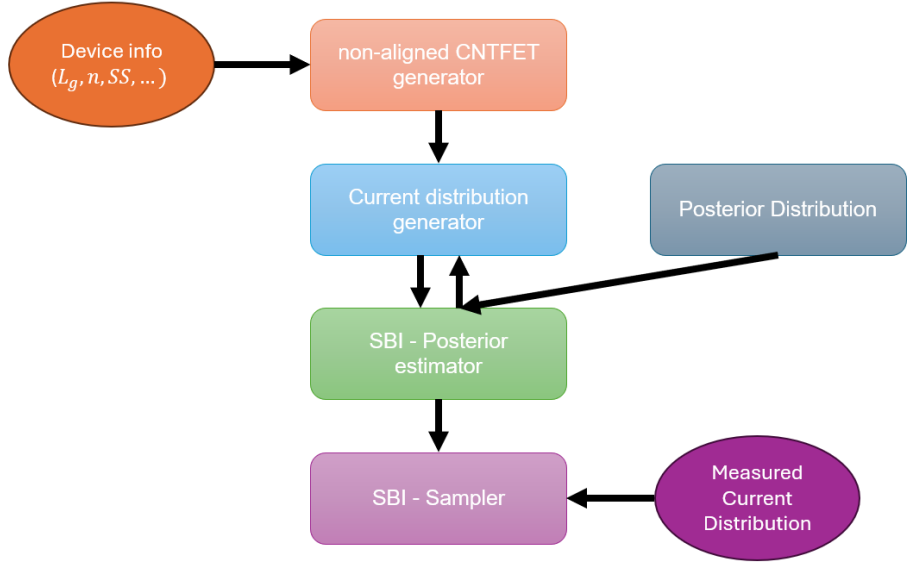

(Fig 5.6: Data flow of the posterior training and sampling)

## 5.6 Results

Here, we have the inferred parameter Divided by the middle of the SS_Vth-induced current ratio. We get that the inferred CNT-metal contact resistance $R_m$ is around 10 $k\Omega$ and CNT-CNT percolation resistance $R_{ex}$ is around 120 $k\Omega$. This is close to the previous experimental studies where $R_m$ is around $5 - 10$ $k\Omega$ and $R_{ex}$ is around 150 $k\Omega$. At the same time, the inferred value of CNT conductance is around 77.3 $k\Omega/\mu m$, and CNT-metal contact resistance is around 150 $k\Omega$. Since the channel length is 0.285um, the resistance of CNT is much smaller than percolation resistance. By inputting a parameter set near the inferred set, we get the current distribution, which fits the measured results. We observed that k is around 1 for $I_{on}$, but tend to be higher for $I_{off}$. At the same time, the metal contact resistance seems to be higher for $I_{off}$.

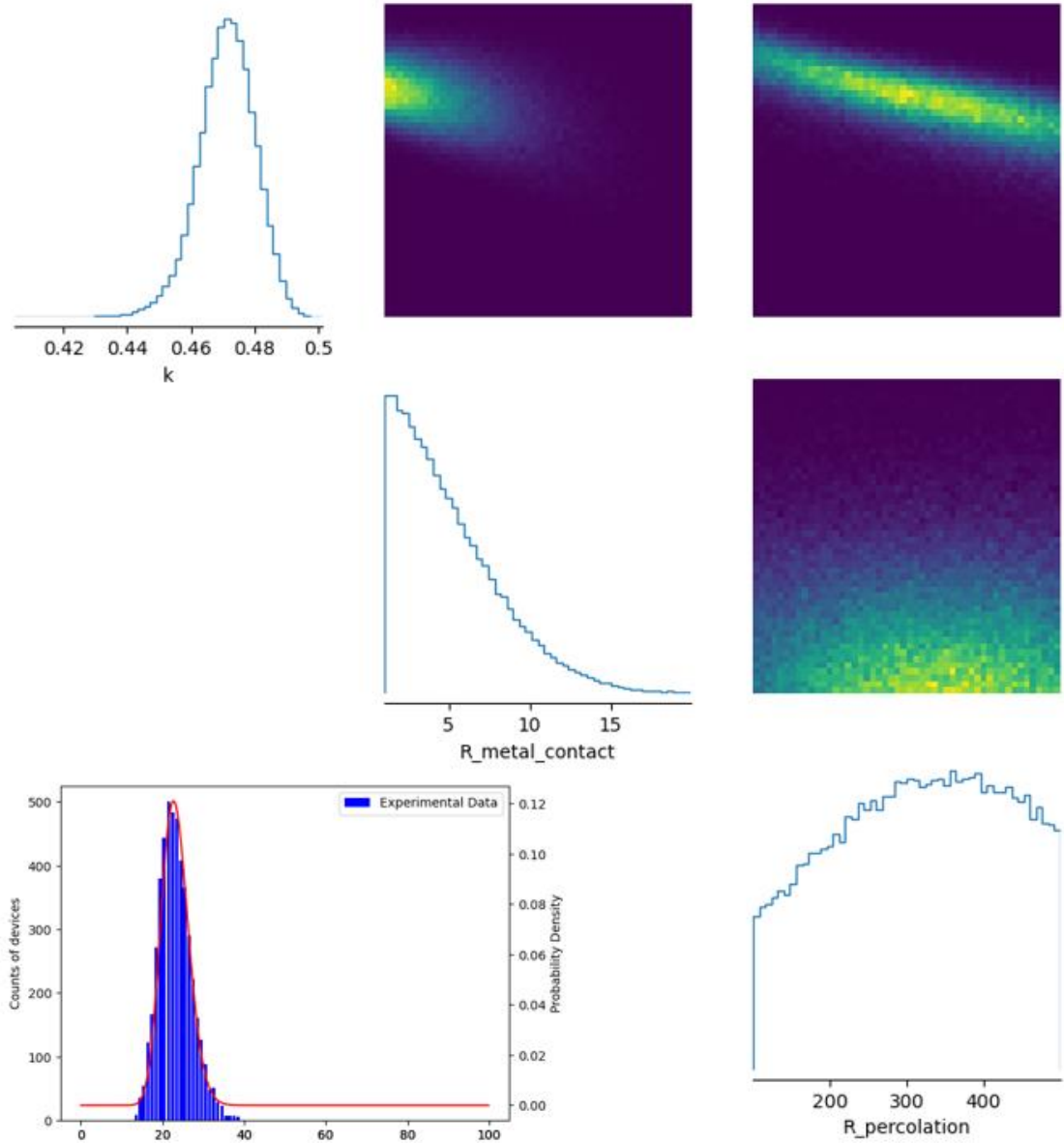

(Fig 5.7: Inferred parameters and fitted distribution for I_on distribution of wafer 1)

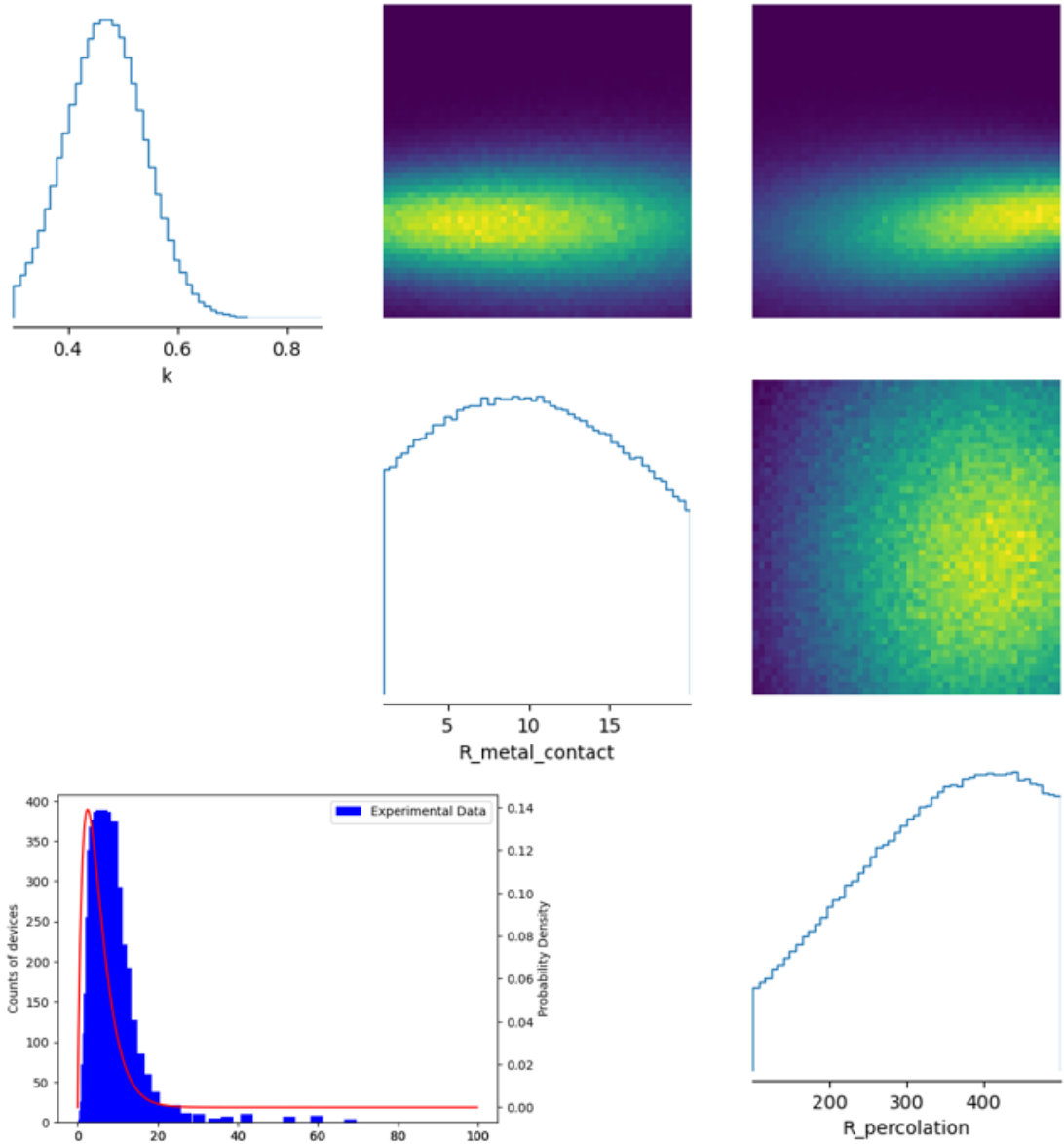

(Fig 5.8: Inferred parameters and fitted distribution for I_off distribution of wafer 1)

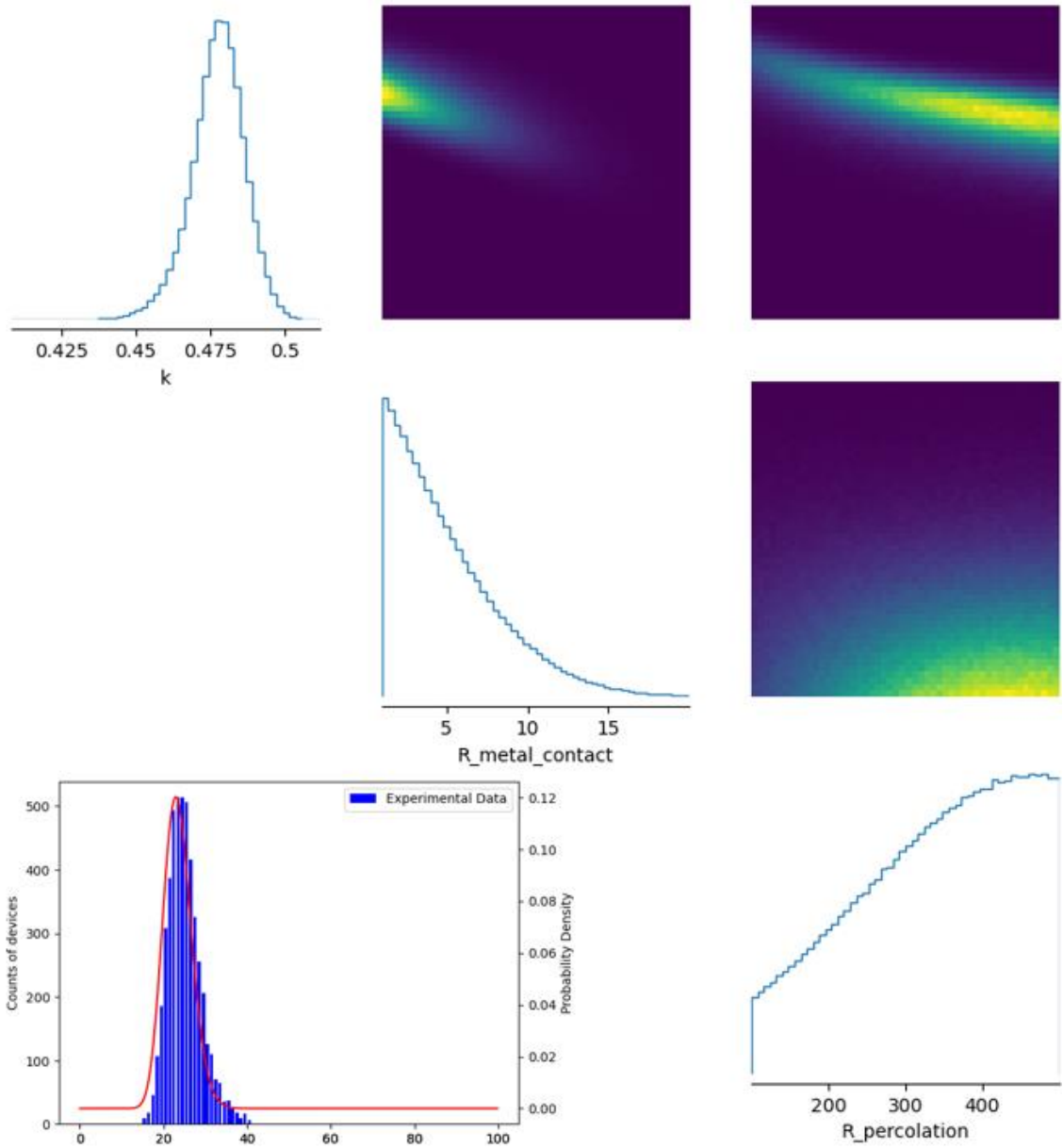

(Fig 5.9: Inferred parameters and fitted distribution for I_on distribution of wafer 2)

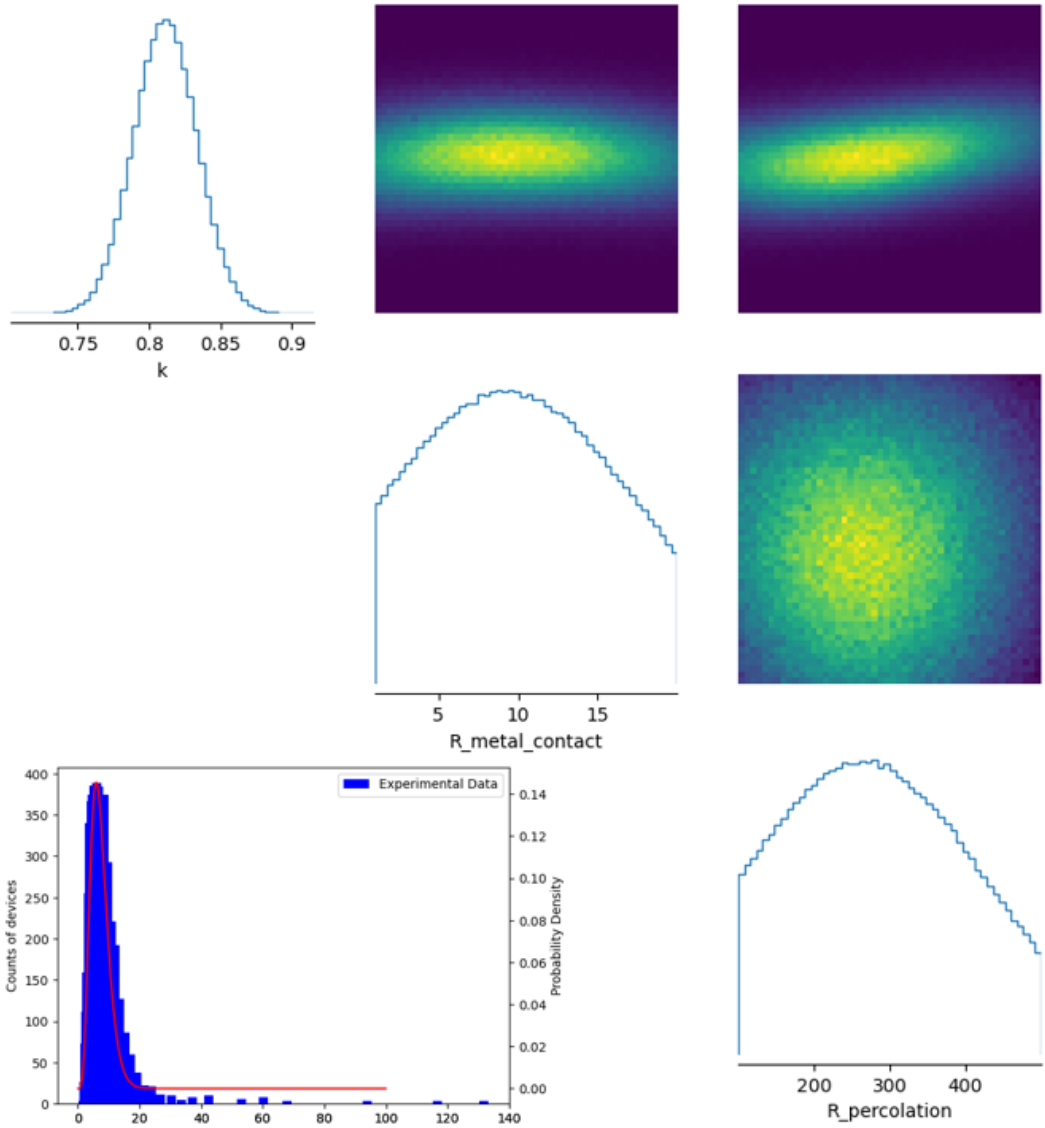

(Fig 5.10: Inferred parameters and fitted distribution for I_off distribution of wafer 2)

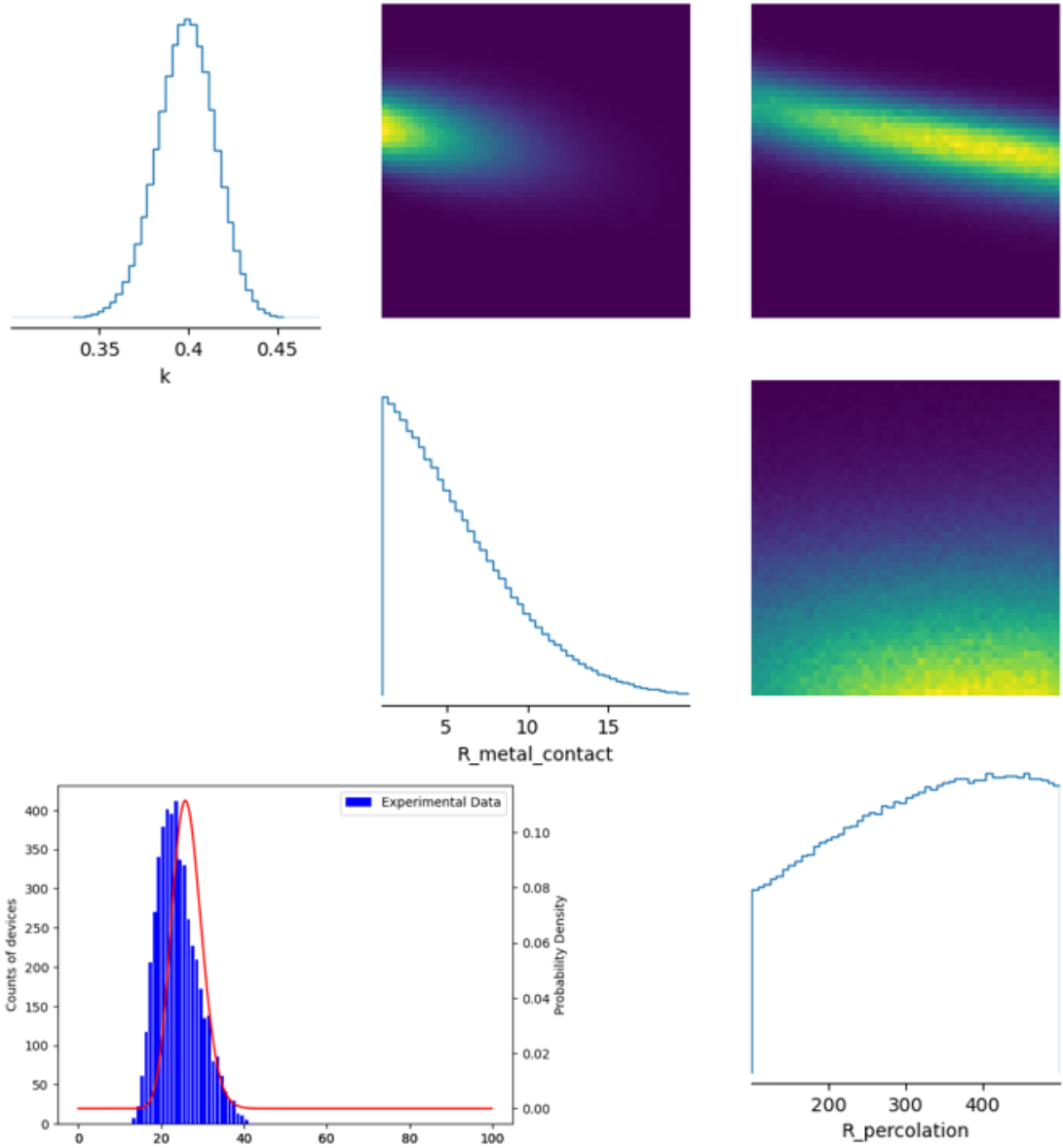

(Fig 5.11: Inferred parameters and fitted distribution for I_on distribution of wafer 3)

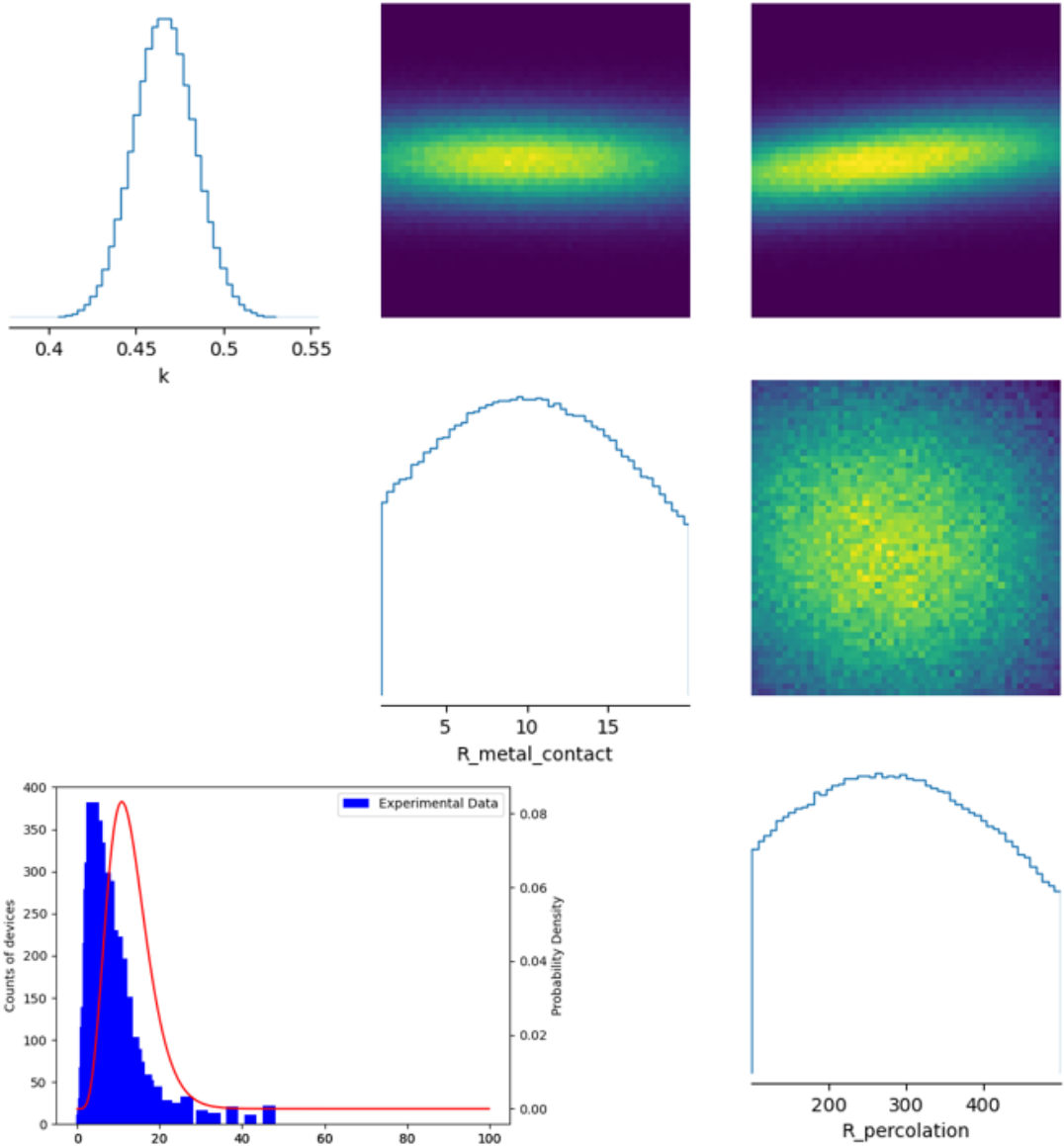

(Fig 5.12: Inferred parameters and fitted distribution for I_off distribution of wafer 3)

Running SBI successfully requires a correct model, and inference may fail if no correct model is provided. We performed a wrong inference by inversing the relationship of resistance change with length. We can see that the SBI agent fails to infer a distribution of CNT resistivity since no parameter combination can give a satisfactory result.

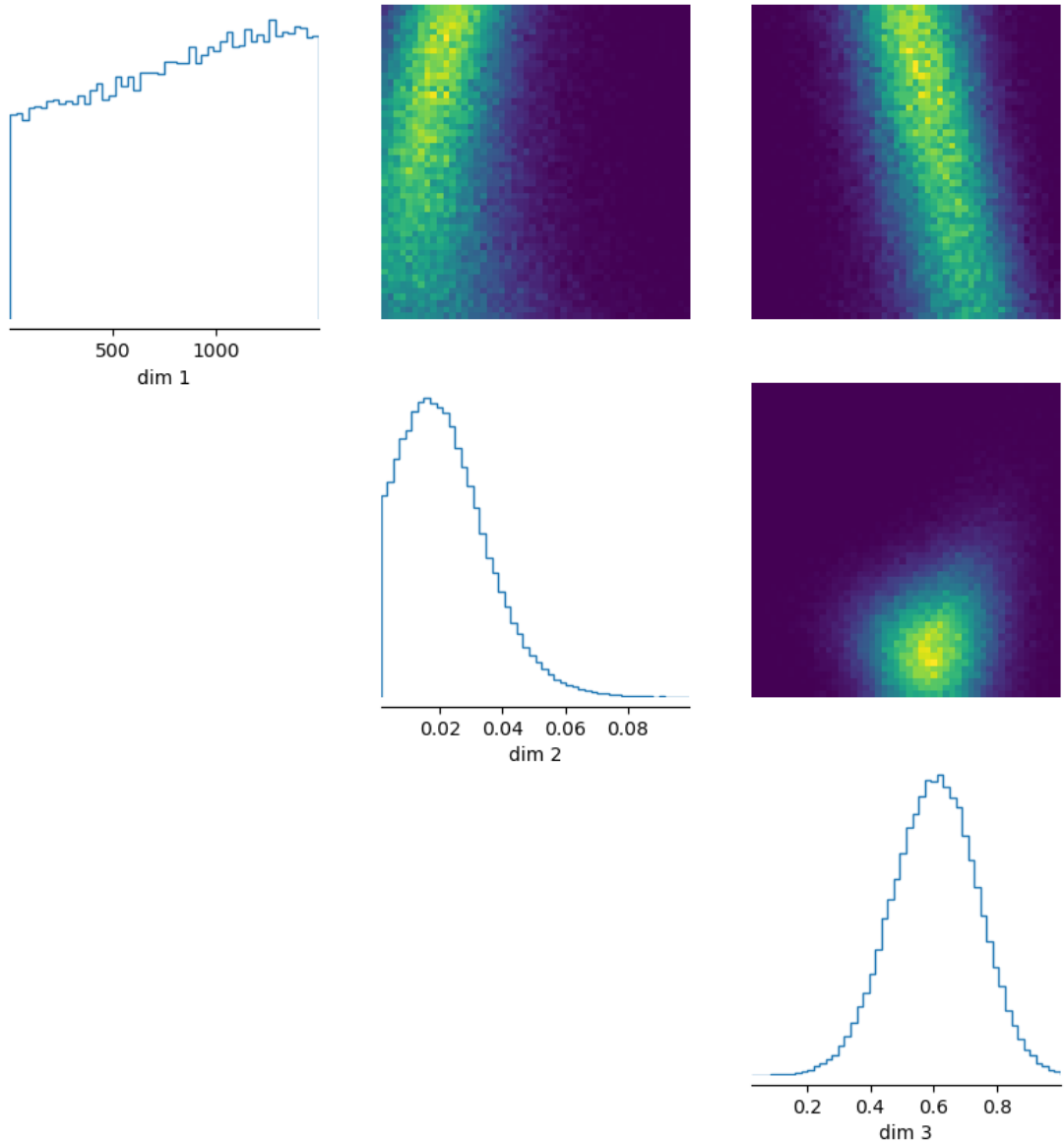

(Fig 5.13: Failed SBI due to wrongly set up model)

We used the inferred parameters to predict the current variation with different gate lengths. The currents are generated with a $V_{gs} = -1.4V$ and $V_{ds} = -1.4V$, CNT density at 45 CNTs per $\mu m$ and $SS = 120$. As is shown in the figure, the current will drop with increasing gate length, but the current drop tends to saturate, which fits with experimental observations.

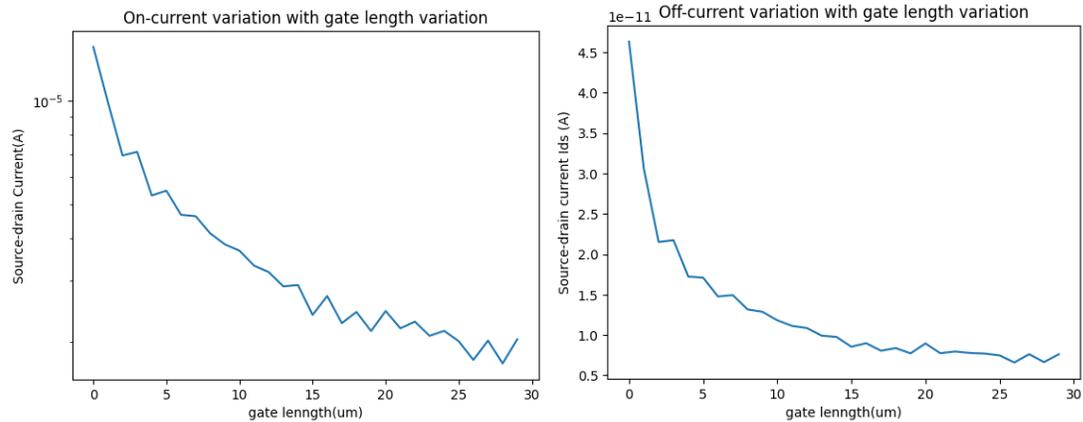

(Fig 5.14: On and off current variation with gate length variation)

We've also analyzed the effect of CNT density on CNTFET performance. As shown in the figure, though increasing CNT density leads to an enhancement in current, the enhancement tends to cease growing with increasing CNT density. This is also shown in the research [98] in which DFT calculation was used. The underlying reason is that the increase in CNT density separates the CNTs into smaller sections, which means more virtual sources.

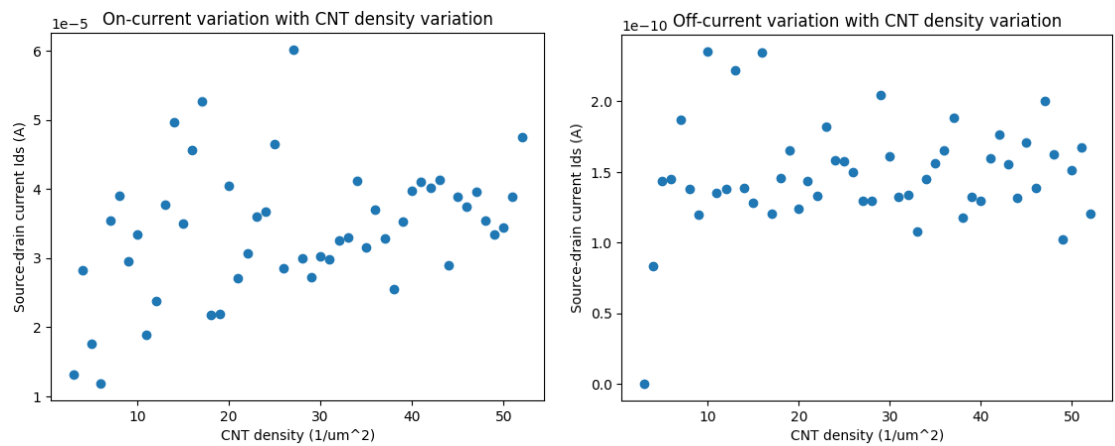

(Fig 5.15: On and off current variation with CNT density variation)

## 5.6 Conclusion and Future Research

We developed a compact model for non-aligned CNT network field effect transistors (CNTFETs). The model calculated the current flow in the CNTs and the current exchanges between two contacting CNTs. By design, this compact model considers the charge accumulation effect of gate bias and the source-drain bias that drives the current. This allows for predicting the performance of non-aligned CNT network field-effect transistors with both on and off-gate bias. We used SBI to extract intersection resistance, metal-contact resistance, and CNT resistivity and successfully found a parameter combo that fits with real-world observations. We believe that this research may open a way for extracting parameters of compact models in cases where device performance has varied.

At the same time, we used the model to explore the effect of CNT density and channel length on non-aligned CNTFETs' performance. We observed that the increase of device current with higher CNT density tends to saturate with higher CNT density, probably due to an increase in associated resistance. The decrease in current due to longer gate length also tend to saturate, which fits with real world observations.

# Appendix

**Approximation of gate approximation**

For the aligned CNTFETs, the total capacitance of the CNTFET $C_{inv}$ is

$$\frac{1}{C_{inv}} = \frac{1}{C_{ox}} + \frac{1}{C_{qe}}$$

where

$$C_{qe} = N \times \left[0.64\sqrt{E_g} + 0.1\right] (fF/\mu m)$$

Is the capacitance of CNTs, and N is the number of CNTs. $C_{ox}$ is the capacitance from the oxide material and is a little bit complicated to express. For a simple cylindrical GAA structure, where oxide material covers CNTs evenly, $C_{ox}$ write as:

$$C_{ox} = N \times \frac{2\pi k_{ox}\varepsilon_0}{\ln[(2t_{ox}+d)/d]}$$

For top-gate structure, $C_{ox}$ is presented in the following steps:

$$C_{gc\_sr} = \frac{4\pi k_{ox}\varepsilon_0}{\ln\left(\frac{s^2+2(h_1-r)\left[h_1+\sqrt{h_1^2-r^2}\right]}{s^2+2(h_1-r)\left[h_1-\sqrt{h_1^2-r^2}\right]}\right) + \lambda_1 \ln\left[\frac{h_1+d_{CNT}}{9r^2+s^2}\right] \cdot \tanh\left(\frac{h_1+r}{s-d}\right)}$$

$$C_{gc\_inf} = \frac{2\pi k_{ox}\varepsilon_0}{\cosh^{-1}\left(\frac{2h_1}{d}\right) + \lambda_1 \ln\left(\frac{2h_1+2d}{3d}\right)}$$

$$r = \frac{d}{2}, h_1 = t_{ox} + r, \lambda_1 = \frac{k_{ox} - k_{sub}}{k_{ox} + k_{sub}}$$

$$C_{gc\_e} = \frac{C_{gc\_inf} \cdot C_{gc\_sr}}{C_{gc\_inf} + C_{gc\_sr}}, C_{gc\_m} = 2C_{gc\_e} - C_{gc\_inf}$$

$$C_{ox} = \begin{cases} C_{gc\_inf} & N = 1 \\ C_{gc\_m}(N-2) + 2C_{gc\_e} & N \geq 2 \end{cases}$$

Here, $C_{gc\_e}$ and $C_{gc\_m}$ denote the capacitances from the gate to the CNTs at the edge and to the CNTs in the middle of the CNT array, respectively. For aligned CNTFETs, $C_{ox}$ can be easily calculated as a linear combination of $C_{gc\_e}$ and $C_{gc\_m}$ since they are parallel to each other. For non-aligned CNTFETs, there's no simple expression of $C_{gc\_e}$, and the gate oxide capacitance is scattered everywhere. However, we can see that with the increase of CNT density, $C_{ox}$ will be dominated by $C_{gc\_m}$. We calculated the deviation of $C_{ox} = C_{gc\_m} * N$ compared to $C_{ox} = C_{gc\_m}(N-2) + 2C_{gc\_e}$ with a CNT diameter of 1nm under the device fabrication condition, and the result is shown in the following figure

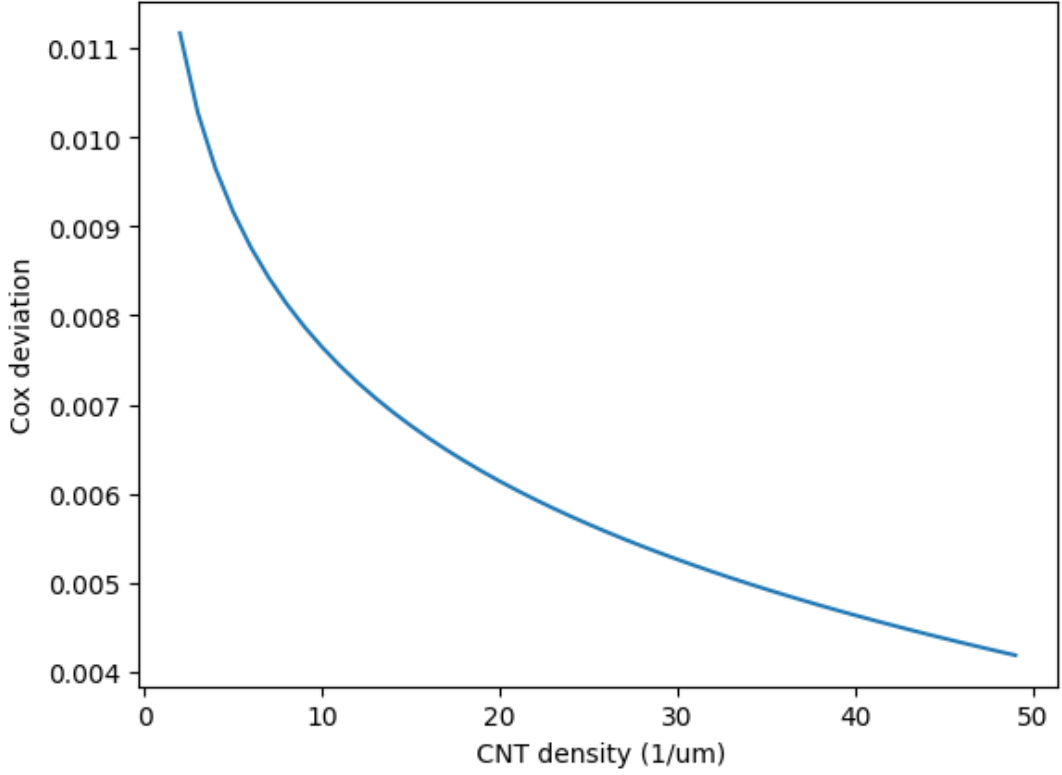

(Fig 5.16: $C_{ox}$ deviation with changing CNT density)

With the experimental CNT density around 45 CNTs per um, the deviation of $C_{ox}$ is around 0.4% from the real one. Therefore, we use the approximation $C_{ox} = C_{gc\_m} * N$ in this research, so the capacitance of the device writes as

$$\frac{1}{C_{inv}} = \frac{1}{N \times C_{gc\_m}} + \frac{1}{N \times [0.64\sqrt{E_g} + 0.1]} = \frac{1}{N} \times \left(\frac{1}{C_{gc\_m}} + \frac{1}{[0.64\sqrt{E_g} + 0.1]}\right)$$

So

$$C_{inv} = N \times \left(\frac{1}{\left(\frac{1}{C_{gc\_m}} + \frac{1}{[0.64\sqrt{E_g} + 0.1]}\right)}\right)$$

Therefore, the capacitance of each individual CNT is

$$C_{inv\_in} = \frac{1}{\left(\frac{1}{C_{gc\_m}} + \frac{1}{[0.64\sqrt{E_g} + 0.1]}\right)}$$

**Effect of CNT diameter on CNTFET performance**

Since the diameter variance of CNT affects the bandgap, $C_{inv}$ (thus $Q_{xo}$) and $v_{xo}$ (thus $V_t$), it should be taken into consideration to achieve a reasonable simulation result. The diameter of CNTs affects the device performance in the following ways:

1. CNT diameter d determines the bandgap of CNT:

$$E_g = \frac{2E_p a_{cc}}{d}$$

Where $E_p = 3eV$ is the tight-binding parameter, and $a_{cc}$ is the carbon-carbon distance in CNTs, 1.44nm. Bandgap affects CNT quantum capacitance $C_{qe}$, which is discussed below. Bandgaps of SWNTs affects the gate capacitance $C_{inv}$ of CNTFET.

2. CNT diameter d also affects the inversion gate capacitance $C_{inv}$ of CNTFET. As discussed in Appendix, the capacitance of one individual CNT is:

$$C_{inv\_in} = \frac{1}{\left(\frac{1}{C_{gc\_m}} + \frac{1}{[0.64\sqrt{E_g} + 0.1]}\right)}$$

Where both $C_{gc\_m}$ and $E_g$ are affected by d.

3. Effect on mobility

Diameter d also affects the mobility of the CNTFET.

$$\mu = \mu_0 \frac{L_g}{\lambda_\mu + L_g} \left(\frac{d}{1nm}\right)^{c_\mu}$$

$\mu_0 = 1350 \frac{cm^2}{V \cdot s}$, $\lambda_\mu = 66.2nm$, $c_\mu = 1.5$. They are empirical extracted.

4. Effect on virtual Source Velocity

$$v_{xo} = \frac{\lambda_v}{\lambda_v + 2L_g} v_B$$

$$v_B = v_{B0}\sqrt{d/d_0}$$

Therefore, we introduced a diameter resistance factor function in the model. If we neglect the small change of $V_{DSAT}$ on $\mu$, the source-drain current $I_{ds}$ is correlated with $C_{inv}$, $v_{xo}$ and $\mu$ in the following way:

$$I_{ds} \propto \frac{C_{inv} \cdot v_{xo}}{\mu}$$

Which means the resistance of each SWNT section is proportional to $\frac{\mu}{C_{inv} \cdot v_{xo}}$.

Therefore, for each SWNT section, we calculate the gate capacity $C_{inv}$, virtual source Velocity $v_{xo}$ and mobility $\mu$, then compare it with those for a SWNT with a diameter of 1nm to obtain a resistance ratio $T_d$.

$$T_d = \frac{\left(\frac{\mu}{C_{inv\_in} \cdot v_{xo}}\right)\Big|_d}{\left(\frac{\mu}{C_{inv\_in} \cdot v_{xo}}\right)\Big|_{1nm}} = \frac{\left(\frac{1}{C_{inv\_in}}\right)\Big|_d}{\left(\frac{1}{C_{inv\_in}}\right)\Big|_{1nm}} \cdot \left(\frac{d}{1nm}\right)$$

Additional thoughts on using resistance networks to characterize CNT network conduction

**Approximation of $V_{th}$ with SS distribution**

CNT diameter distribution

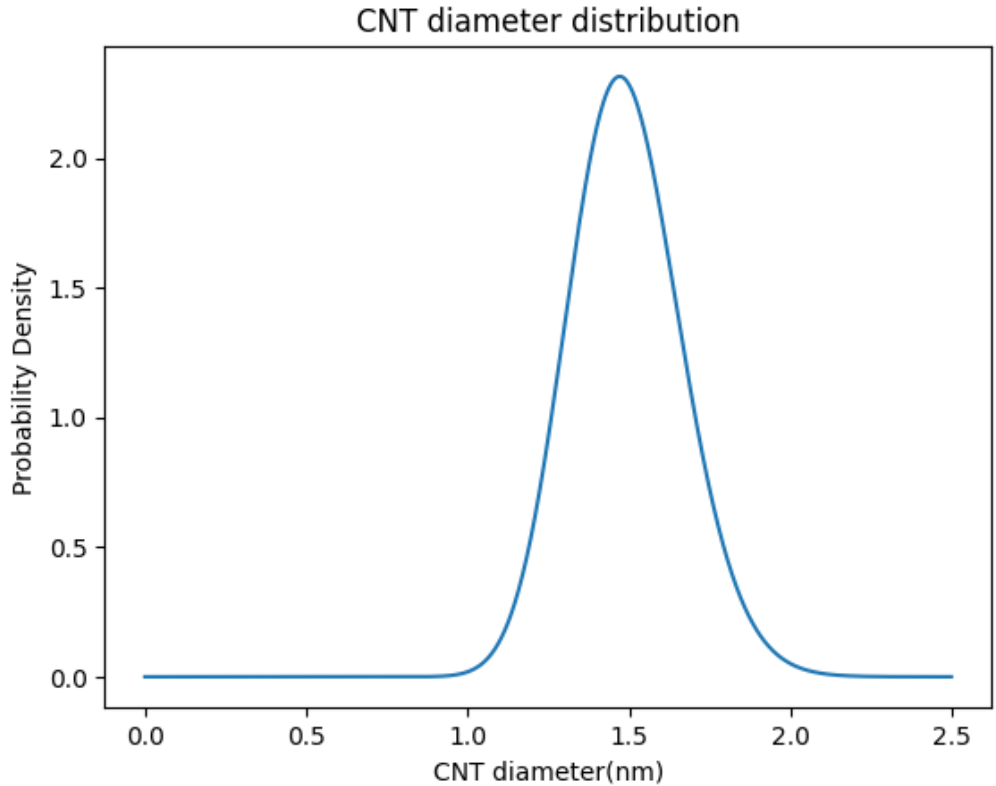

(Fig 5.17: CNT length distribution)

With the current sorting technique, the SWNTs used for fabricating CNTFETs are around 1 – 2 nm. Since the manufacturer does not provide the CNT diameter distribution, we obtained data from similar research using the same sorting technique and used it as the CNT diameter distribution in this research.

We model the CNT length distribution from the technical data sheet of IsoNanotubes-S of NanoIntegris, which is the material used in the experimental research.

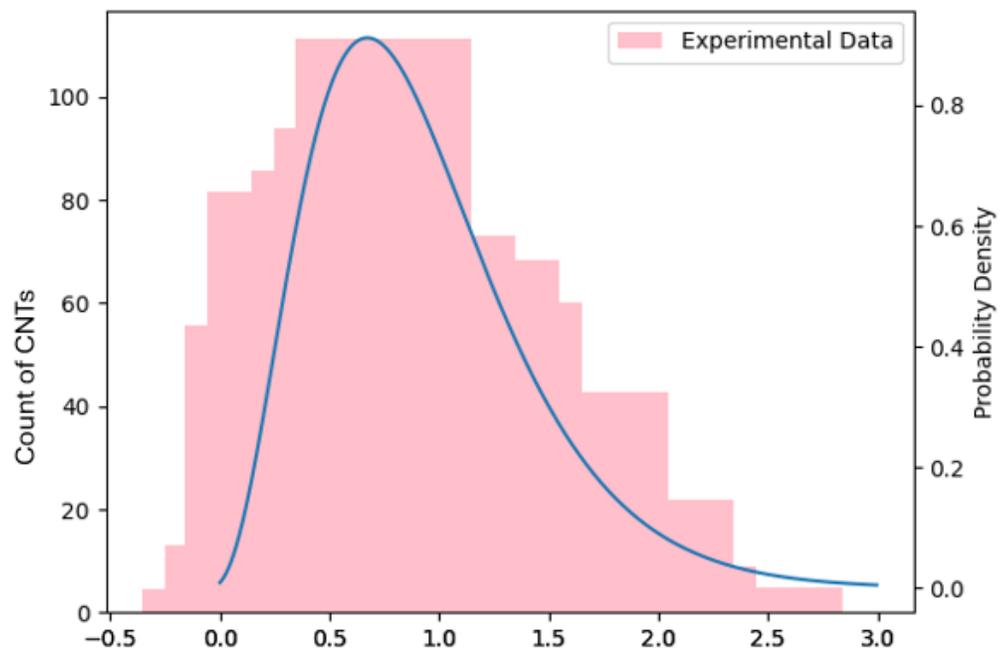

(Fig 5.18: Fitted Distribution of CNT diameter variation)

# Chapter 6

# Generative model for CNTFETs using GFLowNet

## 6.1 Introduction

With the development of CNTFETs, the number of processing methods also increases, making it harder to develop a combination of processing methods and device parameters to achieve target performance. The design of circuits also requires a careful choice of device parameters, which is tedious. Though some models have been prompted to use neural networks to model FETs, few have tried to generate device parameters with target device performance. Currently, detailed device parameters are usually manually selected, so a self-selecting mechanism will surely promote the development of this field. It may also serve as an advisory system for materials science researchers as a tool for accumulating and analyzing past experimental data.

**Choice of generative model**

The recent development of generative models has encouraged people to research in this area. Applications like ChatGPT and Auroa have shown that AI can generate dialogue, images and videos. The success of the generative model has aroused interest in research in generative models. However, the structure of these models may not be meaningful for semiconductor devices. All these models use transformers as their base models, which are proficient in treating sequential data. Transformers consider the sequence of each token and iterate it at each step. However, for semiconductor devices, what matters is the contribution of each parameter to the device performance, not the sequence of data input.

Therefore, transformers may not be a suitable choice for semiconductor generative models. Another generative model structure, generative adversarial network (GAN), uses a convolutional neural network that captures the relation between surrounding data, which is helpful in treating image data where the model tries to recognize patterns of a group of digits near each other but semiconductor device parameters affect device performance individually.

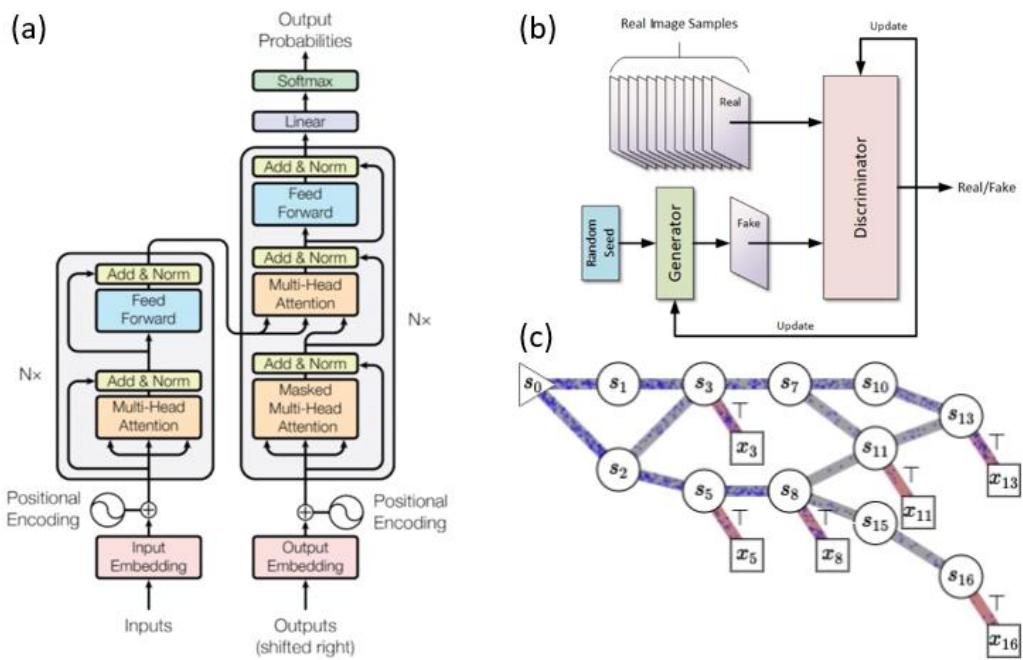

(Fig 6.1: Structures of popular generative models (a): Transformer neural network structure, (b): generative adversarial network (GAN) structure, (c): GFLowNet structure)

As a result, we choose GFlowNet as our technique for the generative model for CNTFETs. The concept GFlowNet used, which treats the effect of each variable as probability, also sounds more reasonable. We designed an environment for generating device parameters and actions for choosing them. Many essential parameters of CNTFETs are continuous, so we used the continuous GFLowNet technique. The target of the experiment is to generate device processing information with a target $I - V_{gs}$ curve since it includes essential information for circuit design like $V_{th}$ and SS. Multi-objective optimization is used here to deal with multiple goals.

**Continuous GFlowNet**

Though GFLowNet was initially designed only for categorical parameters, it can also take continuous values. Continuous GFlowNet represents continuous variables in a σ-finite measure that convert a finite numerical range H into N identities V, so that $H = \bigcup_{n \in N} V^n$, by segregating the continuous space with measure $\mu$. The flow balance for state flow then goes as:

$$\int_{\bar{S}} f(s')\mu(ds') = \iint_{S \times \bar{S}} f(s')\mu(ds)P_F(s, ds')$$

**Multi-Objective Optimization**

Multi-Objective Optimization (MOO) involves finding a set of feasible candidates $x^* \in X$ which simultaneously maximize d objectives $R(x) = [R_1(x), ..., R_d(x)]$. When these objectives are conflicting, there is no single $x^*$ that simultaneously maximizes all objectives. One way to solve MOO problem is scalarization, where a set of weights (preference) $\omega_i$ reassigned to each objective $R_i$, with $\omega_i \geq 0$ and $\sum_{i=1}^{k} \omega_i = 1$. The objective for training can either be a weighted sum scalarization $R(x|\omega) = \sum_{i=1}^{k} \omega_i R_i(x)$ that multiply weights with each objective, and it can be a weighted Tchebycheff that tries to minimize the distance of each objective $R_i$ : $R(x|\omega) = \max_{1 \leq i \leq d} \omega_i |R_i(x) - z_i^*|$, where $z_i^*$ is an ideal value for objective $R_i$.

## 6.2 GFlowNet for device dimensions design

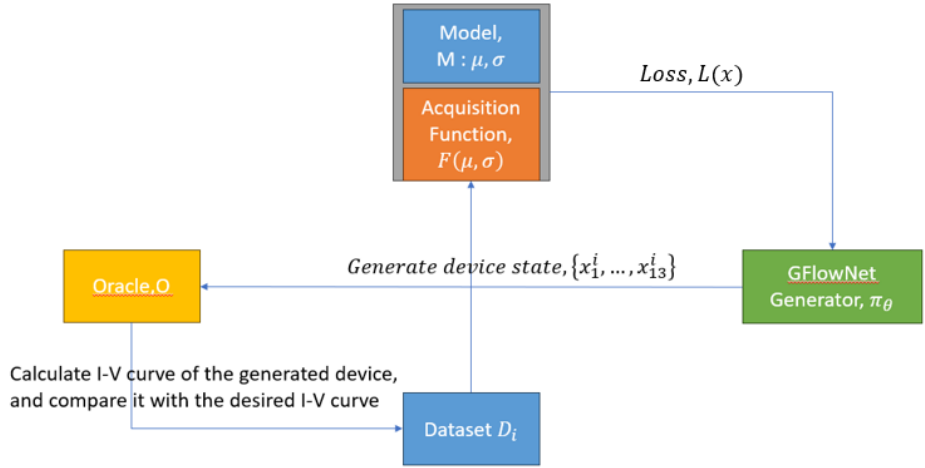

(Fig 6.2: Structure of GFLowNet for CNTFETs generation)

We begin with the basic function that GFlowNet can serve as a model to reproduce parameter distribution. We begin with a simple case that uses CNTFET compact model depicted in chapter 3 as proxy and choose three parameters: gate length $L_g$, CNT density n and oxide thickness $t_{ox}$ to form an action space. The environment is built on continuous GFLowNet.

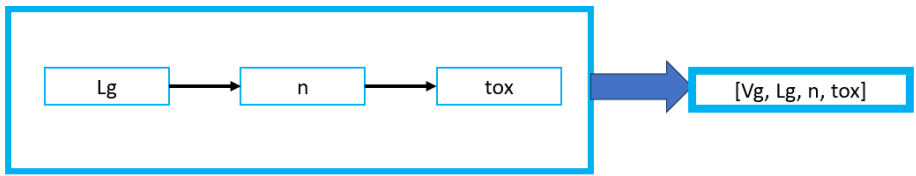

(Fig 6.3: Action space for GFlowNet with compact model)

We designed a reward function to test the generation ability of the model. We use ten $I_{ds}$ values with $V_{ds} = 0.3 - 3V$ under $V_{gs} = 1V$ and SS = 60. Since GFLowNet samples actions proportional to their resulting rewards, we designed a reward function that gives maximum value when the generated $I_{ds}$ values of the result device is the same as the target $I_{ds}$ as

$$Reward_i = 10 * (2 - e^{|I_{target} - I_{generated}|})$$

For each $I_{ds}$ points. The reward will have a maximum value of 10 if $I_{target} = I_{generated}$. We clip the reward value by a minimum of $10^{-4}$ for the ease of training. The target value was generated with the following parameters:

| Lg | CNT_density(n) | t_ox |
|---|---|---|
| 0.5 | 20 | 0.01 |

(Table 6.1: Parameters used for generating target value for GFLowNet with compact model)

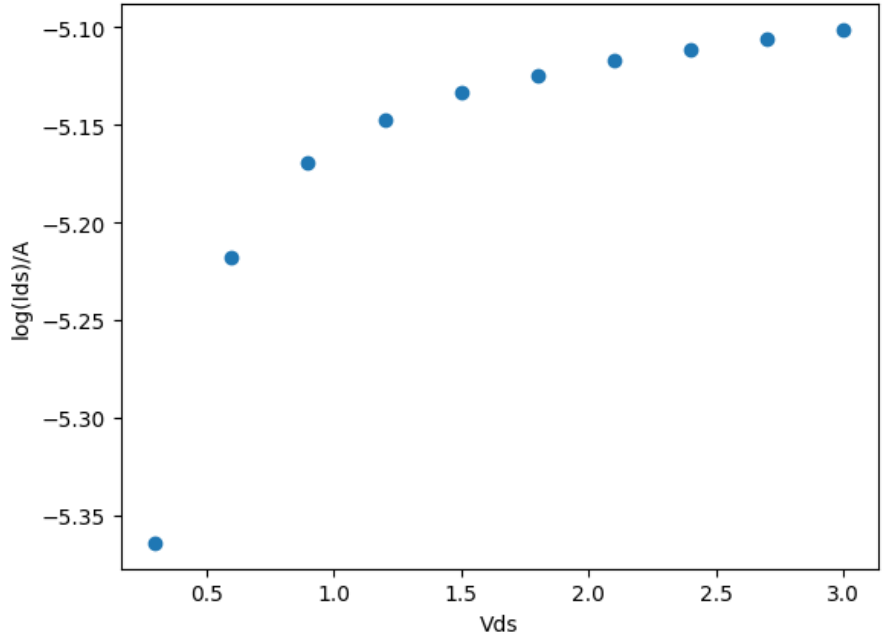

(Fig 6.4: $log(I_{ds})$ generation target for compact model based GFlowNet)

We begin the training with an input range of $L_g : (0.01, 1.5), n: (1, 50)\ t_{ox} : (0.04, 0.5)$. As the results shows, we are able to generate multiple results that can produce the target performance. $(L_g = 0.5, n = 20, t_{ox} = 0.01)$ is not the only choice to achieve the target performance, and combinations like $(L_g = 1.2, n = 45, t_{ox} = 0.04)$ can also generate similar outputs.

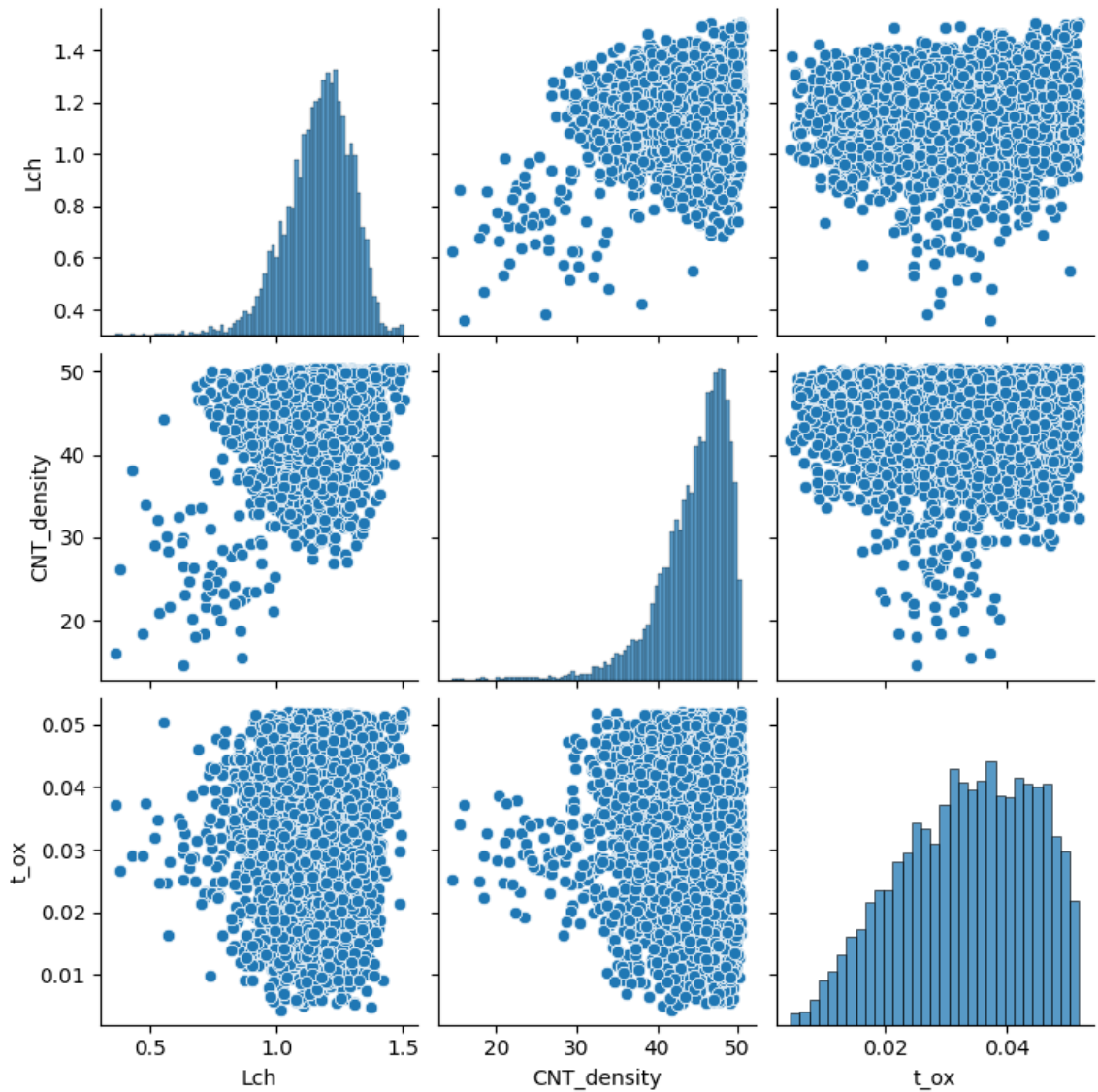

(Fig 6.5 Distribution of generated actions for GFLowNet with compact model)

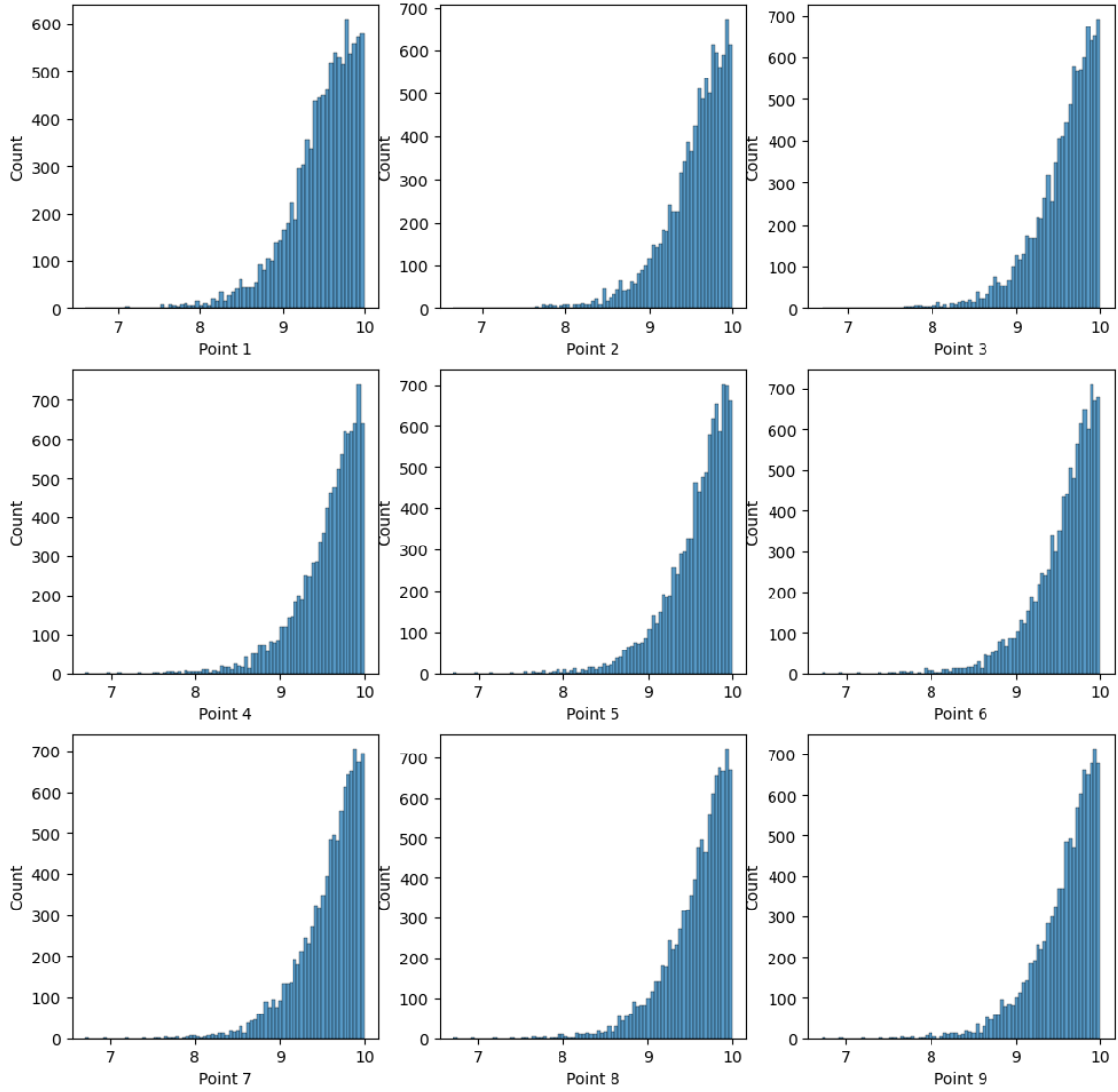

(Fig 6.6 Distribution of generated rewards for GFLowNet with compact model)

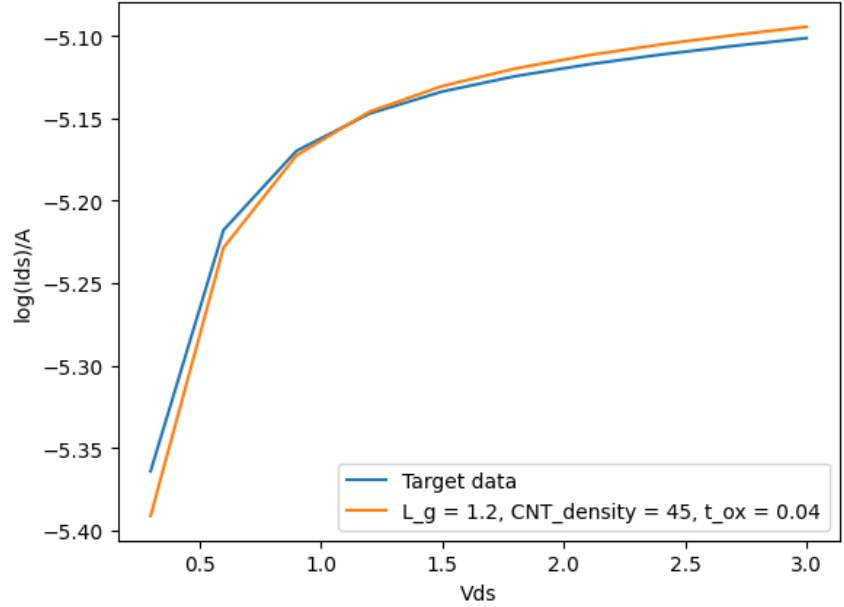

(Fig: 6.7 I-V curve generated by $L_g = 1.2, n = 45, t_{ox} = 0.04$ compared with target)

We also enlarge the training to an input range of $L_g$ : $(0.01, 3), n$: $(1, 100)$ $t_{ox}$ : $(0.04, 1)$. A combination of $(L_g = 2.5, n = 95, t_{ox} = 0.1$ ) is also fits well with the target. We've run the experiment for several times, and the maximum probability always falls near $(L_g = 1.2, n = 45, t_{ox} = 0.04)$, but the optimum solution generated with the smaller range, $(L_g = 1.2, n = 45, t_{ox} = 0.04)$ , is included in the generated results. This could result from the fact that those values near $L_g = 2.5, n = 95, t_{ox} = 0.1$  has a higher probability to generate the target I-V curve.

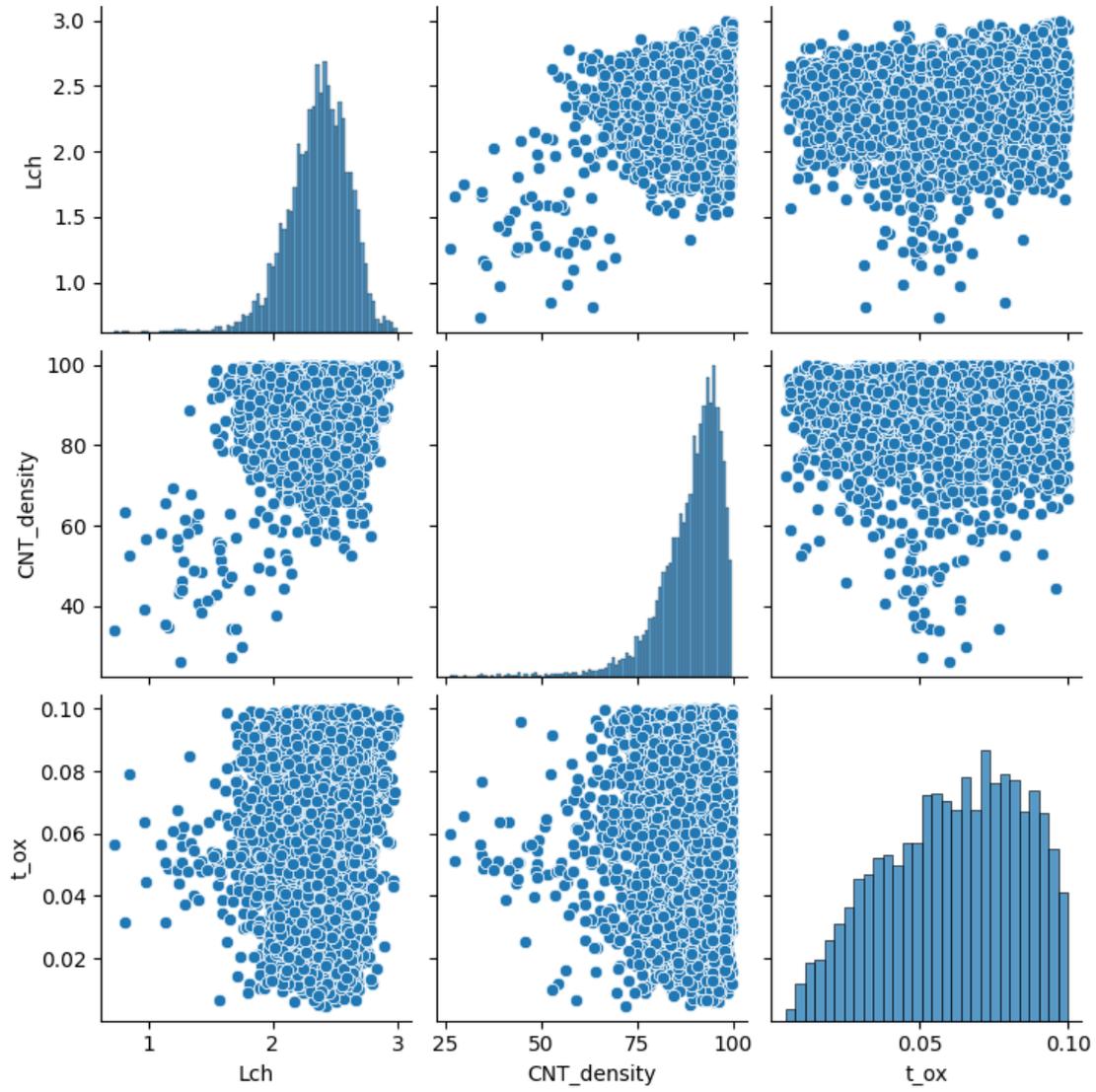

(Fig 6.8 Distribution of generated actions for GFLowNet with compact model for larger range)

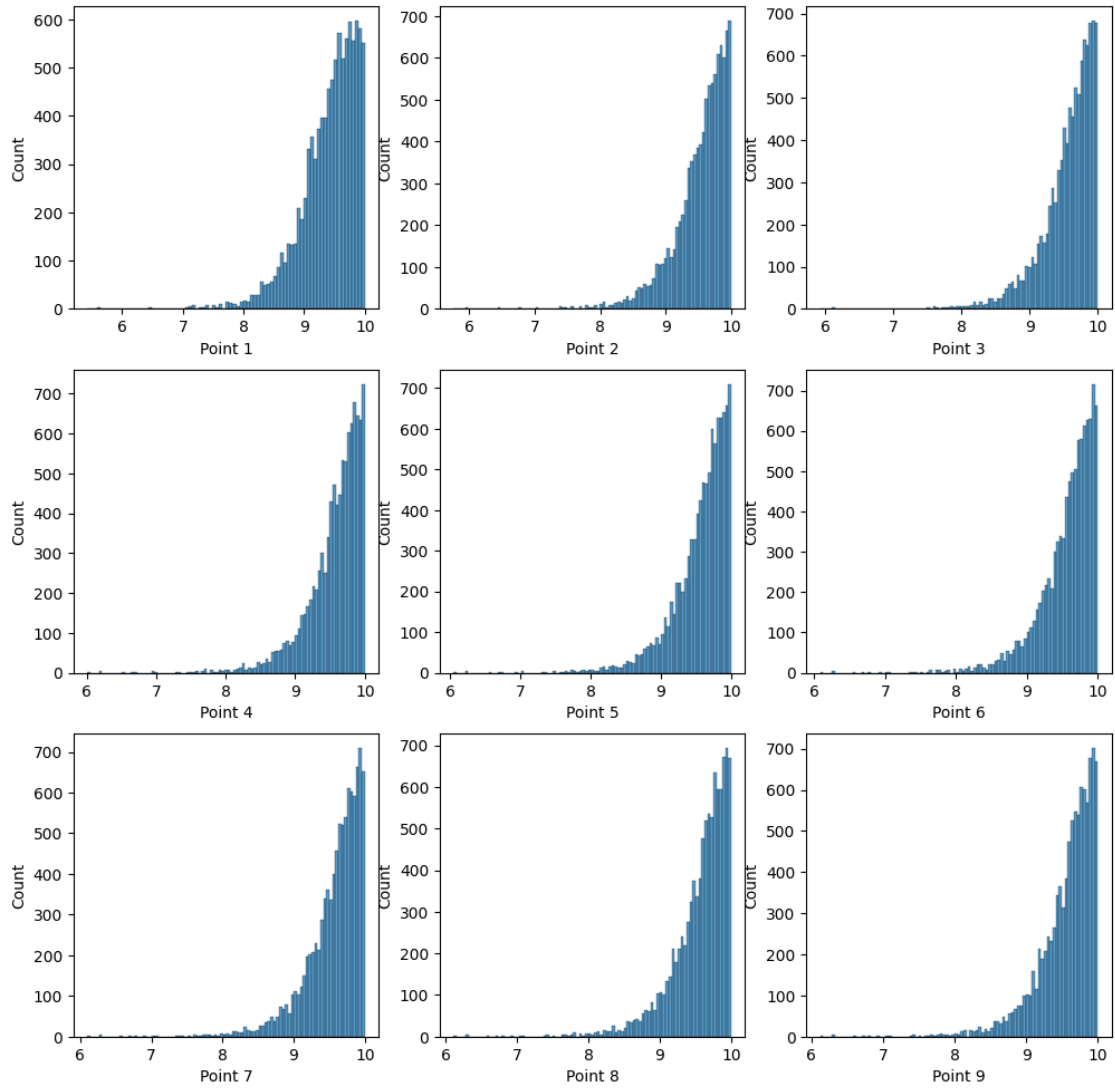

(Fig 6.9 Distribution of generated rewards for GFLowNet with compact model for larger range)

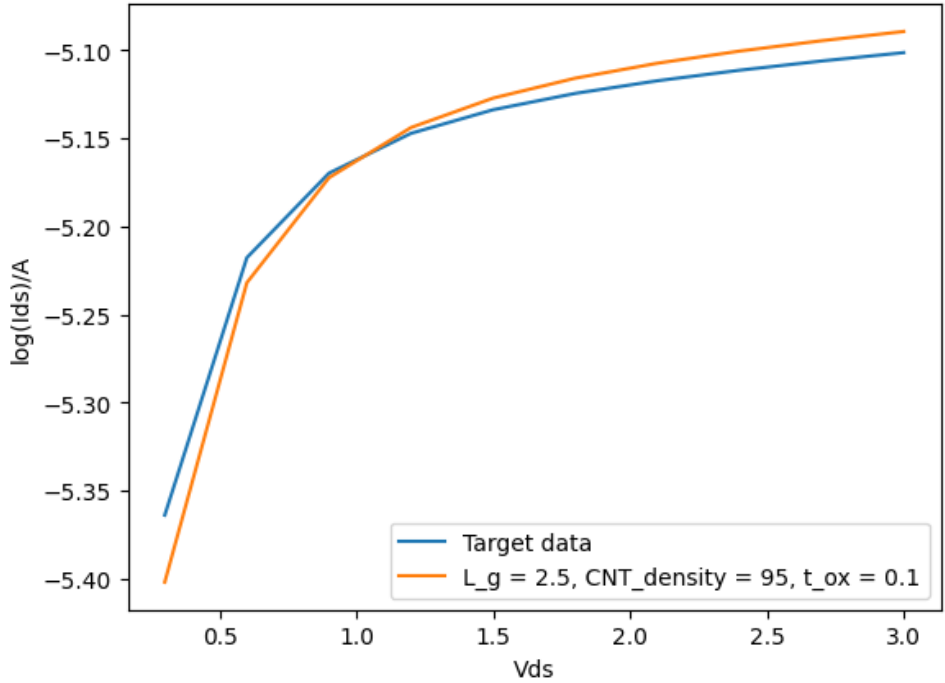

(Fig: 6.10 I-V curve generated by $L_g = 2.5, n = 95, t_{ox} = 0.1$ compared with target)

## 6.3 GFlowNet for CNTFET design incorporating processing information

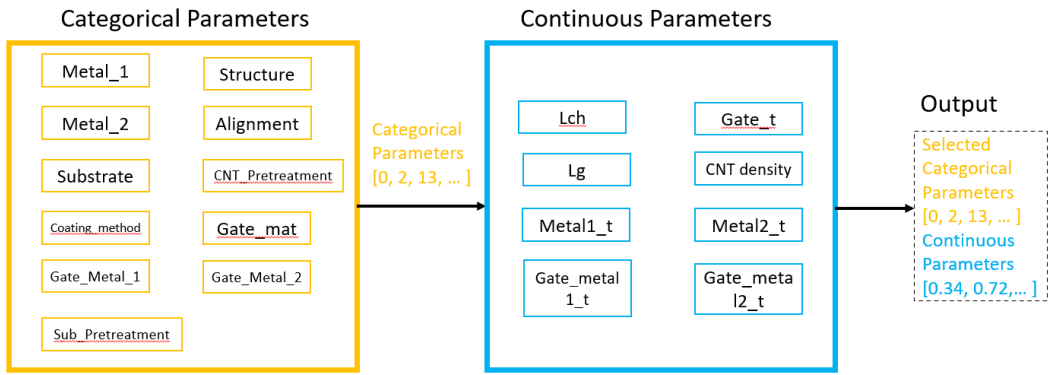

(Fig 6.11: Action space of GFlowNet with compact model)

Since processing information also affects CNTFET performance, we further designed a generative model that can generate both processing methods and device parameters. We used a stack of two GFLowNet environments, the first one to generate processing methods which are categorical data, and the other to generate continuous device parameters. When taking actions, the categorical processing methods are chosen first and then continuous parameters. We build a proxy summarizing $log(I_{ds})$ from 0.15 to 1.5V using the $log(I_{ds})$ model trained in chapter 3. To get a reasonable result, the range of continuous parameters is the range of training data for the $log(I_{ds})$ model. GFlowNet is trained by a trajectory balance model and separate NNs for $P_F$ and $P_B$ are used. As the results show, the reward distribution of the samples generated is larger than that of compact model-based distribution, probably due to the larger action space.

**Experimental setup**

Since categorical and continuous variables affect device performance, we design a stack environment that can take both categorical and continuous data. The categorical data will be selected first; then continuous data will be selected afterward. The sampled action will be a combination of categorical actions and continuous actions. We hope to create a model to generate objective performance device parameters. The range of constant parameters is chosen for the training data since I don't want unphysical conditions to occur. The reward model is built the same way as the compact model, but a ratio is applied for ease of training.

$$Reward_i = 10 * (2 - e^{k|I_{target} - I_{generated}|})$$

We used k = 0.5 during training. The target $I_{ds}$ curve was generated with the following conditions:

| Categorical parameters | substrate | Metal_1 | Metal_2 | Gate_mat | Coating_Method | structure |
|---|---|---|---|---|---|---|
| Value | SiO2 | Pd | Au | HfO2 | DLSA | 1 |
| Categorical parameters | Alignment | Pretreatment | Gate_metal_1 | Gate_metal_2 | Sub_Pretreatment | |
| Value | Aligned | YOCD | Pd | Au | None | |

(Table 6.2 Categorical values used for target generation)

| Continuous parameters | Lch | Lg | CNT_density | Metal_1_t |
|---|---|---|---|---|
| Value | 0.12 | 0.1 | 150 | 0.03 |
| Continuous parameters | Metal_2_t | Gate | Gate_metal_1_t | Gate_metal_2_t |
| Value | 0.05 | 0.0073 | 0.01 | 0.02 |

(Table 6.3 Continuous values for target generation)

**Results**

The model showed some ability to sample actions that can produce a better fitting of the target I-V curve. As is shown in the result, the rewards of point 1, point 4, point 5 and point 6 of the final generated models shows higher distribution towards maximum reward 10, which means that the generated samples have a similar production of $I_{ds}$ at these points. The sampled categorical and continuous parameters show no significant preference. Perhaps multiple combinations can be used to achieve this goal.

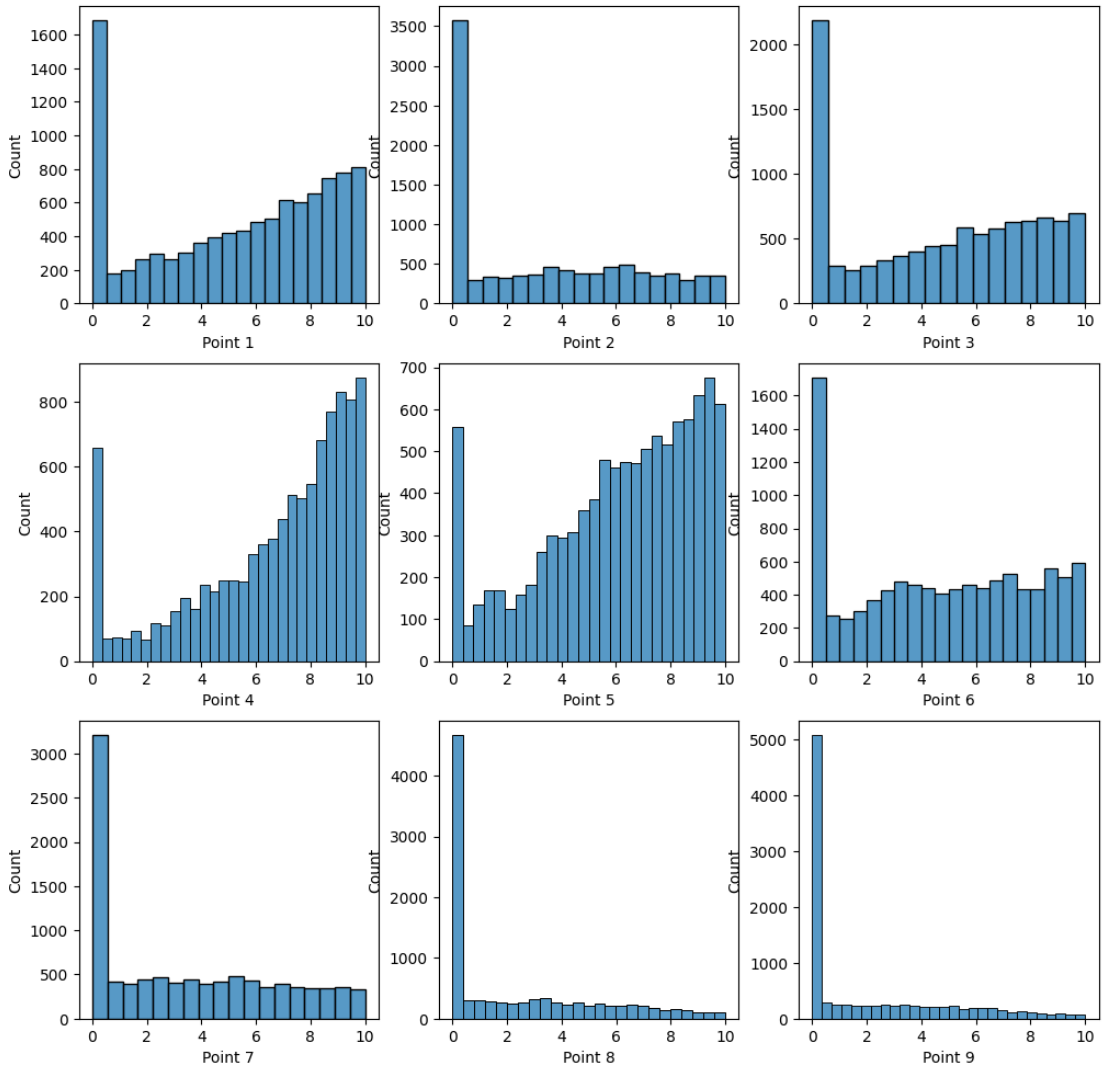

(Fig 6.12: Generate reward distribution of GFLowNet for processing information)

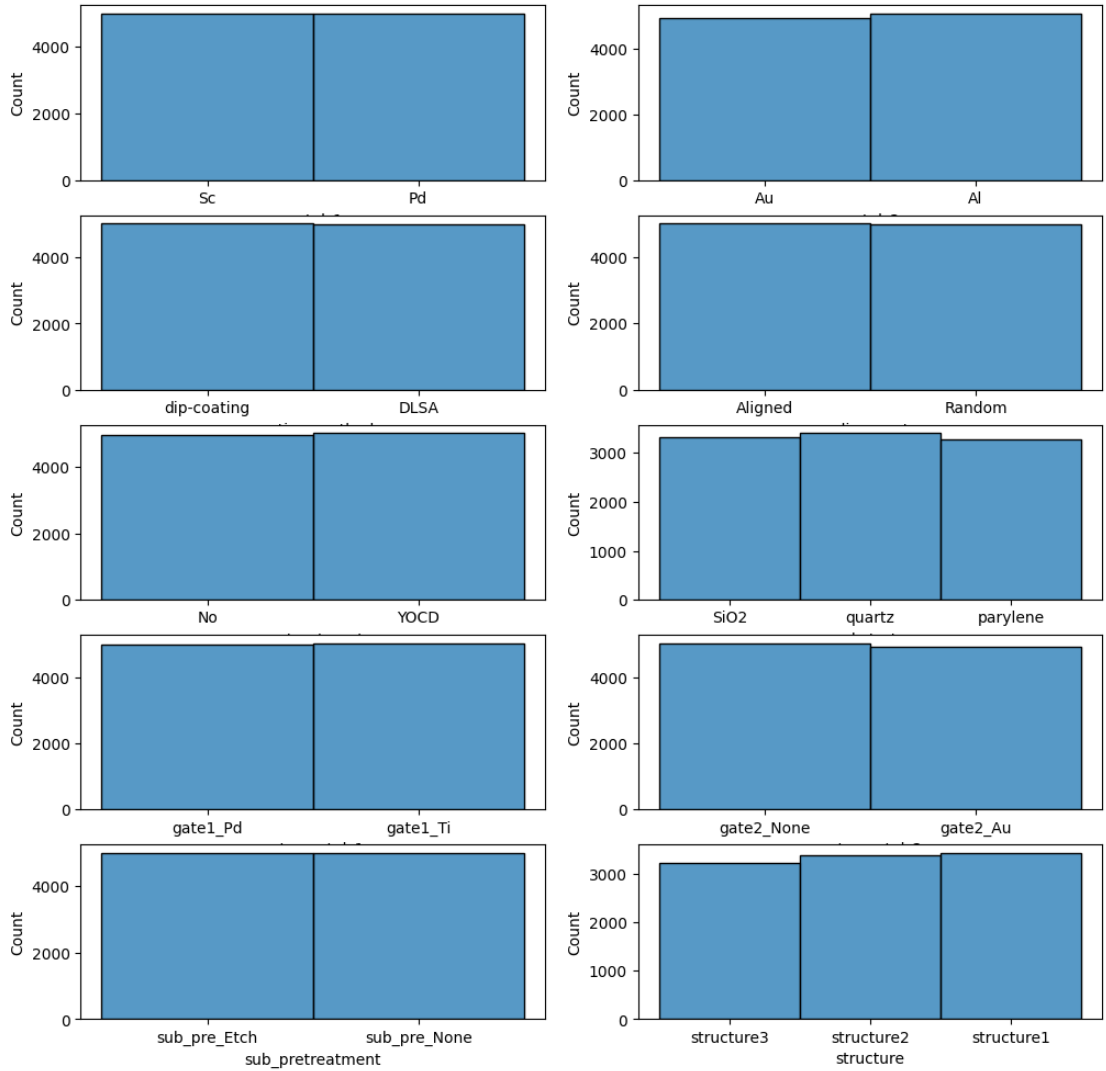

(Fig 6.13: Generate categorical parameters distribution of GFLowNet for processing information)

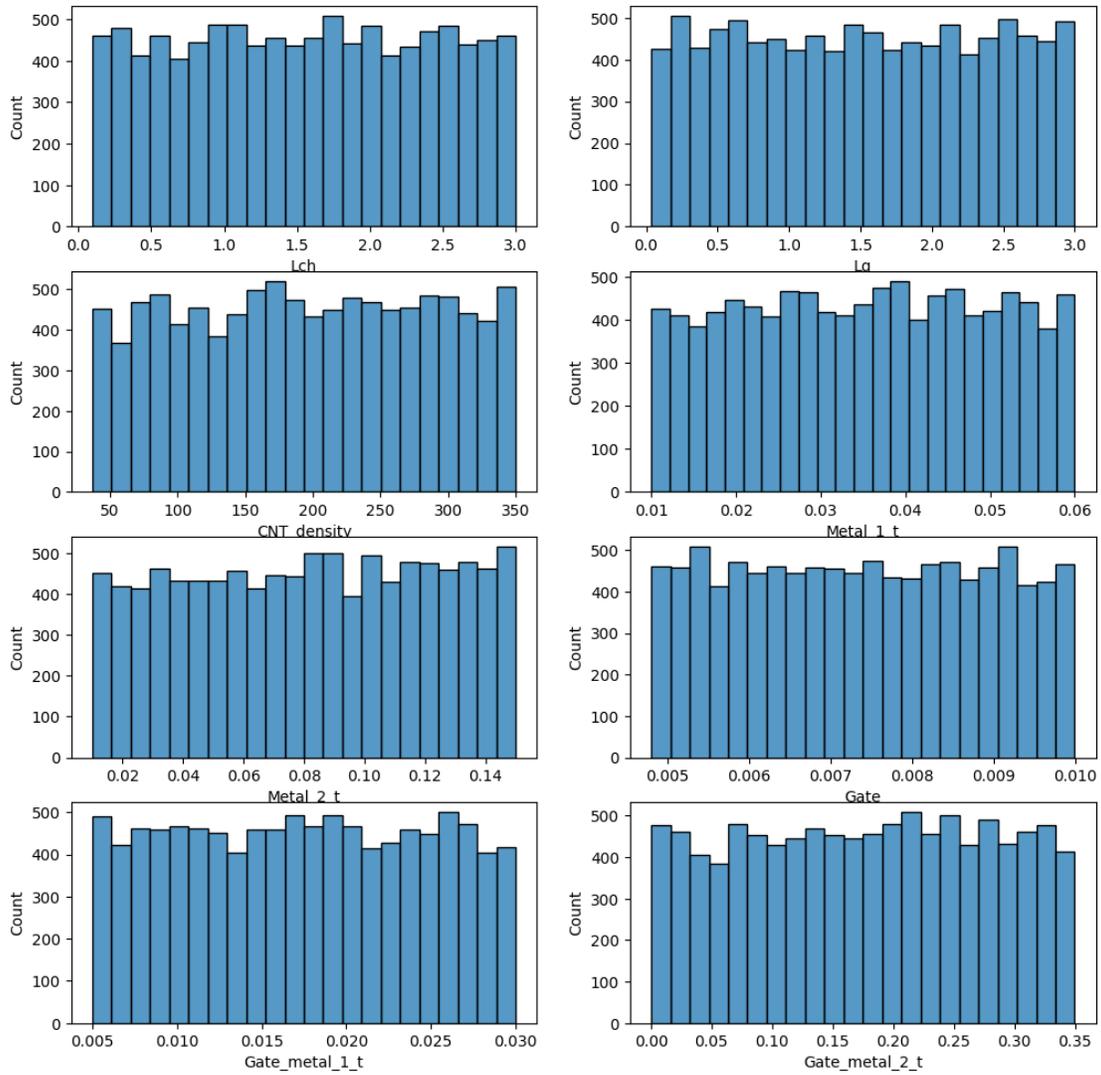

(Fig 6.14: Generate continuous parameters distribution of GFLowNet for processing information)

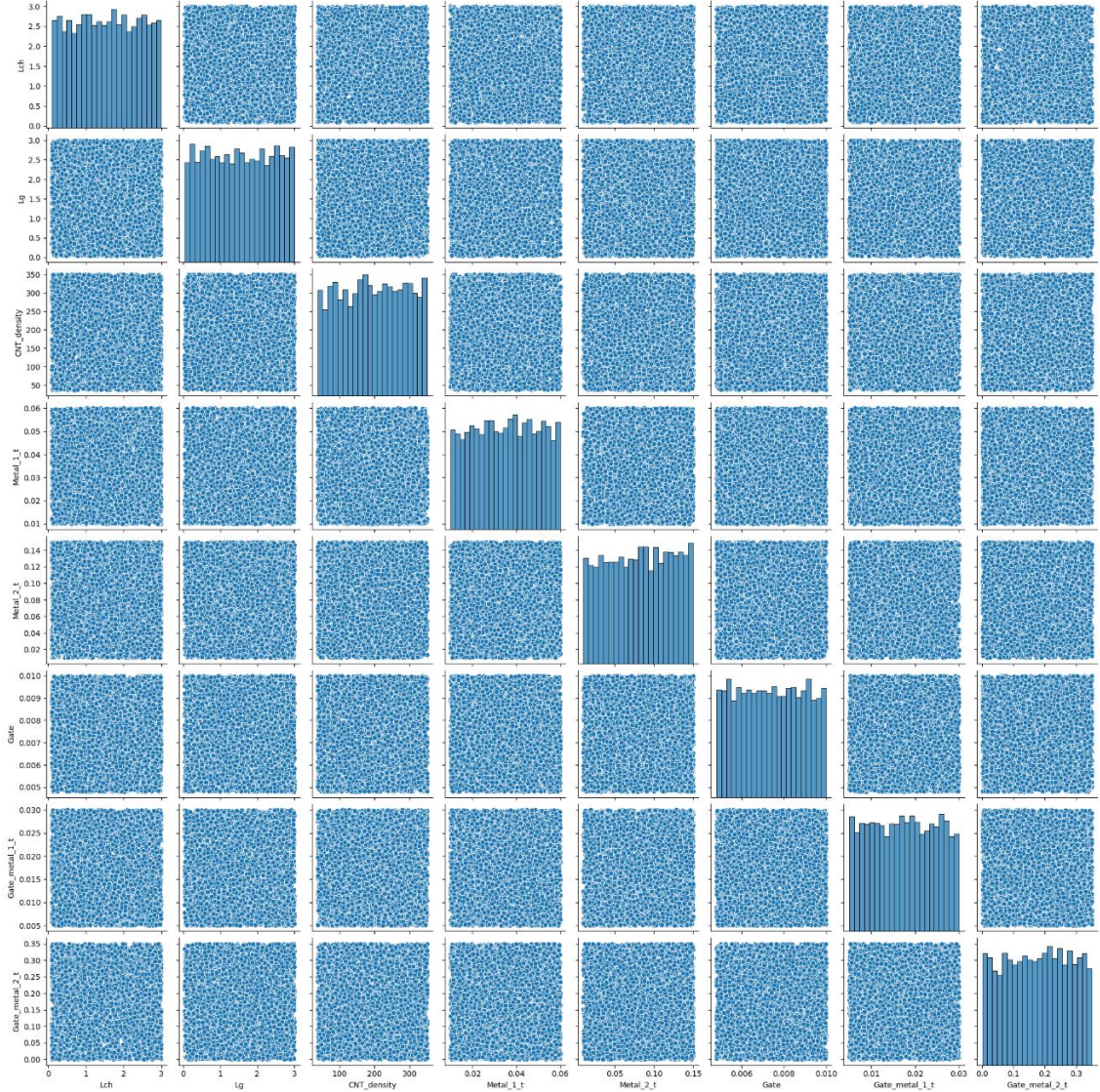

(Fig 6.15: Pair plot of continuous parameters distribution of GFLowNet for processing information)

## 6.4 Conclusion and future work

We created a generative model that can generate device parameters for a target I-V curve using GFLowNet. For the model using compact model, the model successfully generated device parameters that will lead to the target I-V curve. For the model for multiple processing information, the model only achieved some of the targets. A possible way to optimize the performance of the model for multiple processing information could be using Pareto Frontier, which focus more on the fitting of all goals.

# Appendix

Single-value GFLowNet

As a try-out in the start of research, we created a simple reward function to test the environment of GFLowNet. The reward function is a simple sum of ten $log(I_{ds})$ values from 0.3 to 3V, divided by -100 for normalization, since each value falls in [-10,0], as a test out for the ability of GFlowNet to generate current variation. We compare the device performance reward generated by GFLowNet with that of randomly generated device features. We can see that samples generated by GFLowNet have higher probability to have a higher reward, since the probability of states selected in GFLowNet is proportional to reward, so samples with higher reward will be sampled more.

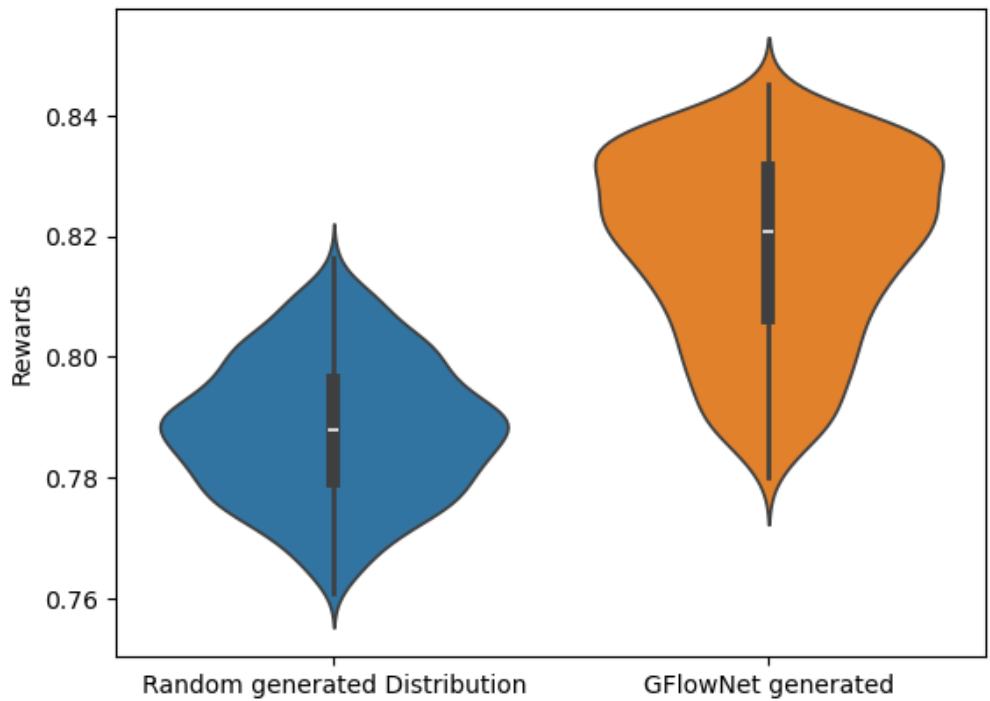

(Fig 6.16: Generated distribution of GFlowNet productions compared with Random generated distribution)

Table 6.4: Action space of GFlowNet for compact model with single value output

| States | Action space | Unit |
|---|---|---|
| Gate length | 0.1 - 3 | um |
| CNT density | 1 - 100 | 1/um |
| gate thickness | 0.004 - 1 | um |

Table 6.5: Action space of GFlowNet for processing information

| States | Action space | Unit |
|---|---|---|
| Metal_1 | Pd: 0, Sc: 1 | |
| Metal_2 | Au: 2, Al: 3 | |
| Gate Material | HfO2:4 | |
| Coating Method | DLSA:5, dip-coating: 6 | |
| Alignment | Aligned:7, Random:8 | |
| Pretreatment | No:9, YOCD:10 | |
| Substrate Material | SiO2:11, parylene:12, quartz:13 | |
| Gate_metal_1 | Pd:14, Ti:15 | |
| Gate_metal_2 | Au:16, None:17 | |
| Sub Pretreatment | Etch:18, None:19 | |
| Device Structure | Structure 1: 20, Structure 2: 21, Structure 3: 22 | |
| Channel Length | 0.08 - 0.8 | um |

| Gate Length | 0.0035 - 0.8 | um |
|---|---|---|
| CNT density | 37 - 350 | 1/um |
| Metal_1 thickness | 0.01 - 0.06 | um |
| Metal_2 thickness | 0.01 - 0.05 | um |
| Gate Thickness | 0.004 - 1 | um |
| Gate_metal_1 Thickness | 0.005 - 0.03 | um |
| Gate_metal_2 Thickness | 0 - 0.35 | um |

Table 6.6: Hyper parameter of GFLowNet for compact model

| Hyperparameters | Values |
|---|---|
| Batch size | 10 |
| GFN temperature parameter β | 15 |
| Number of training steps | 50,000 |
| Number of states embedding layers | 3 |
| Number of $P_F$, $P_B$ NN layers | 2 |
| $P_F$, $P_B$ NN embedding size | 64 |
| Learning rate for GFN's PF | $10^{-4}$ |
| Learning rate for GFN's Z-estimator | $10^{-3}$ |
| Conditioning-vector sampling distribution | w ~ Dirichlet(1) |

Table 6.7: Hyper parameter of GFLowNet for processing information

| Hyperparameters | Values |
|---|---|
| Batch size | 10 |
| GFN temperature parameter β | 15 |
| Number of training steps | 50,000 |
| Number of states embedding layers | 3 |
| Number of $P_F$, $P_B$ NN layers | 2 |
| $P_F$, $P_B$ NN embedding size | 256 |
| Learning rate for GFN's PF | $10^{-4}$ |
| Learning rate for GFN's Z-estimator | $10^{-3}$ |
| Conditioning-vector sampling distribution | $w \sim \text{Dirichlet}(1)$ |

# Chapter 7

## Conclusion and future work

The goal of this thesis is to explore how machine learning can be used in Carbon Nanotube field effect transistor research. Chapter 1 described the structure, fabrication and characterization of CNTs and summarized the current development of CNTFETs and challenges faced. Some of the challenges are hard to solve through traditional methods. In chapter 2 we described the

We have demonstrated that machine learning can be a useful tool to summarize experimental data (chapter 3), build models (chapter 4) and generate experimental conditions (chapter 5). In chapter 3, we developed neural network models for CNTFETs with one single CNT with varying gate length, contact length and. We have also developed a data cleaning method to cope with the noise in experimental observations. The model can successfully predict device performance and predict unseen cases. We further created a model that can take fabrication process into device modeling.

In chapter 4, we explored the use of simulation-based inference to extract key parameters in random CNT network conductance. We build a compact model for random CNT network FETs and use the experimentally observed device performance distribution to extract CNT conductivity, CNT-CNT junction resistance and CNT-metal resistance. We successfully produced a model that can describe the experimental observation.

In chapter 5, we developed a generative model for CNTFETs using GFlowNet structure. We designed environment, action space and proxy reward functions for a CNTFET and shows that GFlowNet can characterize device performance and generate more samples with higher reward values. We've also tried generating device processing information with target I-V curve. The categorical information generated have a good result, but that of continuous parameters needs further improvement.

**Possible future work**

The Neural Network model for CNTFETs needs further improvement. Functions can be designed to incorporate known physical equations into the neural network structure to simplify NN structure and achieve better and more stable training results. A good way may be using NN as a ratio extractor together with some basic models like modulation of Ids with Vgs.

Further work can also be done for the processing information generation of CNTFETs. The generation of categorical parameters shows a good result, but that of continuous data is far from ideal. A way to cope with it could be discretize the continuous dimension into several intervals and use these intervals as categorical parameters.

Future work can also be done on using machine learning for scientific discovery. As is pointed out in chapter 4, using only neural network to model scientific data may likely face the problem of failure in extrapolation. Also, the successful training of a neural network may likely require more data. A better way could be to find a way to incorporate existing knowledge with machine learning to produce a model that both extrapolate well and also fits the reality better. I think this could be done in the following procedure:

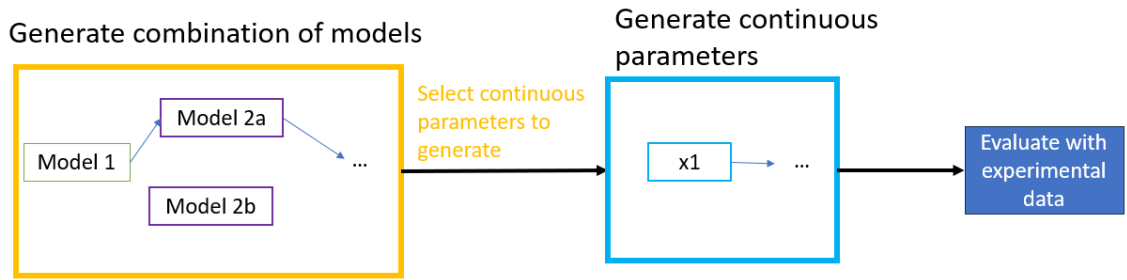

(Fig 7.1: Proposed procedure for auto-scientific discovery)

1. Design action space that contains existing physical knowledge. In each step, physical equations or hypothesis will be included for each unique physical process.
2. Let the agent take actions by choosing which physical equation to use in each step.
3. Choose parameters that is required by these physical equations. The range of parameters should be restricted so that they are reasonable physically or meet experimental observation.
4. Combine the chosen physical equations and parameters. Calculate the results with training data input. The combination of equations and parameters that fit more to reality will have a higher score. If multiple goals need to be achieved, such like the case that multiple experiments were done to justify one case, the multi-goal optimization can be used.
5. We can choose the most likely combination as our hypothesis and do further experiments to test whether the hypothesis works.

Theoretical and experimental work need to be done to justify whether this method would work. The correctness of the model produced will require both correct equations to be included in each step and enough data to train on. Problem may also occur that no model produced can fit all situations and new hypothesis or new combinations of actions might be needed. However, I think it may show some possibility for auto-science discovery and facilitate scientific research.  I hope my idea may give some inspiration to future researchers.

# Bibliography

Reference

[1] Wang, Hanchen, et al. "Scientific discovery in the age of artificial intelligence." *Nature* 620.7972 (2023): 47-60.

[2] Kitano, Hiroaki. "Artificial intelligence to win the nobel prize and beyond: Creating the engine for scientific discovery." *AI magazine* 37.1 (2016): 39-49.

[3] Jain, Moksh, et al. "Gflownets for ai-driven scientific discovery." *Digital Discovery* 2.3 (2023): 557-577.

[4] Lu, Chris, et al. "The ai scientist: Towards fully automated open-ended scientific discovery." *arXiv preprint arXiv:2408.06292* (2024).

[5] Jumper, John, et al. "Highly accurate protein structure prediction with AlphaFold." *nature* 596.7873 (2021): 583-589.

[6] Varadi, Mihaly, et al. "AlphaFold Protein Structure Database: massively expanding the structural coverage of protein-sequence space with high-accuracy models." *Nucleic acids research* 50.D1 (2022): D439-D444.

[7] Mak, Kit-Kay, Yi-Hang Wong, and Mallikarjuna Rao Pichika. "Artificial intelligence in drug discovery and development." *Drug discovery and evaluation: safety and pharmacokinetic assays* (2024): 1461-1498.

[8] Biamonte, Jacob, et al. "Quantum machine learning." *Nature* 549.7671 (2017): 195-202.

[9] Iijima, Sumio. "Helical microtubules of graphitic carbon." *nature* 354.6348 (1991): 56.

[10] Ebbesen, T. W., et al. "Electrical conductivity of individual carbon nanotubes." *Nature* 382.6586 (1996): 54-56.

[11]Tans, Sander J., Alwin RM Verschueren, and Cees Dekker. "Room-temperature transistor based on a single carbon nanotube." *Nature* 393.6680 (1998): 49-52.

[12]Franklin, Aaron D., et al. "Sub-10 nm carbon nanotube transistor." *Nano letters* 12.2 (2012): 758-762.

[13]Franklin, Aaron D. "The road to carbon nanotube transistors." *Nature* 498.7455 (2013): 443-444.

[14]Javey, Ali, et al. "Ballistic carbon nanotube field-effect transistors." *nature* 424.6949 (2003): 654-657.

[15] Cao, Qing, et al. "Carbon nanotube transistors scaled to a 40-nanometer footprint." *Science* 356.6345 (2017): 1369-1372.

[16] Cao, Qing. "Carbon nanotube transistor technology for More-Moore scaling." *Nano Research* 14.9 (2021): 3051-3069.

[17] Avouris, Phaedon, et al. "Carbon nanotube electronics." *Proceedings of the IEEE* 91.11 (2003): 1772-1784.

[18] Bachtold, Adrian, et al. "Logic circuits with carbon nanotube transistors." *Science* 294.5545 (2001): 1317-1320.

[19] Shulaker, Max M., et al. "Carbon nanotube computer." *Nature* 501.7468 (2013): 526-530.

[20] Javey, Ali, et al. "High-κ dielectrics for advanced carbon-nanotube transistors and logic gates." *Nature materials* 1.4 (2002): 241-246.

[21] Avouris, Ph, et al. "Carbon nanotube transistors and logic circuits." *Physica B: Condensed Matter* 323.1-4 (2002): 6-14.

[22] Li, Shengdong, et al. "Carbon nanotube transistor operation at 2.6 GHz." *Nano Letters* 4.4 (2004): 753-756.

[23] Zhong, Donglai, Zhiyong Zhang, and Lian-Mao Peng. "Carbon nanotube radio-frequency electronics." *Nanotechnology* 28.21 (2017): 212001.

[24] Close, Gael F., et al. "A 1 GHz integrated circuit with carbon nanotube interconnects and silicon transistors." *Nano Letters* 8.2 (2008): 706-709.

[25] Zhong, Donglai, et al. "Gigahertz integrated circuits based on carbon nanotube films." *Nature Electronics* 1.1 (2018): 40-45.

[26] Chimot, N., et al. "Gigahertz frequency flexible carbon nanotube transistors." *Applied physics letters* 91.15 (2007).

[27] Wilder, Jeroen WG, et al. "Electronic structure of atomically resolved carbon nanotubes." *Nature* 391.6662 (1998): 59.

[28] Martel, Richard, et al. "Single-and multi-wall carbon nanotube field-effect transistors." *Applied physics letters* 73.17 (1998): 2447-2449.

[29] Cheung, William, et al. "DNA and carbon nanotubes as medicine." *Advanced drug delivery reviews* 62.6 (2010): 633-649.

[30] Dresselhaus, Mildred S., et al. "Raman spectroscopy of carbon nanotubes." *Physics reports* 409.2 (2005): 47-99.

[31] Saito, R., et al. "Raman spectroscopy of graphene and carbon nanotubes." *Advances in Physics* 60.3 (2011): 413-550.

[32] Jorio, A., and RJJoAP Saito. "Raman spectroscopy for carbon nanotube applications." *Journal of Applied Physics* 129.2 (2021).

[33] Rao, A. M., et al. "Diameter-selective Raman scattering from vibrational modes in carbon nanotubes." Science 275.5297 (1997): 187-191.

[34] Duan, Wen Hui, Quan Wang, and Frank Collins. "Dispersion of carbon nanotubes with SDS surfactants: a study from a binding energy perspective." *Chemical Science* 2.7 (2011): 1407-1413.

[35] Tummala, Naga Rajesh. "SDS surfactants on carbon nanotubes: aggregate morphology." *ACS nano* 3.3 (2009): 595-602.

[36] Liu, Fang, et al. "Comparative study of the extraction selectivity of PFO-BPy and PCz for small to large diameter single-walled carbon nanotubes." *Nano Research* 15.9 (2022): 8479-8485.

[37] Arnold, Michael S., et al. "Sorting carbon nanotubes by electronic structure using density differentiation." *Nature nanotechnology* 1.1 (2006): 60-65.

[38] Rastogi, Richa, et al. "Comparative study of carbon nanotube dispersion using surfactants." *Journal of colloid and interface science* 328.2 (2008): 421-428.

[39] Ma, Ze, et al. "Improving the performance and uniformity of carbon-nanotube-network-based photodiodes via yttrium oxide coating and decoating." *ACS applied materials & interfaces* 11.12 (2019): 11736-11742.

[40] Li, Xiaolin, et al. "Langmuir− Blodgett assembly of densely aligned single-walled carbon nanotubes from bulk materials." *Journal of the American Chemical Society* 129.16 (2007): 4890-4891.

[41] Krstic, Vojislav, et al. "Langmuir− Blodgett films of matrix-diluted single-walled carbon nanotubes." *Chemistry of materials* 10.9 (1998): 2338-2340.

[42] Chao, Tzu-Ang, et al. "Small Molecule Additives to Suppress Bundling in Dimensional-Limited Self-Alignment Method for High-Density Aligned Carbon Nanotube Array." *Advanced Materials Interfaces* 11.6 (2024): 2300684.

[43] Zhou, Jianshuo, et al. "Carbon nanotube based radio frequency transistors for K-band amplifiers." *ACS Applied Materials & Interfaces* 13.31 (2021): 37475-37482.

[44] Rinkiö, Marcus, et al. "Effect of humidity on the hysteresis of single walled carbon nanotube field-effect transistors." physica status solidi (b) 245.10 (2008): 2315-2318.

[45]Svensson, Johannes, and Eleanor EB Campbell. "Schottky barriers in carbon nanotube-metal contacts." *Journal of applied physics* 110.11 (2011).

[46] Chen, Zhihong, et al. "The role of metal− nanotube contact in the performance of carbon nanotube field-effect transistors." *Nano letters* 5.7 (2005): 1497-1502.

[47] Lim, Seong Chu, et al. "Contact resistance between metal and carbon nanotube interconnects: Effect of work function and wettability." *Applied Physics Letters* 95.26 (2009).

[48] Goodfellow, Ian. "Deep learning." (2016).

[49] Amari, Shun-ichi. "Backpropagation and stochastic gradient descent method." *Neurocomputing* 5.4-5 (1993): 185-196.

[50] Kingma, Diederik P. "Adam: A method for stochastic optimization." *arXiv preprint arXiv:1412.6980* (2014).

[51] Prince, Simon JD. *Understanding deep learning*. MIT press, 2023.

[52] P. J. Diggle, R. J. Gratton, Monte Carlo methods of inference for implicit statisticalmodels. J. R. Stat. Soc. Ser. B 46, 193–212 (1984).

[53] Cranmer, Kyle, Johann Brehmer, and Gilles Louppe. "The frontier of simulation-based inference." *Proceedings of the National Academy of Sciences* 117.48 (2020): 30055-30062.

[54]Lueckmann, Jan-Matthis, et al. "Benchmarking simulation-based inference." *International conference on artificial intelligence and statistics*. PMLR, 2021.

[55]Glöckler, Manuel, Michael Deistler, and Jakob H. Macke. "Variational methods for simulation-based inference." *arXiv preprint arXiv:2203.04176* (2022).

[56] Deistler, Michael, Pedro J. Goncalves, and Jakob H. Macke. "Truncated proposals for scalable and hassle-free simulation-based inference." *Advances in Neural Information Processing Systems* 35 (2022): 23135-23149.

[57]Papamakarios, George, David Sterratt, and Iain Murray. "Sequential neural likelihood: Fast likelihood-free inference with autoregressive flows." *The 22nd international conference on artificial intelligence and statistics*. PMLR, 2019.

[58] Vaswani, A. "Attention is all you need." *Advances in Neural Information Processing Systems* (2017).

[59] Goodfellow, Ian, et al. "Generative adversarial networks." *Communications of the ACM* 63.11 (2020): 139-144.

[60 ]Bengio, Yoshua, et al. "Gflownet foundations." *The Journal of Machine Learning Research* 24.1 (2023): 10006-10060.

[61] Bengio, Emmanuel, et al. "Flow network based generative models for non-iterative diverse candidate generation." *Advances in Neural Information Processing Systems* 34 (2021): 27381-27394.

[62] Shen, Max W., et al. "Towards understanding and improving gflownet training." *International Conference on Machine Learning*. PMLR, 2023.

[63]Roy, Julien, et al. "Goal-conditioned gflownets for controllable multi-objective molecular design." arXiv preprint arXiv:2306.04620 (2023).

[64] Malkin, Nikolay, et al. "Trajectory balance: Improved credit assignment in gflownets." Advances in Neural Information Processing Systems 35 (2022): 5955-5967.

[65] AI4Science, Mila, et al. "Crystal-gfn: sampling crystals with desirable properties and constraints." *arXiv preprint arXiv:2310.04925* (2023).

[66] Volokhova, Alexandra, et al. "Towards equilibrium molecular conformation generation with GFlowNets." *Digital Discovery* 3.5 (2024): 1038-1047.

[67] Liu, Lijun, et al. "Aligned, high-density semiconducting carbon nanotube arrays for high-performance electronics." *Science* 368.6493 (2020): 850-856.

[68] Long, Guanhua, et al. "Carbon nanotube-based flexible high-speed circuits with sub-nanosecond stage delays." *Nature Communications* 13.1 (2022): 6734.

[69] Liu, Chenchen, et al. "Complementary transistors based on aligned semiconducting carbon nanotube arrays." *ACS nano* 16.12 (2022): 21482-21490.

[70] Lin, Yanxia, et al. "Enhancement-mode field-effect transistors and high-speed integrated circuits based on aligned carbon nanotube films." *Advanced Functional Materials* 32.11 (2022): 2104539.

[71] Zhong, Donglai, et al. "Gigahertz integrated circuits based on carbon nanotube films." *Nature Electronics* 1.1 (2018): 40-45.

[72] Lin, Yanxia, et al. "Improving the performance of aligned carbon nanotube-based transistors by refreshing the substrate surface." *ACS Applied Materials & Interfaces* 15.8 (2023): 10830-10837.

[73] Shi, Huiwen, et al. "Radiofrequency transistors based on aligned carbon nanotube arrays." *Nature Electronics* 4.6 (2021): 405-415.

[74] Zhang, Zhiyong, et al. "Terahertz metal-oxide-semiconductor transistors based on aligned carbon nanotube arrays." (2023).

[75] Janas, Dawid. "Towards monochiral carbon nanotubes: A review of progress in the sorting of single-walled carbon nanotubes." *Materials Chemistry Frontiers* 2.1 (2018): 36-63.

[76] Mesgari, Sara, et al. "Gel electrophoresis using a selective radical for the separation of single-walled carbon nanotubes." *Faraday Discussions* 173 (2014): 351-363.

[77] N. A. Rice, W. J. Bodnaryk, B. Mirka, O. A. Melville, A. Adronov, and B. H. Lessard, "Polycarbazole-sorted semiconducting single-walled carbon nanotubes for incorporation into organic thin film transistors," *Adv. Electron. Mater.*, vol. 5, no. 1, Jan. 2019, Art. no. 1800539, doi: 10.1002/aelm.201800539.

[78] Buldum, Alper, and Jian Ping Lu. "Contact resistance between carbon nanotubes." *Physical Review B* 63.16 (2001): 161403.

[79] Kumar, S., J. Y. Murthy, and M. A. Alam. "Percolating conduction in finite nanotube networks." *Physical review letters* 95.6 (2005): 066802.

[80] Tripathy, Srijeet, et al. "Resistive analysis of scattering-dependent electrical transport in single-wall carbon-nanotube networks." *IEEE Transactions on Electron Devices* 67.12 (2020): 5676-5684.

[81] Zorn, Nicolas F., and Jana Zaumseil. "Charge transport in semiconducting carbon nanotube networks." *Applied Physics Reviews* 8.4 (2021).

[82] Cranmer, Kyle, Johann Brehmer, and Gilles Louppe. "The frontier of simulation-based inference." *Proceedings of the National Academy of Sciences* 117.48 (2020): 30055-30062.

[83] Tejero-Cantero, Alvaro, et al. "SBI--A toolkit for simulation-based inference." *arXiv preprint arXiv:2007.09114* (2020).

[84] Papamakarios, George, and Iain Murray. "Fast ε-free inference of simulation models with bayesian conditional density estimation." *Advances in neural information processing systems* 29 (2016).

[85] Lueckmann, Jan-Matthis, et al. "Flexible statistical inference for mechanistic models of neural dynamics." *Advances in neural information processing systems* 30 (2017).

[86] Bauhofer, Wolfgang, and Josef Z. Kovacs. "A review and analysis of electrical percolation in carbon nanotube polymer composites." *Composites science and technology* 69.10 (2009): 1486-1498.

[87] Hu, Lea, D. S. Hecht, and G. Grüner. "Percolation in transparent and conducting carbon nanotube networks." *Nano letters* 4.12 (2004): 2513-2517.

[88] P. N. Nirmalraj, P. E. Lyons, S. De, J. N. Coleman, and J. J. Boland, Nano Lett. 9(11), 3890 (2009).

[89] Deng, Jie, and H-S. Philip Wong. "A compact SPICE model for carbon-nanotube field-effect transistors including nonidealities and its application—Part I: Model of the intrinsic channel region." *IEEE Transactions on Electron Devices* 54.12 (2007): 3186-3194.

[90] Deng, Jie, and H-S. Philip Wong. "A compact SPICE model for carbon-nanotube field-effect transistors including nonidealities and its application—Part II: Full device model and circuit performance benchmarking." *IEEE Transactions on Electron Devices* 54.12 (2007): 3195-3205.

[91] Lee, C-S., et al. "A compact virtual-source model for carbon nanotube FETs in the sub-10-nm regime—Part I: Intrinsic elements." *IEEE transactions on electron devices* 62.9 (2015): 3061-3069.

[92] Lee, Chi-Shuen, et al. "A compact virtual-source model for carbon nanotube FETs in the sub-10-nm regime—Part II: Extrinsic elements, performance assessment, and design optimization." *IEEE Transactions on Electron Devices* 62.9 (2015): 3070-3078.

[93] Diggle PJ, Gratton RJ (1984) Monte Carlo Methods of Inference for Implicit Statistical Models in Journal of the Royal Statistical Society: Series B (Methodological). Vol. 46, pp. 193–212.

[94] Bishop, Mindy D., et al. "Fabrication of carbon nanotube field-effect transistors in commercial silicon manufacturing facilities." *Nature Electronics* 3.8 (2020): 492-501.

[95] Franklin, Aaron D., Damon B. Farmer, and Wilfried Haensch. "Defining and overcoming the contact resistance challenge in scaled carbon nanotube transistors." *ACS nano* 8.7 (2014): 7333-7339.

[96] A. Gajare et al., "CircuitScribe: Block Diagram based Circuit Simulation Application," in Proc. 2021 IEEE Mysore Sub Section International Conference (MysuruCon), IEEE, 2021.

[97] S. Popov and N. Hinov, "Comparative Analysis of Software Environments for Computer Modeling in Electrical and Electronic Engineering," in Proc. 2023 International Conference on Information Technologies (InfoTech), IEEE, 2023.

[98] V. K. Sangwan, A. Behnam, V. W. Ballarotto, M. S. Fuhrer, A. Ural, and E. D. Williams, "Optimizing transistor performance of percolating carbon nanotube networks," *Appl. Phys. Lett.*, vol. 97, no. 4, p. 43111, Jul. 2010, doi: 10.1063/1.3469930.

[99] Introduction to Random Variables by Nathaniel E. Helwig

[100] Carbon nanotube transistor technology for More-Moore scaling

[101] Meyerson, B. Innovation: The future of silicon technology. In Semico IMPACT Conference, Scottsdale, AZ, USA, 2004.

[102] Liu, Y.; Luisier, M.; Majumdar, A.; Antoniadis, D. A.; Lundstrom, M. S. On the interpretation of ballistic injection velocity in deeply scaled

[103] Ferain, I.; Colinge, C. A.; Colinge, J. P. Multigate transistors as the future of classical metal–oxide–semiconductor field-effect transistors.

[104] Takagi, S.; Koga, J.; Toriumi, A. Subband structure engineering for performance enhancement of Si MOSFETs. In International Electron Devices Meeting. IEDM Technical Digest, Washington, DC, USA, 1997, pp 219–222.

[105] Nature 2011, 479, 310–316. MOSFETs. IEEE Trans. Electron Devices 2012, 59, 994–1001.

[106] Cao, Q.; Han, S. J.; Tersoff, J.; Franklin, A. D.; Zhu, Y.; Zhang, Z.; Tulevski, G. S.; Tang, J. S.; Haensch, W. End-bonded contacts for carbon nanotube transistors with low, size-independent resistance. Science 2015, 350, 68–72.

[107] Anantram, M. P., and F. Leonard. "Physics of carbon nanotube electronic devices." *Reports on progress in physics* 69.3 (2006): 507.

[108] Farmer, Damon B. "Metallization considerations for carbon nanotube device optimization." Journal of Applied Physics 132.10 (2022).

[109] Wahab, Muhammad Abdul. "Interpolation and extrapolation." *Proc. Topics Syst. Eng. Winter Term* 17 (2017): 1-6.

[110] Steven C. Chapra, "Applied Numerical Methods with MATLAB® for Engineers and Scientists",3rd Edition, New York: McGraw-Hill, 2002, ch.17-18

[111] Sanchez, A. OrtiZ-Conde FJ Garcia, P. E. Schmidt, and A. Sa-Noto. "ON THE CHARGE-SHEET MODEL OF THE THIN-FILM MOSfET."